\documentclass[a4paper,11pt]{article}
\pdfoutput=1

\usepackage{jheppub}

\usepackage{preprintcover}
\PreprintCoverPaperTitle{Search for top squark pair production in final states with one isolated
lepton, jets, and missing transverse momentum in $\sqrt{s}=8$\,\TeV\  $pp$
collisions with the ATLAS detector}
\PreprintIdNumber{CERN-PH-EP-2014-143}
\PreprintCoverAbstract{The results of a search 
for top squark (stop) pair production in final states with one isolated lepton, jets, and missing transverse
momentum are reported.
The analysis is performed with proton--proton collision data at $\sqrt{s} = 8$\,\TeV\ collected with the 
ATLAS detector at the LHC in 2012 corresponding to an integrated luminosity of $20$\,\ifb.
The lightest supersymmetric particle (LSP) is taken to be the lightest neutralino
which only interacts weakly and is assumed to be stable. 
The stop decay modes considered are those to a top quark and the LSP
as well as
to a bottom quark and the lightest chargino, where the chargino decays to the LSP by emitting
a $W$ boson.
A wide range of scenarios with different mass splittings between the stop, the lightest neutralino and 
the lightest chargino are considered, including cases where the $W$ bosons or the top quarks are off-shell.
Decay modes involving the heavier charginos and neutralinos are addressed using
a set of phenomenological models of supersymmetry.
No significant excess over the Standard Model prediction is observed.
A stop with a mass between $210$ and $640$\,\GeV\ decaying directly to a top quark and a massless LSP
is excluded at $95$\% confidence level, and in models where the mass of the lightest chargino is twice that of the LSP, 
stops are excluded at $95$\% confidence level up to a mass of $500$\,\GeV\ for an LSP mass in the range of
$100$ to $150$\,\GeV.
Stringent exclusion limits are also derived for all other stop decay modes considered,
and model-independent upper limits are set on the visible cross-section for processes beyond the 
Standard Model.}
\PreprintJournalName{JHEP}

\abstract{The results of a search 
for top squark (stop) pair production in final states with one isolated lepton, jets, and missing transverse
momentum are reported.
The analysis is performed with proton--proton collision data at $\sqrt{s} = 8$\,\TeV\ collected with the 
ATLAS detector at the LHC in 2012 corresponding to an integrated luminosity of $20$\,\ifb.
The lightest supersymmetric particle (LSP) is taken to be the lightest neutralino
which only interacts weakly and is assumed to be stable. 
The stop decay modes considered are those to a top quark and the LSP
as well as
to a bottom quark and the lightest chargino, where the chargino decays to the LSP by emitting
a $W$ boson.
A wide range of scenarios with different mass splittings between the stop, the lightest neutralino and 
the lightest chargino are considered, including cases where the $W$ bosons or the top quarks are off-shell.
Decay modes involving the heavier charginos and neutralinos are addressed using
a set of phenomenological models of supersymmetry.
No significant excess over the Standard Model prediction is observed.
A stop with a mass between $210$ and $640$\,\GeV\ decaying directly to a top quark and a massless LSP
is excluded at $95$\% confidence level, and in models where the mass of the lightest chargino is twice that of the LSP, 
stops are excluded at $95$\% confidence level up to a mass of $500$\,\GeV\ for an LSP mass in the range of
$100$ to $150$\,\GeV.
Stringent exclusion limits are also derived for all other stop decay modes considered,
and model-independent upper limits are set on the visible cross-section for processes beyond the 
Standard Model.}

\usepackage{atlasphysics}
\usepackage{placeins}
\usepackage{subfigure}
\usepackage{engord}
\usepackage{amssymb,amsmath}
\usepackage{bm}
\usepackage{verbatim}
\usepackage{units}
\usepackage{graphicx}
\usepackage{multirow}
\usepackage{pdflscape}
\usepackage{hyperref}
\usepackage{pifont}
\usepackage{afterpage}

\input{boostedsymbols.tex}

\newcommand{\br}{\ensuremath{\mathcal{BR}}}
\newcommand{\Wjets}{$W$+jets}
\newcommand{\mvetmiss}{\ensuremath{\vec{p}_{\rm T}^{\,\rm miss}}} %

\def\Ptmiss{\mvetmiss}             
\newcommand{\tone}{\ensuremath{{\tilde{t}^{}_{1}}}}
\newcommand{\ttwo}{\ensuremath{{\tilde{t}^{}_{2}}}}
\newcommand{\tleft}{\ensuremath{\tilde{t}^{}_{\rm L}}}
\newcommand{\tright}{\ensuremath{\tilde{t}^{}_{\rm R}}}
\newcommand{\geant}{{\sc GEANT4}}

\newcommand{\acermc}{{\sc ACERMC}}
\newcommand{\acermcv}{{\sc ACERMC-3.8}}
\newcommand{\madgraph}{{\sc MadGraph}}
\newcommand{\madgraphv}{{\sc MadGraph-5.1.4.8}}
\newcommand{\powheg}{{\sc POWHEG}}
\newcommand{\powhegv}{{\sc POWHEG-r2129}}
\newcommand{\herwig}{{\sc Herwig}}

\newcommand{\herwigpp}{{\sc Herwig++}}
\newcommand{\herwigppv}{{\sc Herwig++~2.5.2}}
\newcommand{\pythia}{{\sc PYTHIA}}
\newcommand{\pythiav}{{\sc PYTHIA-6.426}}
\newcommand{\pythiaEightv}{{\sc PYTHIA-8.160}}

\newcommand{\jimmy}{{\sc Jimmy}}

\newcommand{\sherpa}{{\sc SHERPA}}
\newcommand{\sherpav}{{\sc SHERPA-1.4.1}}

\newcommand{\figref}[1]{figure~\ref{#1}}
\newcommand{\figsref}[1]{figures~\ref{#1}}
\newcommand{\Figref}[1]{Figure~\ref{#1}}
\newcommand{\Figsref}[1]{Figures~\ref{#1}}
\newcommand{\secref}[1]{section~\ref{#1}}
\newcommand{\secsref}[1]{sections~\ref{#1}}

\newcommand{\tabref}[1]{table~\ref{#1}}
\newcommand{\tabsref}[1]{tables~\ref{#1}}
\newcommand{\Tabref}[1]{Table~\ref{#1}}

\newcommand{\mt}{\ensuremath{m_\mathrm{T}}}
\newcommand\mtTwo{\ensuremath{m_{\mathrm{T}2}}}
\newcommand\amtTwo{\ensuremath{am_{\mathrm{T}2}}}
\newcommand\mtTwoTau{\ensuremath{m_{\mathrm{T}2}^\tau}}

\newcommand\dR{\Delta R}

\def\tauh{\ensuremath{\tau_{\mathrm{had}}}}
\def\metsig{\ensuremath{E_\mathrm{T}^{\mathrm{miss}}/\sqrt{ \HT }}}

\def\HTmissSig{\ensuremath{H_\mathrm{T,sig}^{\mathrm{miss}}}}
\def\mtophad{\ensuremath{m_\mathrm{had-top}}}

\def\HTsig{\ensuremath{H_{\rm T}^{\rm sig}}}
\def\meff{\ensuremath{m_{\rm eff}}}
\def\HTtwo{\ensuremath{H_{{\rm T},2}}}

\def\topness{\textit{topness}}

\def\CLs{\ensuremath{\mathrm{CL}_s}}

\def\bChargino{\ensuremath{\tone \to b \chinoonepm}}
\def\topLSP{\ensuremath{\tone \to t \ninoone}}
\def\threeBody{\ensuremath{\tone \to b W \ninoone}}
\def\fourBody{\ensuremath{\tone \to b f f' \ninoone}}
\def\charmDecay{\ensuremath{\tone \to c \ninoone}}

\def\ninoOneTwoThreeFour{\ensuremath{\mathchoice%
      {\displaystyle\raise.4ex\hbox{$\displaystyle\tilde\chi^0_{1,2,3,4}$}}%
         {\textstyle\raise.4ex\hbox{$\textstyle\tilde\chi^0_{1,2,3,4}$}}%
       {\scriptstyle\raise.3ex\hbox{$\scriptstyle\tilde\chi^0_{1,2,3,4}$}}%
 {\scriptscriptstyle\raise.3ex\hbox{$\scriptscriptstyle\tilde\chi^0_{1,2,3,4}$}}}}
\def\chinoOneTwopm{\ensuremath{\mathchoice%
      {\displaystyle\raise.4ex\hbox{$\displaystyle\tilde\chi^\pm_{1,2}$}}%
         {\textstyle\raise.4ex\hbox{$\textstyle\tilde\chi^\pm_{1,2}$}}%
       {\scriptstyle\raise.3ex\hbox{$\scriptstyle\tilde\chi^\pm_{1,2}$}}%
 {\scriptscriptstyle\raise.3ex\hbox{$\scriptscriptstyle\tilde\chi^\pm_{1,2}$}}}}

\def\SRtNonep{\texttt{tN\_diag}}
\def\SRtNtwox{\texttt{tN\_med}}
\def\SRtNthreep{\texttt{tN\_high}}
\def\SRtNboost{\texttt{tN\_boost}}
\def\SRbCzero{\texttt{bCc\_diag}}
\def\SRbCone{\texttt{bCd\_bulk}}
\def\SRbCfour{\texttt{bCd\_high1}}
\def\SRbCfive{\texttt{bCd\_high2}}
\def\SRbCvW{\texttt{bCb\_med2}}
\def\SRtNbC{\texttt{tNbC\_mix}}
\def\SRbWN{\texttt{3body}}
\def\SRoneLoneBa{\texttt{bCa\_med}}
\def\SRoneLoneBc{\texttt{bCa\_low}}
\def\SRoneLtwoBa{\texttt{bCb\_med1}}
\def\SRoneLtwoBc{\texttt{bCb\_high}}

\def\softLepton{soft lepton}
\def\softLeptonHyphen{soft-lepton}

\title{Search for top squark pair production in final states with one isolated
lepton, jets, and missing transverse momentum in $\sqrt{s}=8$\,\TeV\  $pp$
collisions with the ATLAS detector}

\author{The ATLAS Collaboration}

\date{\today}

\begin{document}

\maketitle
\flushbottom

\clearpage

 \section{Introduction}\label{sec:introduction}

The hierarchy problem~\cite{Weinberg:1975gm,Gildener:1976ai,Weinberg:1979bn,Susskind:1978ms}
has gained additional attention with the observation of a new particle
consistent with the Standard Model (SM) Higgs boson~\cite{atlas_higgs_observation,cms_higgs_observation} at the
LHC~\cite{LHC:2008}.
Supersymmetry (SUSY)~\cite{Miyazawa:1966,Ramond:1971gb,Golfand:1971iw,Neveu:1971rx,Neveu:1971iv,Gervais:1971ji,Volkov:1973ix,Wess:1973kz,Wess:1974tw}, which extends the SM by introducing supersymmetric partners for all SM particles, provides an elegant solution to the hierarchy problem. 
The partner particles have identical quantum numbers except for a half-unit difference in spin. 
The superpartners of the left- and right-handed top quarks, \tleft\ and \tright, mix to form the two mass eigenstates \tone\ and \ttwo, where \tone\ (top squark or stop) is the lighter one. 
If the supersymmetric partners of the top quarks have masses $\lesssim$ 1 \TeV, loop diagrams involving top quarks, which are the dominant contribution to the divergence of the Higgs boson mass, can be largely cancelled~\cite{Dimopoulos:1981zb,Witten:1981nf,Dine:1981za,Dimopoulos:1981au,Sakai:1981gr,Kaul:1981hi,Barbieri:1987fn,deCarlos:1993yy}.
Significant mass splitting between \tone\ and \ttwo\ is possible due to the large 
top Yukawa coupling\footnote{The masses of the \tone\ and \ttwo\ are given by the eigenvalues of the stop mass matrix. The stop mass matrix involves the top-quark Yukawa coupling in the off-diagonal elements, which typically induces a large mass splitting. The stop mass matrix is diagonalised by the stop mixing matrix, which gives the \tleft\ and \tright\ components of the mass eigenstates \tone\ and \ttwo.}.
Furthermore, effects of the renormalisation group equations are strong for the third generation squarks, usually driving their masses significantly lower than those of the other generations. 
These considerations suggest a light stop which, together with the stringent LHC limits excluding other coloured supersymmetric particles up to masses at the \TeV\ level, motivates 
dedicated stop searches.

SUSY models can violate the conservation of baryon number and lepton number, resulting in a proton lifetime shorter than current experimental limits~\cite{Regis:2012sn}. This is commonly solved by introducing a multiplicative quantum number called $R$-parity, which is $1$ and $-1$ for all SM and SUSY particles, respectively. A generic $R$-parity-conserving minimal supersymmetric extension of the SM
(MSSM)~\cite{Fayet:1976et,Fayet:1977yc,Farrar:1978xj,Fayet:1979sa,Dimopoulos:1981zb} predicts pair production of SUSY particles and the existence of a stable lightest supersymmetric particle (LSP).
In a large variety of SUSY models, the lightest neutralino\footnote{
The charginos \chinoOneTwopm\ and neutralinos \ninoOneTwoThreeFour\ are the mass eigenstates 
formed from the linear superposition of the charged and neutral SUSY partners of the Higgs and electroweak gauge bosons 
(higgsinos, winos and binos).} (\ninoone) is the LSP, which
is also the assumption throughout this paper.
Since the \ninoone\ interacts only weakly it is a candidate for dark matter.

\begin{figure}
\center
\includegraphics[width=0.8\textwidth]{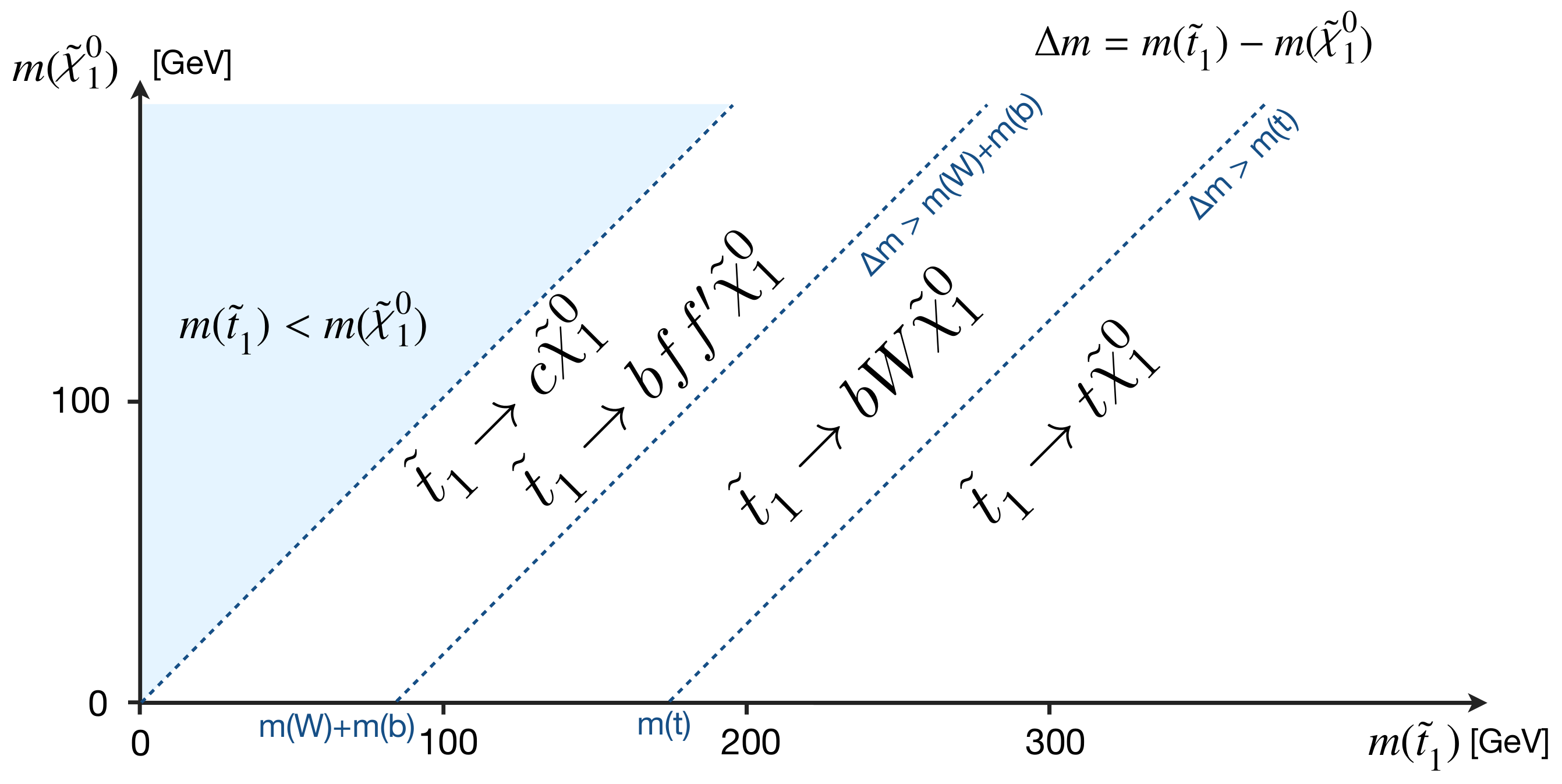}
\caption{
Illustration of stop decay modes in the plane spanned by the masses of the stop (\tone) and the lightest neutralino (\ninoone), where the latter is assumed to be the lightest supersymmetric particle. Stop decays to supersymmetric particles other than the lightest supersymmetric particle are not displayed.
\label{fig:stopDecays}}
\end{figure}

The stop can decay into a variety of final states, depending amongst other things 
on the SUSY particle mass spectrum, in particular on the masses of the stop and the lightest neutralino.  
\Figref{fig:stopDecays} illustrates the simplest decay modes as a function of the stop and LSP masses.  In the rightmost wedge, the stop mass is greater than the sum of the top quark and the LSP masses, hence the decay \topLSP\ is kinematically allowed. A lighter stop can undergo a three-body decay \threeBody\ if the stop mass is still above the $b+W+\ninoone$ mass. 
For an even lighter stop, the decay proceeds via a four-body process \fourBody, where $f$ and $f'$ 
are two distinct fermions,
or flavour-changing neutral current (FCNC) processes, such as the loop-suppressed \charmDecay.
If supersymmetric particles other than the \ninoone\ are lighter than the stop, then additional decay modes can open up.
The stop decay to a bottom quark and the lightest chargino (\bChargino) is an important example, where
the \chinoonepm\ can decay to the lightest neutralino by emitting an on- or off-shell $W$ boson ($\chinoonepm \to W^{(*)} \ninoone$). 
The \bChargino\ decay is considered for a stop mass above around $100$\,\GeV\
since the LEP limit on the lightest chargino is $m_{\chinoonepm} > 103.5$\,\GeV~\cite{lepsusy_web_chargino}.

\begin{figure}
\center
\subfigure[]{\includegraphics[width=0.4\textwidth]{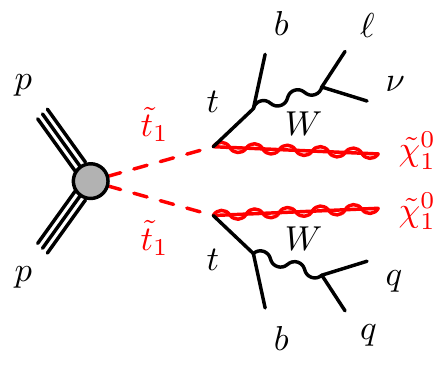}}
\hfill
\subfigure[]{\includegraphics[width=0.4\textwidth]{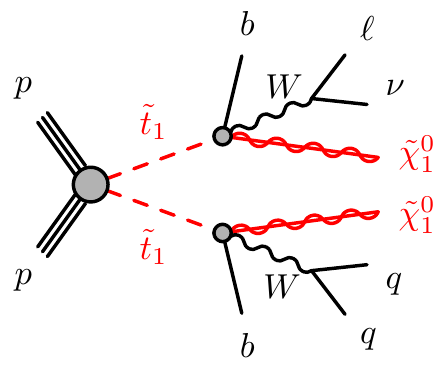}}
\subfigure[]{\includegraphics[width=0.4\textwidth]{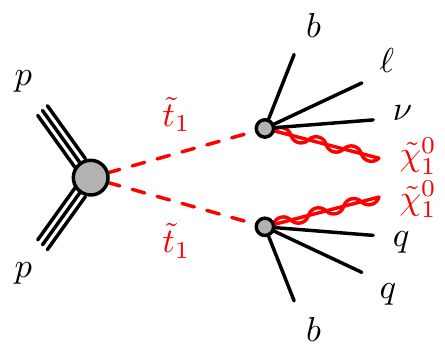}}
\hfill
\subfigure[]{\includegraphics[width=0.4\textwidth]{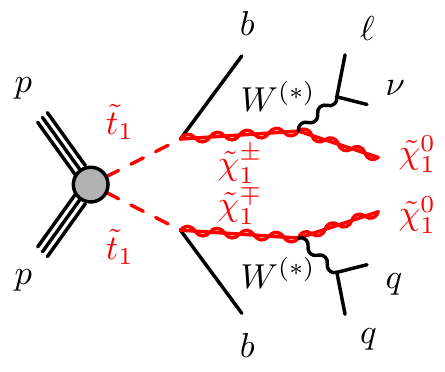}}
\caption{
Diagrams illustrating the considered signal scenarios, which are referred to as (a) \topLSP, (b) \threeBody\ (three-body), (c) \fourBody\ (four-body), (d) \bChargino. Furthermore, a non-symmetric decay mode where each \tone\ can decay via either \topLSP\ or \bChargino\ is considered (not shown). 
In these diagrams, the charge-conjugate symbols are omitted for simplicity; all scenarios begin with a top squark--antisquark pair. 
The three-body and four-body decays are assumed to proceed through an off-shell top quark, and an off-shell top quark followed by an off-shell $W$ boson, respectively.
\label{fig:stopFeynmanDiagrams}}
\end{figure}

This article presents a search for direct \tone\ pair production in final states with exactly one isolated charged lepton (electron or muon,\footnote{Electrons and muons from $\tau$ decays are included.} henceforth referred to simply as `leptons'), several jets, and a significant amount of missing transverse momentum, the magnitude of which is referred to as \met.
The lepton arises from the decay of either a real or a virtual $W$ boson, and the potentially large \met\ is generated by the two undetected LSPs and neutrino(s).
All stop decay modes described above except for the FCNC modes are considered, as illustrated in \figref{fig:stopFeynmanDiagrams}. 
With several decay modes kinematically available, the \tone\  decay branching ratio is determined by factors including the 
stop mixing matrix and the field content of the neutralino/chargino sector.
Results are mainly based on simplified models that have 
100\% branching ratio to one or a pair of these specific decay chains. 
In addition, phenomenological MSSM (pMSSM)~\cite{Djouadi:1998di} models
are used to study the sensitivity to realistic scenarios where more 
complex decay chains are present alongside the simpler ones.

Searches for direct \tone\ pair production have previously been reported
by the 
ATLAS~\cite{stop-0lep,stop-1lep,stop-verylight,stop_florida,stop-2lep,sbottom0L_8TeV,stop2L_8TeV} 
and CMS~\cite{Chatrchyan:2012uea,Chatrchyan:2013lya,Chatrchyan:2013xna,Chatrchyan:2013mya,Khachatryan:2014doa} collaborations, as well as by the CDF and D\O\ collaborations 
(for example refs.~\cite{PhysRevLett.104.251801, D0_stopSearch}) and the LEP collaborations~\cite{lepsusy_web_stop}.
Indirect searches for stops, mediated by gluino pair production, have been
reported by the 
ATLAS~\cite{ATLAS:2012ah,EPJC-gluino,Aad:2013wta,Aad:2014pda} 	
and
CMS~\cite{Chatrchyan:2012uea,Chatrchyan:2013lya,cmsMT2,Chatrchyan:2013wxa,Chatrchyan:2013iqa,Chatrchyan:2013fea,Chatrchyan:2014lfa}
collaborations.

 \section{Analysis strategy}\label{sec:strategy}

Searching for \tone\ pair production in the various decay modes
and over a wide range of stop masses requires different analysis approaches. 
The \tone\ pair production cross-section falls rapidly with increasing stop mass $m_{\tone}$:
for the range targeted by this search, $m_{\tone} \sim 100$--$700$\,\GeV, the cross-section 
at $\sqrt{s} = 8$\,\TeV\ proton--proton ($pp$) collisions decreases from $560$\,pb to $8$\,fb. 
While the various \tone\ decay modes considered all have identical final state objects ---
one electron or muon accompanied by one neutrino (or more for a leptonic $\tau$ decay), two 
jets originating from bottom quarks ($b$-jets), two light-flavour jets, and two LSPs -- 
their kinematic properties change significantly for the different decay modes and as a function of the masses of the stop, LSP, and lightest chargino (if present). 
The search presented in this paper is based on 15 dedicated analyses that target the various scenarios. 
The identification of $b$-jets ($b$-tagging) is utilised in the 
event selections and for constructing kinematic variables.
The search for a heavy stop exploits a specialised technique, which reconstructs several decay products in a single large-radius (\largeR) jet.
Low-momentum leptons (referred to as {\softLepton}s) are reconstructed and identified to enhance the sensitivity for \bChargino\ decays 
where the \ninoone\ and \chinoonepm\ states are close in mass.
These and other tools and variables to discriminate signal from background, described in \secref{sec:object_defs}, are used to design sets of requirements for the event selection.
Each of these sets of requirements is referred to as a signal region (SR), and is optimised to target one or more signal scenarios. 
Furthermore, two different analysis techniques are employed, which are referred to as `cut-and-count' and `shape-fit'. 
The former is based on counting events 
in a single region of phase space (bin),
while the latter employs several bins.
By utilising different signal-to-background ratios in the various bins, shape-fits
enhance the search sensitivity in challenging scenarios, where it is particularly difficult to separate signal from background.
All SRs are described in \secref{sec:signals}.

The dominant background in most SRs arises from top quark pair production (\ttbar) where both $W$ bosons decay leptonically (dileptonic \ttbar) but one of the leptons is not identified, is outside the detector
acceptance, or is a hadronically decaying $\tau$ lepton.  
The sub-leading background for most SRs stems from \Wjets\ production. 
As part of each analysis, the \ttbar\ and \Wjets\ backgrounds are estimated using dedicated control regions (CRs), making the analysis more robust 
against potential mis-modelling effects in simulated events and reducing the uncertainties on the background estimates.
Other small backgrounds are estimated using simulation only. 
Dedicated samples are used to validate the background predictions.
The background estimation including the definition of all CRs is detailed in \secref{sec:backgrounds}.

The analysis results are based on maximum likelihood fits, which include the CRs to simultaneously normalise the \ttbar\ and \Wjets\ backgrounds. 
Systematic uncertainties due to theoretical and experimental effects are considered for all background and signal processes, and
are described in \secref{sec:systematics}.
The final results and interpretations, both in terms of model-dependent exclusion limits on the masses of relevant SUSY particles 
and model-independent upper limits on the number of beyond-SM events, are presented in \secref{sec:results}.

 \section{The ATLAS detector}\label{sec:detector}

The ATLAS experiment~\cite{Aad:2008zzm} is a multi-purpose particle physics detector with nearly 4$\pi$~steradian coverage in solid 
angle.
It consists of an inner detector of tracking devices surrounded by a thin superconducting solenoid, 
electromagnetic and hadronic calorimeters, and a muon spectrometer
in a toroidal magnetic field.
The inner detector, in combination with the $2$\,T axial field from the solenoid, provides precision tracking and momentum measurement of charged particles up to
$|\eta|=2.5$  and allows efficient $b$-jet identification.\footnote{ATLAS uses a right-handed
  coordinate system with its origin at the nominal interaction point
  in the centre of the detector and the $z$-axis along the beam
  pipe. Cylindrical coordinates $(r,\phi)$ are used in the transverse
  plane, $\phi$ being the azimuthal angle around the beam pipe. The
  pseudorapidity $\eta$ is defined in terms of the polar angle
  $\theta$ by $\eta=-\ln\tan(\theta/2)$, and the angular separation $\Delta R$ in the $\eta$--$\phi$ space is defined as $\mathrm{\Delta} R =
  \sqrt{(\mathrm{\Delta} \eta)^2 + (\mathrm{\Delta} \phi)^2}$.} 
It consists of a silicon pixel detector, a semiconductor microstrip detector and a straw-tube tracker
which also provides transition radiation measurements for electron identification.
High-granularity liquid-argon (LAr) sampling electromagnetic calorimeters cover the pseudorapidity range $|\eta|<3.2$.
The hadronic calorimeter system is based on two different technologies, a scintillator-tile sampling calorimeter ($|\eta|<1.7$)
and a LAr sampling calorimeter ($1.5 < |\eta| < 3.2$). LAr calorimeters in the most forward region ($3.1 < |\eta| < 4.9$)
provide electromagnetic and hadronic measurements.
The muon spectrometer has separate trigger and high-precision tracking chambers, the former provide trigger coverage up to $|\eta| = 2.4$ while the latter provide muon identification and momentum measurements for $|\eta| < 2.7$.
Events are selected by a three-level trigger system~\cite{Aad:2012xs},
the first level (L1) is implemented in customised hardware while the two high-level triggers (HLT) are software-based.

 \section{Trigger and data collection}\label{sec:trigger}

The data used in this analysis were collected from March to December 2012 
with the LHC operating at a $pp$ centre-of-mass energy of $\sqrt{s}=8\,\TeV$.
After application of beam, detector and data quality requirements, the total integrated luminosity 
is 20.3\,\ifb\ with an uncertainty of $2.8\%$.
The uncertainty is derived, following the methodology detailed in ref.~\cite{Aad:2013ucp}, from a preliminary calibration of the luminosity scale from beam-separation scans performed in November 2012.

The dataset was recorded using three different types of triggers based on requiring either an electron, a muon, or large \met.
The single-electron trigger identifies electrons based on the presence of an energy cluster in the electromagnetic calorimeter with a shower shape consistent with that of an electron, low hadronic leakage, and a matching track in the inner detector. The HLT threshold\footnote{The trigger thresholds refer to lower requirements on the given quantity, and the HLT thresholds are always more stringent than the corresponding L1 thresholds.} on the energy deposit transverse to the beam (\ET) is $24$\,\GeV.
An electron isolation criterion at the HLT requires
the scalar sum of the transverse momenta (\pT) of tracks within a cone of radius $\Delta R = 0.2$ around the electron (excluding the electron itself) to be less than $10\%$ of the electron \ET. 
The single-muon trigger identifies muons using tracks reconstructed in the muon spectrometer and inner detector. The \pT\ threshold at the HLT is $24$\,\GeV. An isolation criterion at the HLT
requires the scalar sum of the \pT\ of tracks within a cone of radius $\Delta R = 0.2$ around the muon (excluding the muon itself) to be less than $12\%$ of the muon \pT.
To recover some of the small efficiency loss for high-\pT\ leptons, 
events were also collected using complementary single-lepton triggers. 
These triggers have less stringent shower-shape requirements and no hadronic leakage criterion for electrons, and no isolation criteria,
but have an increased \ET\ (\pT) threshold of $60$\,\GeV\ ($36$\,\GeV) for electrons (muons).
Corrections are applied to the simulated samples to account for small differences between data and simulation in the lepton trigger efficiencies.

The \met\ trigger is based on the vector sum of the transverse energies deposited in projective calorimeter trigger towers.
A more refined calculation based on the vector sum of all calorimeter cells above noise is made at the HLT. The trigger \met\ threshold at the HLT is $80$\,\GeV, and it is fully efficient for offline-calibrated $\met\ > 150$\,\GeV\ in signal-like events.
At the beginning of the 2012 data-taking, the \met\ trigger used in this analysis was disabled for the first three bunch crossings of every bunch train, causing a loss of $0.2$\,\ifb\ in integrated luminosity. 

Candidate events in the electron (muon) channel were collected using a logical-OR combination of the single-electron (single-muon) and \met\ triggers. 
Since the single-lepton trigger thresholds are too high for the \softLeptonHyphen\ selection,
these candidate events were recorded using only the \met\ trigger.
Consequently, the effective dataset for the \softLeptonHyphen\ selections amounts to an integrated luminosity of 
$20.1$\,\ifb.
All results quote a rounded value of $20$\,\ifb,
while inside the analysis the appropriate integrated luminosity values are used.
The efficiency of the \met\ and lepton triggers are measured with 
$W \to \mu \nu$ and $Z \to \ell\ell$ data samples, respectively.
In all cases the combined trigger efficiency is greater than 98\% for
simulated signal events satisfying the selection criteria for the analyses described in
\secsref{sec:signals}--\ref{sec:backgrounds}.

 \section{Simulated samples}\label{sec:samples}

Samples of Monte Carlo (MC) simulated events are used for the description of the background and to model the SUSY signals.
As detailed below, the samples are generated with either \powhegv~\cite{Frixione:2007vw}, \acermcv~\cite{Kersevan:2004yg}, 
\madgraphv~\cite{Alwall:2011uj}, \sherpav~\cite{Gleisberg:2008ta} or \herwigppv~\cite{Bahr:2008pv}.
All \powheg\ and \sherpa\ samples use the next-to-leading-order (NLO) parton distribution function (PDF) set 
CT10~\cite{Lai:2010vv}, 
while samples generated with \acermc, \madgraph\ or \herwigpp\ use
the CTEQ6L1~\cite{Pumplin:2002vw} PDF set.
The ATLAS Underlying Event Tune 2B~\cite{ATLAS:2011gmi} is used for all \madgraph\ samples, while the samples generated with
\powheg\ or \acermc\ use the {\sc Perugia} 2011C~\cite{Skands:2010ak} tune and samples generated with \herwigpp\ use
{\sc UEEE3}~\cite{Gieseke:2012ft}. The \sherpa\ generator has an integrated underlying event tune.
The fragmentation and hadronisation for the \powheg, \acermc, and \madgraph\ samples is performed with 
\pythiav~\cite{Sjostrand:2006za}, while \sherpa\ and \herwigpp\ use their own built-in models.

The samples are processed with either a detector simulation~\cite{:2010wqa} using \geant~\cite{Agostinelli:2002hh} or a fast simulation framework where the showers in the electromagnetic and hadronic calorimeters are simulated using a parameterised description~\cite{ATLAS:2010bfa} and  \geant\ is used for the rest of the detector.
The fast simulation has been validated against full \geant\ simulation for several signal models and for the main background, \ttbar\ production. All samples are produced with varying numbers of simulated minimum-bias interactions, generated with \pythiaEightv~\cite{Sjostrand:2007gs}, overlaid on the hard-scattering event to account for multiple $pp$ interactions in the same or nearby bunch crossings (pileup). The average number of interactions per bunch crossing is reweighted to match the distribution in data that varies between approximately $10$ and $30$.

\subsection{Background samples}\label{sec:samples_background}
A sample of \ttbar\ events is generated with \powheg\ using a top quark mass $m_{t} = 172.5$\,\GeV.
The same top quark mass is used when simulating other signal or background processes involving top quarks. 
To account for discrepancies between data and simulated \ttbar\ events, the simulated sample is reweighted as a function of the \pt\ of the \ttbar\ system; 
the weights are based on the ATLAS measurement of the differential \ttbar\ cross-section at $7$\,\TeV, following the method described in ref.~\cite{atlasttdifferential2013}.
Single top quark production in the $s$-channel and the $Wt$ mode are also generated with \powheg, while the $t$-channel process is generated with \acermc.
In $Wt$ production, the interference with \ttbar\ at NLO in quantum chromodynamics (QCD) is treated by the diagram removal scheme~\cite{Frixione:2008yi}.
Associated production of \ttbar\ and vector bosons ($W$, $Z$ and $WW$) as well as single top production in association with a $Z$ boson, are generated with \madgraph\ with up to two additional partons.
Samples of \Wjets\ and $Z/\gamma^*$ + jets are produced with \sherpa, containing up to four additional partons and the correct treatment of bottom and charm quark masses.
The diboson processes ($WW$, $ZZ$ and $WZ$) are also generated with \sherpa.

The processes are normalised using theoretical inclusive cross-sections, including higher-order QCD corrections where available.
The \ttbar\ production cross-section 
is calculated at next-to-next-to-leading order (NNLO) in QCD including resummation of next-to-next-to-leading logarithmic (NNLL) soft gluon terms with top++2.0~\cite{Czakon:2013goa,Czakon:2012pz,Czakon:2012zr,Baernreuther:2012ws,Cacciari:2011hy,Czakon:2011xx}.
Cross-sections for single top quark production are calculated to approximate NLO+NNLL precision~\cite{Kidonakis:2011wy,Kidonakis:2010ux,Kidonakis:2010tc}. The production of \ttbar\ in association with vector bosons is calculated at NLO~\cite{Campbell:2012dh,Garzelli:2012bn}, while the production of a single top quark in association with a $Z$ boson is normalised to the LO cross-sections from the generator, because NLO calculations are only available for $t$-channel production~\cite{Campbell:2013yla}.
The cross-sections for the production of $W$ and $Z$ bosons are calculated with {\sc DYNNLO}~\cite{Catani:2009sm}. 
The production cross-sections for electroweak diboson production are calculated at NLO with MCFM~\cite{Campbell:1999ah,Campbell:2011bn}.
The \ttbar, single top, $W$, $Z$, and diboson calculations 
use the MSTW2008 NNLO PDF set~\cite{Martin:2009iq}, while 
the cross-sections for \ttbar\ in association with a vector boson use
the MSTW2008 NLO ($W$) or CTEQ6.6M~\cite{Nadolsky:2008zw} ($Z$) PDF set. 
The cross-sections for \ttbar\ and $W$ production are used for the optimisation of the selections, while for the final results the 
two processes are normalised to data in control regions.

\subsection{Signal samples}\label{sec:samples_signal}
Signal samples of top squark--antisquark pairs
are generated with different stop decay and mass configurations. 
The first scenario assumes the \topLSP\ decay with a branching ratio (\br) of $100\%$. The samples are generated with \herwigpp\ 
in a grid across the plane of \tone\ and \ninoone\ masses with a spacing of  
$50$\,\GeV\ for most of the plane; the grid is more finely sampled towards the diagonal region where $m_{\tone}$ approaches $m_{t} + m_{\ninoone}$.
The $\tone$ is chosen to be mostly the partner of the right-handed top quark\footnote{
The \tright\ component is given by the the off-diagonal entry of the stop mixing matrix. Here, 
this matrix is set with (off-) diagonal entries of approximately ($\pm 0.83$) $0.55$.} and the \ninoone\ to be almost pure bino. 
This choice is consistent with a large \br\ for the given \tone\ decay.
Different hypotheses on the left/right mixing in the stop sector and the nature of the neutralino lead to different acceptance values. The acceptance is affected because the polarisation of the top quark changes as a function of the field content of the supersymmetric particles, which impacts the boost of the lepton in the top quark decay. A subset of models where the \tone\ is purely \tleft\ are studied to quantify this effect.

The second signal scenario assumes the $\bChargino \to b W^{(*)} \ninoone$ decay with a $\br$ of $100\%$.\footnote{
All possible decays of the (possibly virtual) $W$ boson are considered.}
The stop pairs are always generated with \madgraph, while for the 
\tone\ decay either \madgraph\ or \pythia\ is employed.
For models where the $W$ boson is on-shell, the full \tone\ decay is performed by \madgraph, while \pythia\ is used to decay the \tone\ in models where the $W$ is off-shell. 
In the latter case, generating the full event with \madgraph\ would be computationally too expensive.
Seven two-dimensional planes
are defined to probe the three-dimensional parameter space of the \tone, \chinoonepm, and \ninoone\ masses. 
The typical grid spacing is $50$\,\GeV; higher grid densities are generated in regions where a rapid change of sensitivity is expected. 
The boundary conditions are derived from the LEP chargino mass limit of $103.5$\,\GeV~\cite{lepsusy_web_chargino}, and by requiring the 
 \chinoonepm\ mass to be below the \tone\ mass. 
Six out of the seven planes span the \tone\ and \ninoone\ masses.
The first plane sets the chargino mass to twice the LSP mass ($m_{\chinoonepm} = 2 m_{\ninoone}$), motivated by the pattern in GUT-scale models with gaugino universality. 
The second and third planes fix the chargino mass to be above ($m_{\chinoonepm} = 150$\,\GeV) or 
close to ($m_{\chinoonepm} = 106$\,\GeV) the chargino mass limit, respectively.
In the fourth and fifth planes the chargino mass and neutralino mass are 
relatively close, $m_{\chinoonepm} - m_{\ninoone}= 5$\,\GeV\ and  $m_{\chinoonepm} - m_{\ninoone}= 20$\,\GeV, respectively;
small mass differences are motivated by higgsino-like states.
The sixth plane sets the chargino mass to be slightly below the stop mass, $m_{\chinoonepm} = m_{\tone} - 10$\,\GeV. 
The last plane fixes the stop mass, $m_{\tone} = 300$\,\GeV, while varying the \chinoonepm\ and \ninoone\ masses. 
The samples in all planes assume that the \tone\ is a \tleft\ state. The \bChargino\ branching ratio might not reach $100\%$ in the MSSM if the $\tone \to t+\ninoone/\ninotwo$ decays are kinematically allowed, but high branching ratios can occur in the allowed parameter space, such as for the above choices of particle field content.

The $\br = 100\%$ assumption is relaxed in a third signal scenario where a stop can decay either via \topLSP\ or via \bChargino. 
For this purpose, `asymmetric' samples are generated where in each event one stop is forced to decay via one and the second stop via the other decay mode.
The signal plane as a function of the \br\ can be probed by combining, with appropriate reweighting, the asymmetric samples with the two $\br = 100\%$ samples for the \topLSP\ and \bChargino\ decays.
The asymmetric samples are generated with the same generator settings used for the other \bChargino\ samples, except for using the maximum stop mixing angle (yielding equal components of \tleft\ and \tright) since the stop mixing is directly related to the \br. 
The mass points generated are identical to those for the $m_{\chinoonepm} = 2 m_{\ninoone}$ scenario.

The three- and four-body stop decay modes, \threeBody\ and \fourBody\ respectively, are relevant for 
a relatively light stop, as shown in \figref{fig:stopDecays}. 
Samples for each scenario are generated with the assumption of $\br = 100\%$.
The three-body samples are produced with \herwigpp, which performs the full matrix element calculation 
of the three-body decay, using the same settings as for the \topLSP\ decay mode.
The four-body decay mode is generated with \madgraph\ interfaced with \pythia\ for the \tone\ decay and for parton showering, and with up to one additional parton. The four-body decay itself is forced to proceed via a virtual $W$ boson. 
The \tone\ and \ninoone\ mass parameters are varied with a grid spacing between $25$ and $50$\,\GeV.

Signal cross-sections are calculated in the MSSM at 
NLO, including the resummation of soft
gluon emission at next-to-leading-logarithmic accuracy
(NLO+NLL)~\cite{Beenakker:1997ut,Beenakker:2010nq,Beenakker:2011fu}.
The nominal cross-section and the uncertainty are taken from an
envelope of cross-section predictions using different PDF sets and
factorisation and renormalisation scales, as described in
ref.~\cite{Kramer:2012bx}. 
The $\tone$ pair production cross-section obtained using this prescription 
is $(5.6 \pm 0.8)$\,pb for $m_{\tone}  = 250$\,\GeV, 
and $(0.025 \pm 0.004)$\,pb for $m_{\tone}  = 600$\,\GeV.

Although the simplified models described above can probe large regions of the allowed SUSY parameter space, more realistic SUSY models can feature more complex stop decays involving the heavier charginos and neutralinos.
To study the sensitivity of the various analyses to these well-motivated scenarios, the pMSSM models described in ref.~\cite{Cahill-Rowley:2013yla} are used.
These models produce a Higgs boson in the mass range ($m_h = 126 \pm 3$\,\GeV), saturate the WMAP relic density~\cite{Hinshaw:2012aka} and produce values of fine-tuning no worse than 1 part in 100 
using the measure proposed by Barbieri, Giudice and Ellis et al.~\cite{Ellis:1986yg,Barbieri:1987fn}.
In all models the \ninoone\ is the LSP.
To investigate the impact of varying parameters other than the stop and LSP mass
while at the same time avoiding the processing of a large number of events, 
only three different \tone\ and \ninoone\ mass regions are considered.
Only models where both \topLSP\ and \bChargino\  are kinematically allowed are used.
This serves to remove the models which have a branching ratio of
$100$\% for only one decay mode, as these regions of parameter space are already probed by the simplified models.
This results in a total of 27 models, for which top squark--antisquark pair events are generated 
with \herwigpp\ and processed with the fast simulation.
Some details of the models are given in \tabref{table:pMSSM_info}.
By keeping the stop and LSP masses fixed, the
impact on the sensitivity from varying other parameters can be studied, such as 
the branching ratios to the heavier charginos and neutralinos.
The sensitivity for pMSSM models can then be compared to that obtained in the simplified models 
with the corresponding stop and LSP masses.

\begin{table}[h]
\begin{center}
{\small
\begin{tabular}{|cccccc|ccccc|cc|}
\hline
\multicolumn{6}{|c|}{Mass [GeV]} & \multicolumn{5}{c|}{Branching ratio $\tilde{t}_{1} \to$} & \multicolumn{2}{c|}{} \\
$\tilde{t}_{1}$ & $\tilde{\chi}_{1}^{0}$ & $\tilde{\chi}_{2}^{0}$ & $\tilde{\chi}_{3}^{0}$ & $\tilde{\chi}_{1}^{\pm}$ & $\tilde{\chi}_{2}^{\pm}$ & 
$t\tilde{\chi}_{1}^{0}$ & $t\tilde{\chi}_{2}^{0}$ & $t\tilde{\chi}_{3}^{0}$ & $b\tilde{\chi}_{1}^{\pm}$ 
& $b\tilde{\chi}_{2}^{\pm}$ & $[T_{11}]^2$ & $[N_{11}]^{2}$ \\ 
\hline
404 & 40 & 221 & 230 & 220 & 1073 & 0.09 & 0.01 & 0.09 & 0.81 & 0.00 & 0.53 & 0.96 \\
404 & 44 & 324 & 445 & 325 & 471 & 0.16 & 0.00 & 0.00 & 0.84 & 0.00 & 0.98 & 0.99 \\
407 & 46 & 368 & 372 & 367 & 1515 & 0.74 & 0.00 & 0.00 & 0.26 & 0.00 & 0.02 & 0.98 \\
408 & 49 & 187 & 207 & 188 & 376 & 0.02 & 0.31 & 0.23 & 0.41 & 0.04 & 0.97 & 0.95 \\
409 & 39 & 211 & 212 & 206 & 1768 & 0.05 & 0.24 & 0.02 & 0.68 & 0.00 & 0.56 & 0.95 \\
409 & 49 & 180 & 190 & 179 & 795 & 0.02 & 0.22 & 0.17 & 0.59 & 0.00 & 0.99 & 0.94 \\
410 & 40 & 232 & 253 & 234 & 427 & 0.11 & 0.25 & 0.00 & 0.64 & 0.00 & 0.96 & 0.97 \\
410 & 43 & 387 & 396 & 386 & 889 & 0.88 & 0.00 & 0.00 & 0.12 & 0.00 & 0.01 & 0.99 \\
413 & 42 & 197 & 367 & 197 & 385 & 0.03 & 0.10 & 0.00 & 0.85 & 0.02 & 0.95 & 0.98 \\
413 & 45 & 373 & 406 & 374 & 508 & 0.32 & 0.00 & 0.00 & 0.68 & 0.00 & 0.99 & 0.99 \\
414 & 45 & 194 & 440 & 195 & 453 & 0.03 & 0.14 & 0.00 & 0.83 & 0.00 & 0.96 & 0.99 \\
416 & 45 & 394 & 397 & 393 & 1975 & 0.90 & 0.00 & 0.00 & 0.10 & 0.00 & 0.99 & 0.99 \\
417 & 46 & 333 & 350 & 335 & 573 & 0.65 & 0.00 & 0.00 & 0.35 & 0.00 & 0.96 & 0.98 \\
418 & 39 & 206 & 209 & 202 & 1779 & 0.09 & 0.05 & 0.28 & 0.59 & 0.00 & 0.47 & 0.95 \\
\hline
546 & 46 & 292 & 310 & 292 & 520 & 0.02 & 0.28 & 0.24 & 0.44 & 0.01 & 0.98 & 0.98 \\
547 & 46 & 346 & 374 & 346 & 500 & 0.12 & 0.49 & 0.00 & 0.22 & 0.16 & 0.93 & 0.98 \\
550 & 40 & 225 & 235 & 225 & 760 & 0.02 & 0.28 & 0.24 & 0.46 & 0.00 & 0.98 & 0.96 \\
551 & 43 & 351 & 366 & 351 & 621 & 0.07 & 0.38 & 0.21 & 0.35 & 0.00 & 0.98 & 0.99 \\
552 & 41 & 249 & 275 & 252 & 420 & 0.02 & 0.20 & 0.21 & 0.44 & 0.13 & 0.98 & 0.97 \\
552 & 42 & 332 & 337 & 331 & 1496 & 0.05 & 0.47 & 0.35 & 0.13 & 0.00 & 0.99 & 0.98 \\
552 & 43 & 346 & 350 & 344 & 1501 & 0.08 & 0.27 & 0.52 & 0.13 & 0.00 & 0.97 & 0.98 \\
552 & 43 & 385 & 397 & 385 & 731 & 0.36 & 0.00 & 0.00 & 0.64 & 0.00 & 0.97 & 0.99 \\
554 & 44 & 439 & 445 & 439 & 1007 & 0.21 & 0.00 & 0.00 & 0.79 & 0.00 & 0.99 & 0.99 \\
555 & 47 & 279 & 287 & 280 & 933 & 0.04 & 0.54 & 0.38 & 0.04 & 0.00 & 0.97 & 0.97 \\
\hline
553 & 147 & 169 & 444 & 168 & 455 & 0.31 & 0.12 & 0.00 & 0.27 & 0.30 & 0.07 & 0.93 \\
554 & 151 & 195 & 207 & 191 & 1969 & 0.09 & 0.35 & 0.43 & 0.12 & 0.00 & 0.88 & 0.68 \\
546 & 154 & 210 & 213 & 200 & 434 & 0.07 & 0.40 & 0.34 & 0.05 & 0.14 & 0.86 & 0.70 \\
\hline
\end{tabular}
}
\end{center}
\caption{Properties of the 27 selected pMSSM models. The table contains the masses of the stop, 
of neutralinos and of the charginos, the branching ratios of the stop decays, 
the \tleft\ content of the \tone\ ($[T_{11}]^2$, with $T$ being the stop mixing matrix)
and the bino content of the $\chi_{1}^{0}$ ($[N_{11}]^{2}$, with 
$N$ being the neutralino mixing matrix). 
\label{table:pMSSM_info}}
\end{table}

 \section{Physics object reconstruction and discriminating variables}
\label{sec:object_defs}

\subsection{Physics object reconstruction}\label{sec:physics_object_selection}

The reconstruction and identification of all final state objects used in this search, such as vertices, jets, leptons, and missing transverse momentum, is described in the following. 
Two sets of lepton identification criteria are utilised. One set defines the leptons used in the overlap removal procedure with jets and other objects and to veto events with more than one lepton. The second set imposes tighter identification criteria, and is used to select the primary lepton in the event.

The reconstructed primary vertex is required to be consistent with the beam diamond envelope
and to have at least five associated tracks with $\pT > 0.4$\,\GeV~\cite{ATLAS-CONF-2010-069}. If there are multiple primary vertices in an event, 
 the vertex with the largest summed $\pT^2$ of the associated tracks is chosen. Relevant quantities such as the track impact parameters are calculated 
 with respect to the selected primary vertex.

Jets are reconstructed from three-dimensional noise-suppressed calorimeter energy clusters~\cite{topoclusters} using the anti-$k_t$ jet
clustering algorithm~\cite{Cacciari:2008gp,Cacciari:2005hq} with a radius parameter ($R$) of $0.4$. 
The impact of pileup is statistically subtracted based on the jet area method~\cite{Cacciari:2008gn}. 
To calibrate the reconstructed energy, jets are corrected for the effects of calorimeter response and inhomogeneities using energy- and $\eta$-dependent calibration factors based on simulation and validated with extensive test-beam and collision-data studies. 
In the simulation, this procedure calibrates the jet energies to those of the corresponding jets constructed from stable simulated particles (particle-level jets).
In-situ measurements are used to further correct the data to match the energy scale in simulated events~\cite{Aad:2011he,Aad:2014bia}. 
Events containing jets that are likely to have arisen from detector noise, cosmic-ray muons, or 
machine-induced backgrounds such as beam-gas interactions and beam-halo particles, are removed~\cite{Aad:2014bia}.
Only jets with $\pT > 20$\,\GeV\ are considered. 
After the overlap removal procedure (described below), jets are required to have $|\eta| < 2.5$.

A second collection of anti-$k_t$ jets reconstructed with $R=1.0$ is used to collect collimated decay products of high-\pt\ top quarks and $W$ bosons; these jets are referred to as \largeR\ jets~\cite{Aad:2013gja}.
The energy calibration is based on the same strategy as used for the jets with $R=0.4$~\cite{Aad:2013gja}.
Jet trimming~\cite{Krohn:2009th} -- a procedure to remove contributions from pileup and from the underlying event by discarding softer components of the jet -- is applied with a $k_t$ sub-jet size $\Rsub = 0.3$ 
and a transverse momentum of the sub-jet relative to the \largeR\ jet, $\fcut$, greater than $0.05$.
\LargeR\ jets are required to have $\pt > 150$\,\GeV\ and $| \eta | < 2.0$. The invariant mass of \largeR\ jets is obtained from the energy and momentum of the jet constituents (themselves treated as massless) after the trimming procedure. In addition to the energy calibration, a mass calibration is applied to both data and simulation that accounts for differences between the jet masses derived at particle- and reconstruction-level.
\LargeR\ jets may overlap with other physics objects such as jets or leptons; no overlap removal procedure between \largeR\ jets and other objects is applied. Consequently, \largeR\ jets are neither an input to the calculation of the missing transverse momentum, nor considered for the identification of $b$-jets.

The identification of $b$-jets uses the `MV1' $b$-tagging algorithm (defined in refs.~\cite{ATLAS-CONF-2011-102,ATLAS-CONF-2011-089}),
which is based on a neural network and exploits both impact parameter and secondary vertex information.
It is trained to assign high weights to $b$-jets and low weights to jets originating from light-flavour quarks or gluons.
Three working points are chosen to maximise the search sensitivity for the various selections.
They correspond to an average $b$-tagging efficiency of $60\%$, $70\%$ and $80\%$ for
$b$-jets with $\pt > 20$~\GeV\ and $\left|\eta\right| < 2.5$ in simulated \ttbar\ events.
For these three working points, the average rejection factors for light-quark or gluon jets are approximately 600, 140, and 25 in the same simulated \ttbar\ events~\cite{ATLAS-CONF-2012-040}, respectively.
In the simulated samples, the efficiency of identifying $b$-jets and the probability for mis-identifying (mis-tagging) jets from light-flavour quarks, gluons and charm quarks are corrected to match those found in data.

Electron candidates are reconstructed from energy clusters in the electromagnetic calorimeter
matched to a track measured in the inner detector~\cite{Aad:2011mk,Aad:2014fxa}. 
They are required to have $\pt >10\,\GeV$, $|\eta| < 2.47$, and
to satisfy the `loose' shower shape and track selection criteria (defined in ref.~\cite{ATLAS-CONF-2014-032}).
The energy is corrected in data to match simulation, while the reconstruction efficiency is scaled in simulated samples to match that observed in data.
Muons are reconstructed and identified either as a combined track in the muon spectrometer and inner detector systems, or as an inner detector track matched with a muon spectrometer segment~\cite{Collaboration:muon,ATLAS-CONF-2013-088,Aad:2014zya}.
Candidate muons are required to have $\pt > 10\,\GeV$ and $|\eta| < 2.4$.
Corrections are applied to the momentum and to the reconstruction efficiency in simulation to match the data.
For the \softLeptonHyphen\ selections, the thresholds are lowered to $\pt > 7\,\GeV$ (electrons) and $\pt > 6\,\GeV$ (muons), and
electron candidates are required to satisfy the `medium' identification criteria (defined in ref.~\cite{ATLAS-CONF-2014-032}).

Potential ambiguities between overlapping candidate objects are resolved based on their 
angular separation.
If an electron candidate and a non-$b$-tagged jet (using the $70\%$ efficiency $b$-tagging working point) overlap 
within $\Delta R < 0.2$ of each other, then the object is considered to be an electron and the jet is dropped. If an electron candidate and any jet overlap within $0.2 < \Delta R < 0.4$ of each other, or if an electron candidate and a $b$-tagged jet overlap within $\Delta R < 0.2$ of each other, then the electron is dropped and the jet is retained.
If a muon candidate and any jet overlap within $\Delta R < 0.4$ of each other, then the muon is not considered and the jet is kept. 
For the analysis exploiting a \largeR-jet, the last requirement is changed to $\Delta R < 0.1$, still between the muon and the $R=0.4$ jets, to recover efficiency losses in boosted topologies.
The remaining leptons are referred to as `baseline' leptons, and are used to veto events with more than one lepton.

Photons are not used in the main selections in this analysis, but they are used to select events for one validation sample.
Photon candidates must satisfy the `tight' quality criteria with $\pt > 20 \GeV$ and $|\eta| < 2.47$~\cite{ATLAS-CONF-2012-123,phoPaper2011}. For the validation sample selection only, jets close to a photon, with $\Delta R < 0.2$, are dropped.

An event-veto based on 
identifying hadronically decaying $\tau$ leptons (\tauh) is used in some selections to reject \ttbar\ background.
The \tauh\ candidates are reconstructed in the same way as jets with $\pt > 15 \GeV$ and $|\eta| < 2.47$, 
but calibrated separately to account for a different calorimeter response. 
The $\tau$-identification is performed with a boosted decision tree (BDT) discriminator~\cite{Collaboration:tau1,Collaboration:tau2}, which combines 
tracking information and the transverse and longitudinal shapes of the energy deposits in the calorimeter.
If a \tauh\ candidate overlaps with any baseline lepton within $\Delta R < 0.2$, the \tauh\ is not counted.

The missing transverse momentum vector \Ptmiss\
is the negative vector sum of the \pt\ of reconstructed objects in the event: 
jets with $\pt > 20\,\GeV$,  charged lepton (electron and muon) and photon candidates with $\pt > 10\,\GeV$,
and calibrated calorimeter clusters not assigned to these physics objects~\cite{Aad:2012re,ATLAS-CONF-2013-082}.

The lepton identification criteria are tightened for the selection of the primary electron or muon in the event. 
The lepton \pt\ is required to be above $25$\,\GeV, except for the \softLeptonHyphen\ selections where the baseline thresholds 
of $7$\,\GeV\ (electron) or $6$\,\GeV\ (muon) are kept. Electrons are required to satisfy the `tight' selection criteria (defined in ref.~\cite{ATLAS-CONF-2014-032}), and are required to satisfy a track-isolation criterion. The scalar sum of the \pt\ of tracks associated with the primary vertex and found within a cone of radius 
$\dR = 0.2$ around the electron (excluding the electron itself) is required to be less than $10\%$ of the electron \pt.
Similarly, a muon isolation criterion is imposed: the track isolation is required to be less than $1.8$\,\GeV\ in a cone of radius $\dR = 0.2$.
A less stringent muon isolation criterion is used for the analysis using a \largeR\ jet: the track isolation is required to be less than $12\%$ of the muon \pt. This helps to recover signal efficiency losses in boosted topologies. 
For the analyses based on a \softLepton, 
the `tight' electron selection is omitted (keeping the `medium' criteria from the baseline selection), and a modified version of the track-isolation is applied to electrons and muons: the scalar sum of the \pt\ of tracks within a cone of radius 
$\dR = 0.3$ around the lepton (excluding the lepton itself) is required to be less than $16\%$ ($12\%$) of the electron (muon) \pt.
Furthermore, the impact parameters along the beam direction ($z_0$) and in the transverse plane ($d_0$) are used to impose additional \softLeptonHyphen\ requirements: $| z_0 \sin\theta | < 0.4 (0.4)$\,mm and $| d_0 / \sigma_{d_0} | < 5 (3)$ for electrons (muons), where $\sigma_{d_0}$ is the uncertainty on $d_0$.
The modified criteria of the \softLeptonHyphen\ selection are specifically optimised to suppress low-\pt\ jets mis-identified as isolated leptons. 

\subsection{Tools to discriminate signal from background}\label{sec:variables}

Requiring one isolated lepton ($\ell$), several jets, and \met\ selects a sample enriched in semi-leptonic \ttbar\ and \Wjets\ events. 
Both backgrounds are reduced by requiring the transverse mass (\mt) to be above the $W$ boson mass, where \mt\ is defined by

\begin{equation*}
	\mt\ = \sqrt{2 \cdot \pt^{\ell} \cdot \met \, \bigl(1-\cos \Delta\phi (\vec{\ell},\,\Ptmiss) \bigr) }.
\end{equation*}
Here $\pt^{\ell}$ is the lepton \pT, and $\Delta\phi (\vec{\ell},\,\Ptmiss)$ is the azimuthal angle between the lepton and 
the \Ptmiss\ directions.\footnote{This formula of \mt\ makes the assumption that the lepton mass is negligible.}
The dominant background after this requirement stems from dileptonic \ttbar\ events, where one lepton 
is not identified, or is outside the detector acceptance, or is a hadronically decaying $\tau$ lepton. 
In all of these cases, the \ttbar\ decay products include two or more high-\pt\ neutrinos, 
resulting in large \met\ and large \mt\ values.
Requiring one or more $b$-tagged jets further removes \Wjets\ events, while a $b$-tag veto reduces the \ttbar\ background but also the stop signal in most models. All but one SR require at least one or two $b$-tagged jets.

A number of variables and tools have been developed specifically to suppress the different types of dileptonic \ttbar\ events. 
The detailed definitions of the variables are provided in appendix~\ref{sec:appendixeq}.

\newcounter{itemcounter}
\begin{list}
{\textbf{\arabic{itemcounter}.}}
{\usecounter{itemcounter}\leftmargin=1.4em}

\item[-] \amtTwo\ and \mtTwoTau\ are two variants of the variable $\mtTwo$~\cite{mT21}, which is a generalisation of the transverse mass applied to signatures with two particles that are not directly detected.
\Figref{fig:mt2} illustrates the \ttbar\ event topologies targeted by the two variables. 

The first variant is a form of
asymmetric \mtTwo\ (\amtTwo)~\cite{mT22,mT23,mT24} in which the
undetected particle is the $W$ boson for the branch with the lost lepton
and the neutrino is the missing particle for the branch with the observed charged lepton. 
For dileptonic \ttbar\ events with a lost lepton, \amtTwo\ is bounded from above by the top quark mass, 
whereas new physics can exceed this bound. 

The second \mtTwo\ variant (\mtTwoTau) is designed for
events with a hadronically decaying $\tau$ lepton by using 
the `$\tau$-jet' as a visible particle for one branch and
the observed lepton for the other branch.  
For \ttbar\ events where one $W$ boson decays leptonically and the other to a $\tauh$, the endpoint is the $W$ boson mass in the limit of collinear neutrinos.

\begin{figure}
\center
\includegraphics[width=0.52\textwidth]{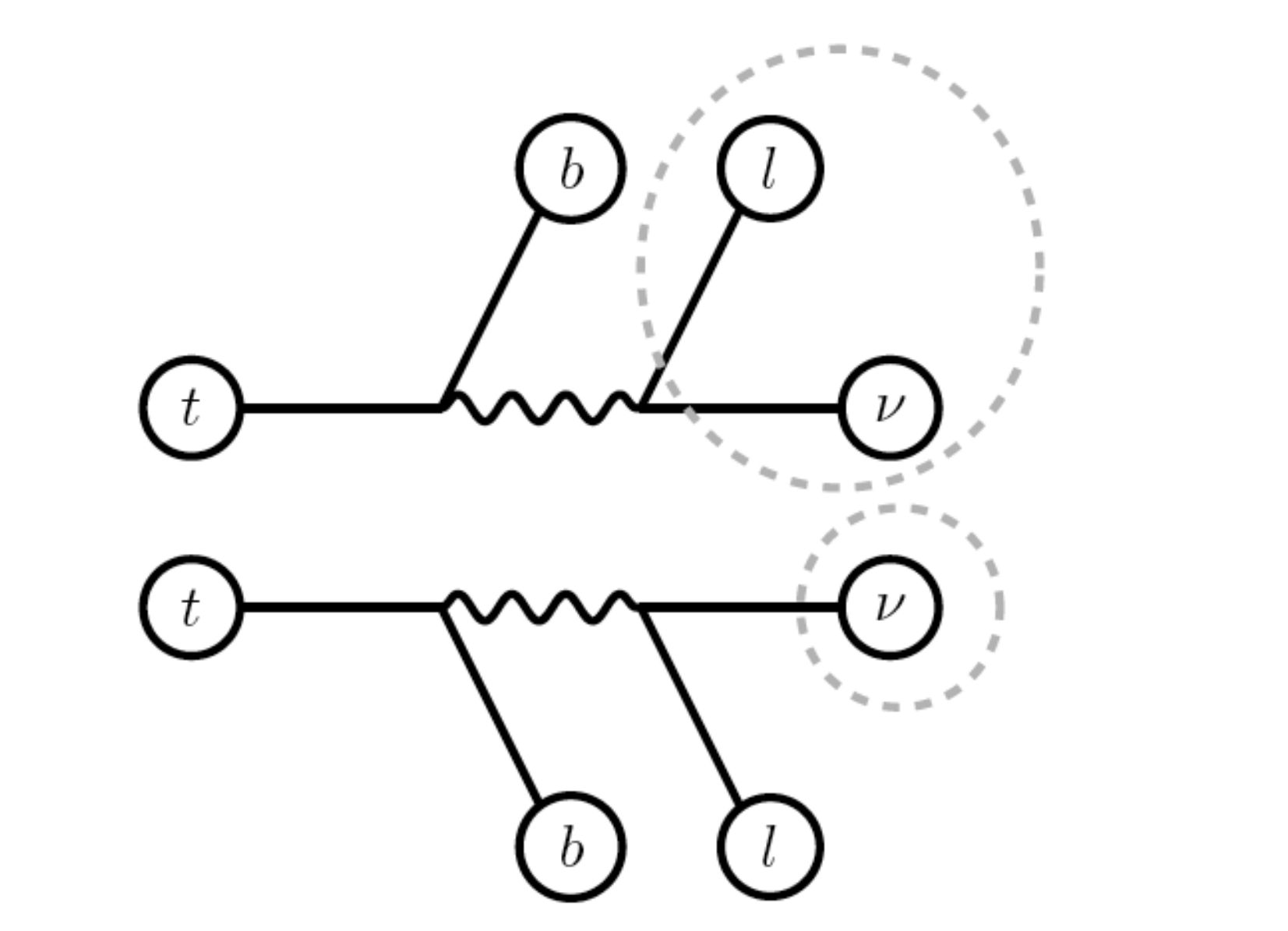}
\includegraphics[width=0.45\textwidth]{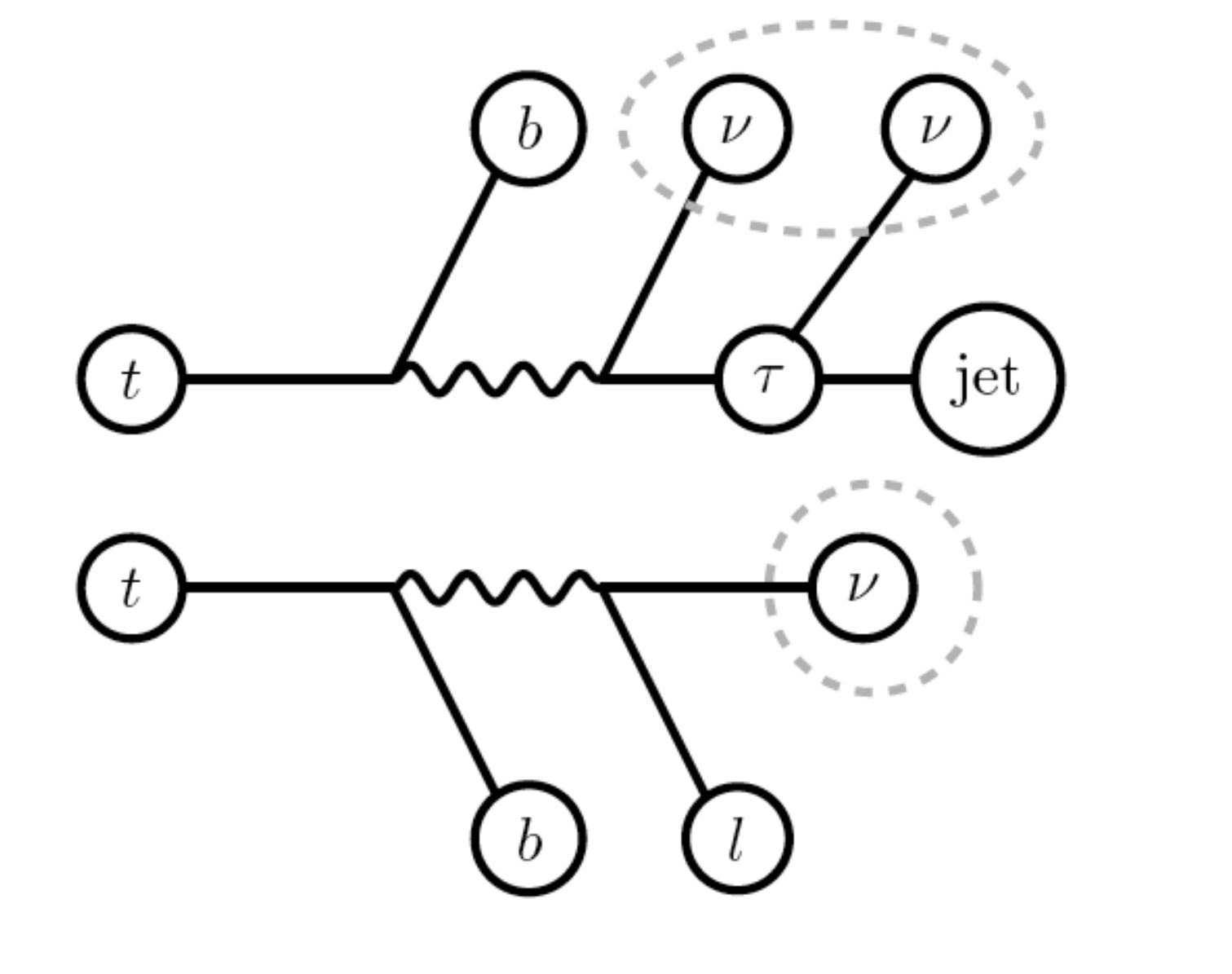}
\caption{Illustration of 
the construction of the \amtTwo\ (left) and \mtTwoTau\ (right) variables, which are used to discriminate against dileptonic \ttbar\ background
with one lost lepton (left) or with a hadronically decaying $\tau$ (right). The dashed lines indicate 
the objects that are assumed to be undetected (`lost') for the purpose of defining the two variables.
}
\label{fig:mt2}
\end{figure}

\item[-] \topness\ is a variable designed to identify and suppress partially reconstructed dileptonic \ttbar\ events, as proposed in ref.~\cite{Graesser:2012qy}.
The \topness\ variable is based on minimising a $\chi^2$-type function indicating the similarity of the event to dileptonic \ttbar\ events. 
Similar to the \amtTwo\ variable, one lepton
is assumed to be lost.

\item[-] A hadronic top mass, \mtophad, 
is designed to reject dileptonic \ttbar\ events while retaining signal events that contain a hadronically decaying on-shell top quark, as in the \topLSP\ decay mode. The \mtophad\ variable is a three-jet invariant mass, where the jets are selected by minimising a $\chi^2$-distribution taking into account the jet momenta and energy resolutions~\cite{Aad:2011he,Aad:2012ag}.

\item[-] 
Dedicated $\tau$-identification criteria are used to reject \ttbar\ events which contain a hadronic $\tau$ decay.
For the construction of the $\tau$-veto, the reconstructed \tauh\ candidates, as previously defined, are subject to further selection requirements
(described in appendix~\ref{sec:appendixeq}). 
Three $\tau$-veto working points are defined: loose, tight, and extra-tight.

\item[-] A track-veto is designed to reject events which contain an isolated track not associated with a baseline lepton. This complements the second-lepton veto, and helps to reject \ttbar\ events with a one-prong \tauh. The selection criteria are detailed in appendix~\ref{sec:appendixeq}.

\end{list}

Multijet events can pass the event selection if 
a jet is mis-identified as a lepton or when a real lepton from a heavy-flavour decay satisfies the isolation criteria, and if large \met\ occurs due to mis-measured jets. 
The former is suppressed by the lepton isolation criteria, 
while the following variables are used to 
reduce the latter effect.

\begin{list}
{\textbf{\arabic{itemcounter}.}}
{\usecounter{itemcounter}\leftmargin=1.4em}

\item[-] $\Delta\phi(\text{jet}_i, \vec{p}_\text{T}^\text{miss})$, 
the azimuthal opening angle between jet $i$ and $\vec{p}_\text{T}^\text{miss}$,  is used to suppress multijet events 
where $\vec{p}_\text{T}^\text{miss}$ is aligned with a jet. 

\item[-] \metsig, where \HT\ is defined as the scalar \pt\ sum of the four leading jets,
is an approximation of the \met\ significance. 

\item[-] $\met / \meff$, where $\meff = \HT + \pt^{\ell} + \met$. 

\item[-] \HTmissSig\ is an object-based missing transverse momentum, divided by the per-event resolution of the jets. 
The object-based missing transverse momentum is the negative sum of the jets and lepton vectors. A detailed description is given in appendix~\ref{sec:appendixeq}.

\end{list}

\section{Signal selections}\label{sec:signals}

Signal selections are optimised using simulated samples only. 
The metric of the optimisation is to maximise the exclusion sensitivity for the various decay modes, and for different regions of SUSY simplified model parameter space. 
A set of signal benchmark models, selected to cover the various stop scenarios, was used for the optimisation 
considering all studied discriminating variables and including statistical and systematic uncertainties. 
The shape-fits employ multiple bins in one or two discriminating variables, which were selected considering
the signal and background separation potential, inter-variable correlations, systematic uncertainties, and modelling 
of the data.

\Tabref{tab:signalOverview} summarises all 15 SRs with a brief description of the targeted signal scenarios, the exclusion analysis techniques, and forward references to the tables which list the event selection details.
Four SRs target the \topLSP\ decay. The corresponding SR labels begin with \texttt{tN}, which is an acronym for `top neutralino'; additional text in the label describes the stop mass region, for example \SRtNonep\ targets the `diagonal' of the \tone--\ninoone\ mass plane. 
Nine SRs target the \bChargino\ decay, where the SR labels follow the same logic:  the first two characters \texttt{bC} stand for `bottom chargino', a third letter (`a' to `d') 
denotes the four different mass hierarchies illustrated in \figref{fig:bChargino_mass_hierarchies}, and the last piece
 of text describes the stop mass region. 
Furthermore, two SRs labelled \SRbWN\ and \SRtNbC\ are dedicated to the three-body decay mode (\threeBody), and the mixed scenario where \topLSP\ and \bChargino\ decays both occur, respectively.
The SRs are not mutually exclusive.

\begin{table}
\begin{center}
\renewcommand{\arraystretch}{1.5}
{\small
\begin{tabular}{| l | l | l | c |}
\hline
SR & Signal scenario &  Exclusion technique & Table \\
\hline
\SRtNonep & \topLSP, $m_{\tone} \gtrsim m_{t} + m_{\ninoone}$ & shape-fit (\met\ and \mt) & \ref{tab:SRs_tN} \\
\SRtNtwox & \topLSP, $m_\tone \sim 550$\,\GeV, $m_{\ninoone} \lesssim 225$\,\GeV & cut-and-count & \ref{tab:SRs_tN} \\
\SRtNthreep & \topLSP, $m_\tone \gtrsim 600$\,\GeV & cut-and-count & \ref{tab:SRs_tN} \\  
\SRtNboost & \topLSP, $m_\tone \gtrsim 600$\,\GeV, with a \largeR\ jet & cut-and-count & \ref{tab:SRs_tN} \\
\hline
\SRoneLoneBc & \bChargino, $\Delta M \lesssim 50$\,\GeV & shape-fit (lepton \pt) & \ref{tab:SRs_softLeptons} \\
& \fourBody\ & & \\
\SRoneLoneBa & \bChargino, $50$\,\GeV $\lesssim \Delta M \lesssim 80$\,\GeV & shape-fit (lepton \pt) & \ref{tab:SRs_softLeptons} \\
& \fourBody\ & & \\
\SRoneLtwoBa & \bChargino, $\Delta m  \lesssim 25$\,\GeV, $m_{\tone} \lesssim 500$\,\GeV & shape-fit (\amtTwo) & \ref{tab:SRs_softLeptons} \\
\SRoneLtwoBc & \bChargino, $\Delta m \lesssim 25$\,\GeV, $m_{\tone} \gtrsim 500$\,\GeV & shape-fit (\amtTwo) & \ref{tab:SRs_softLeptons} \\
\hline
\SRbCvW & \bChargino, $\Delta m  \lesssim 80$\,\GeV, $m_{\tone} \lesssim 500$\,\GeV & shape-fit (\amtTwo\ and \mt) & \ref{tab:overview_SRbC} \\
\SRbCzero & \bChargino, $m_{\tone} \gtrsim m_{\chinoonepm}$ & cut-and-count & \ref{tab:overview_SRbC} \\
\SRbCone & \bChargino, ($\Delta M, \Delta m$) $\gtrsim 100$\,\GeV, $m_{\tone} \lesssim 500$\,\GeV & shape-fit (\amtTwo\ and \mt) & \ref{tab:overview_SRbC} \\ 
\SRbCfour & \bChargino, ($\Delta M, \Delta m$) $\gtrsim 100$\,\GeV, $m_{\tone} \gtrsim 500$\,\GeV & cut-and-count & \ref{tab:overview_SRbC} \\
\SRbCfive & \bChargino, $\Delta M$ $\gtrsim 250$\,\GeV, $m_{\tone} \gtrsim 500$\,\GeV & cut-and-count & \ref{tab:overview_SRbC} \\
\hline
\SRbWN & \threeBody, $m_{\tone} \lesssim 300$\,\GeV & shape-fit (\amtTwo\ and \mt) &  \ref{tab:SRs_alternative} \\
\SRtNbC & non-symmetric (\topLSP, \bChargino) & cut-and-count & \ref{tab:SRs_alternative} \\
\hline
\end{tabular}
}
\end{center}

\caption{Overview of all signal regions (SRs) together with the targeted signal scenario, the analysis technique used for model-dependent exclusions, and a reference to the table with the event selection details.
For the \bChargino\ decay mode, the mass splittings $\Delta M=m(\tone)-m(\ninoone)$ and $\Delta m = m(\chinoonepm)-m(\ninoone)$ are used to characterise the mass hierarchies, as illustrated in \figref{fig:bChargino_mass_hierarchies}. The SRs \SRoneLoneBc, \SRoneLoneBa, \SRoneLtwoBa\ and \SRoneLtwoBc\ employ selections based on a {\softLepton}.
\label{tab:signalOverview}
}
\end{table}

\begin{figure}
\center
\includegraphics[width=0.8\textwidth]{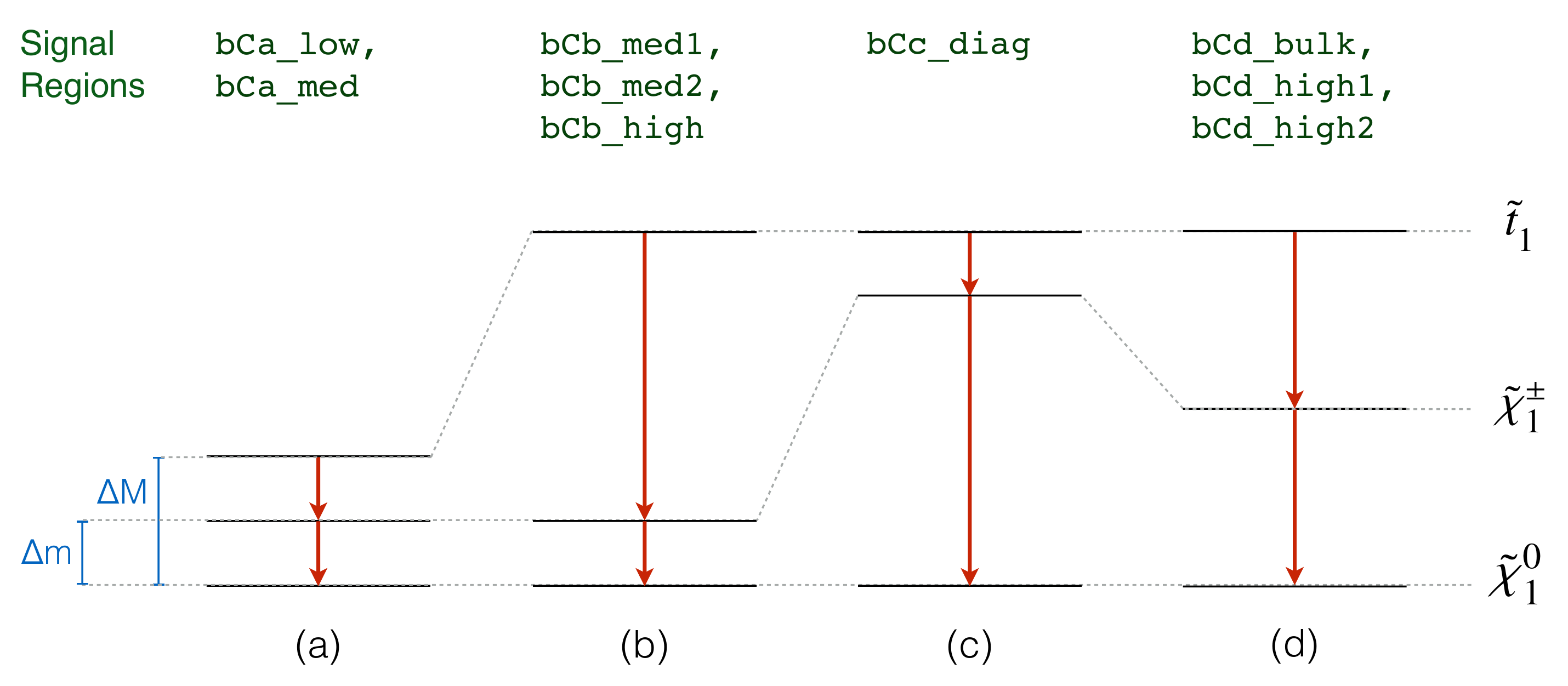}
\caption{
Schematic diagram of the mass hierarchies for the \bChargino\ decay mode, with a subsequent 
$\chinoonepm \to W^{(*)} \ninoone$ decay. 
A list of the corresponding signal regions is given above the diagram.
 }
\label{fig:bChargino_mass_hierarchies}
\end{figure}

All SRs employ selection requirements to suppress the multijet background, and most SRs  use the tools described in \secref{sec:variables} to reduce the dileptonic \ttbar\ background.  Shape-fit techniques are employed to derive model-dependent exclusion limits where useful, while for all model-independent results a simple cut-and-count approach is used. This procedure implies that for SRs using shape-fits one bin is probed at a time to extract the model-independent results. Only a single bin, or the four bins with highest signal-to-background ratio are included; these are referred to as signal-sensitive bins. 
The model-dependent and model-independent selections are defined in this section, and the corresponding fit configurations are described in \secref{sec:results}.

\subsection{Event preselection}\label{sec:signal:trigger_event_preselection}
Common preselection criteria are employed as follows.
Events are required to contain a reconstructed primary vertex. 
Furthermore, a set of quality requirements to avoid badly reconstructed jets, mismeasured-\met\ and detector-related problems is imposed on all events. Events with a bad quality muon or with a cosmic-ray muon candidate\footnote{
Defined as a muon candidate with a transverse or longitudinal impact parameter
of $|d_0| > 0.2$\,mm or $|z_0| > 1$\,mm.
}  are rejected.

Exactly one isolated lepton is required with $\pT > 25$\,\GeV\ except for the \softLeptonHyphen\ selections which employ a $\pT > 6 (7)$\,\GeV\ requirement for muons (electrons). The common lepton isolation criteria are tightened for the \softLeptonHyphen\ selections while they are relaxed for the analysis exploiting a \largeR\ jet (cf. \secref{sec:object_defs}). Events containing additional baseline leptons
are rejected. 
A minimum number of jets ranging between 2 and 4,
and $\met > 100\,\GeV$ are common requirements amongst all analyses. 
\Tabref{tab:preselection} summarises the preselection criteria.

\begin{table}
\begin{center}
\renewcommand{\arraystretch}{1.5}
{\small
\begin{tabular}{| l | l |}
\hline
Preselection & Description \\
\hline
\textbf{Trigger} & logical-OR combination of single-lepton and \met\ triggers; \\
  & \softLepton: \met\ trigger only.\\
\hline
\textbf{Data quality} & jet and \met\ cleaning, cosmic-ray muon veto, primary vertex.\\
\hline
\textbf{Lepton} & one isolated electron or muon with $\pT > 25$\,\GeV;  \\
& \softLepton: the \pT\ threshold is $6 (7)$\,\GeV\ for muons (electrons).\\
\hline
\textbf{2\textsuperscript{nd}-lepton veto}& No additional baseline lepton with $\pT > 10$\,\GeV; \\
& \softLepton: no further baseline soft muon (electron) with $\pT > 6 (7)$\,\GeV.\\
\hline
\textbf{Jets} & The minimum jet multiplicity requirement varies between $2$ and $4$.\\
\hline
$\bm{\met}$ & $\met > 100$\,\GeV\ or tighter is required in all selections.\\
\hline
\end{tabular}
}
\end{center}

\caption{Preselection criteria common to all signal selections.
\label{tab:preselection}
}
\end{table}

\Figref{fig:datamc_preselection} illustrates the separation power for a selection of discriminating variables.
For these distributions, events are required to pass the preselection (\tabref{tab:preselection}), to have at least four jets with $\pT > 25$\,\GeV, one of which above $60$\,\GeV, and with at least one of them 
$b$-tagged using the 70\% working point, and to have $\met > 100\,\GeV$,
$\mt > 60\,\GeV$ and $\met / \sqrt{\HT} > 5\,\GeV^{1/2}$. The \Wjets\ background is normalised to match data in a sample selected in the same way, except that a $b$-veto is imposed. The other processes are normalised to their theoretical cross-sections. Data and background estimation are seen to be in good agreement.

\begin{figure}
\begin{center}
\includegraphics[width=0.49\textwidth]{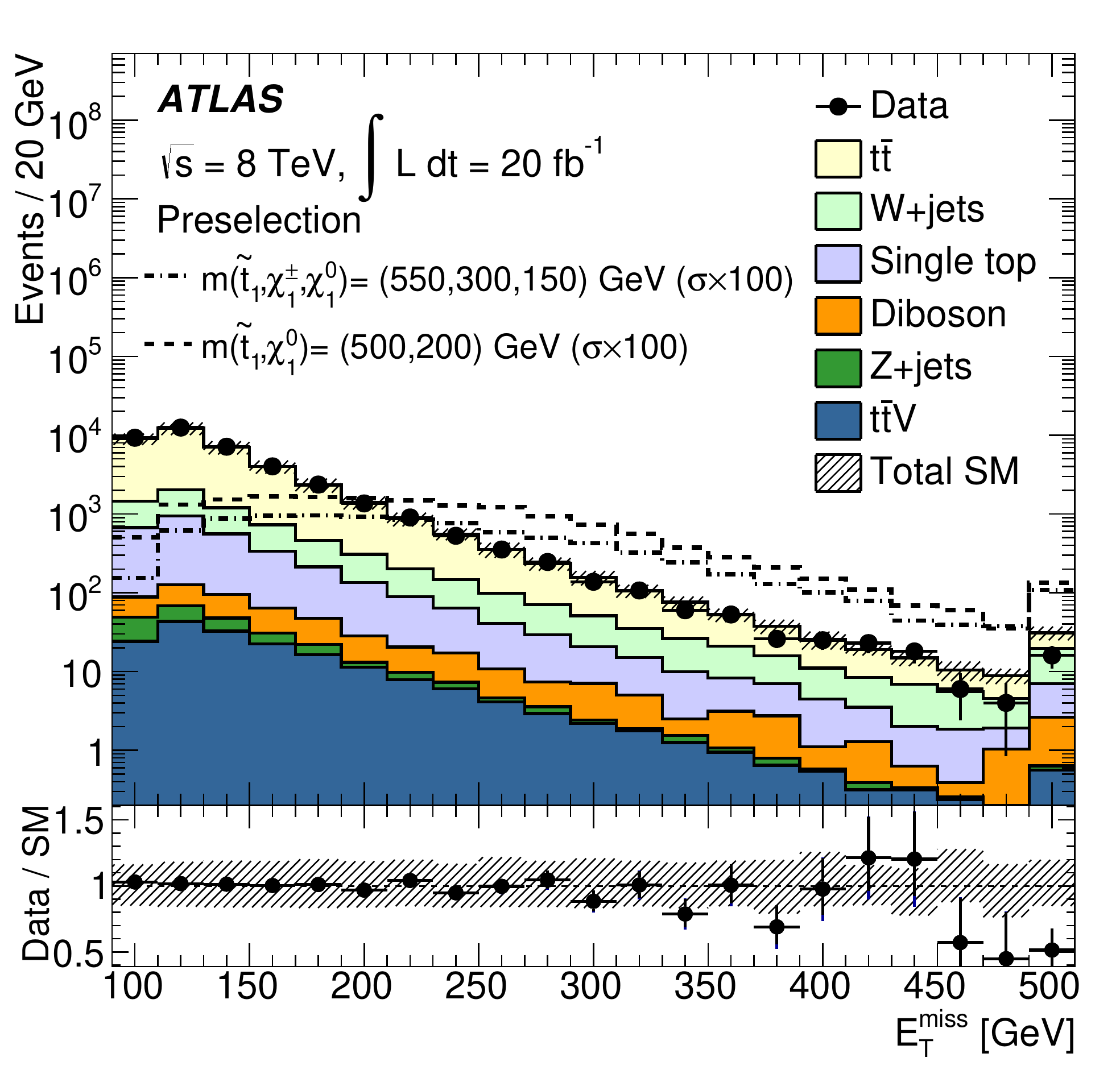} 
\includegraphics[width=0.49\textwidth]{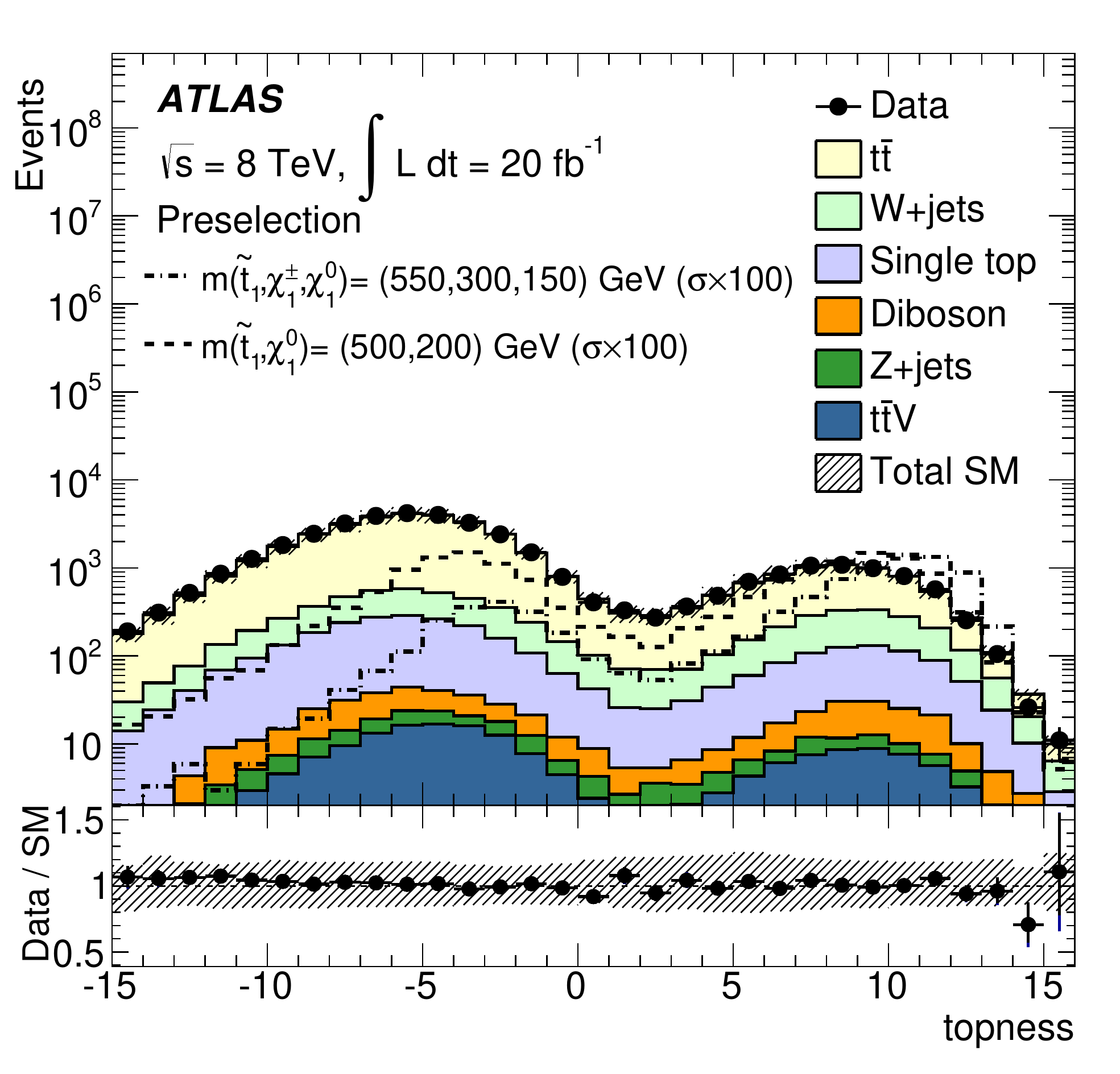} 
\includegraphics[width=0.49\textwidth]{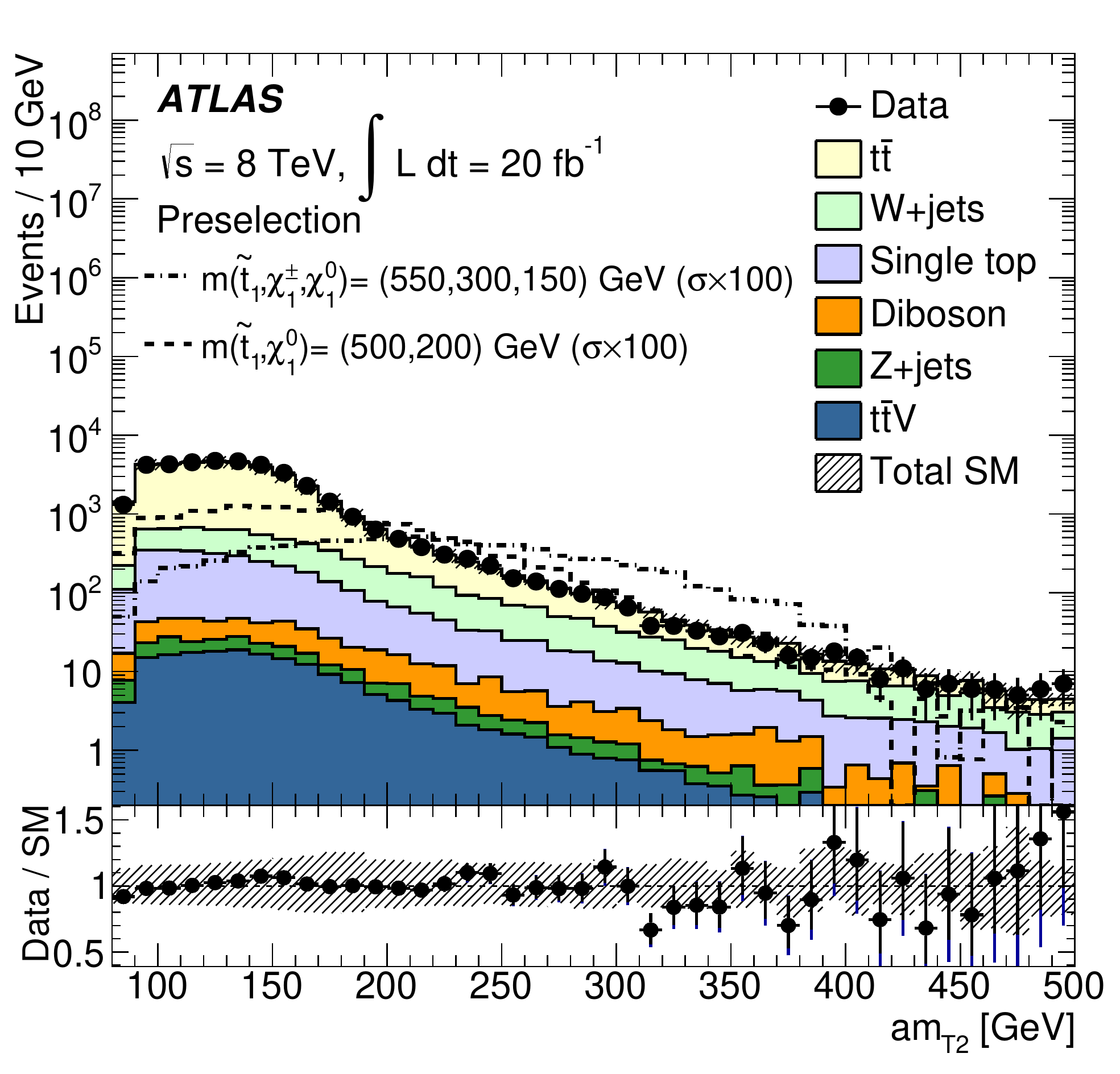} 
\includegraphics[width=0.49\textwidth]{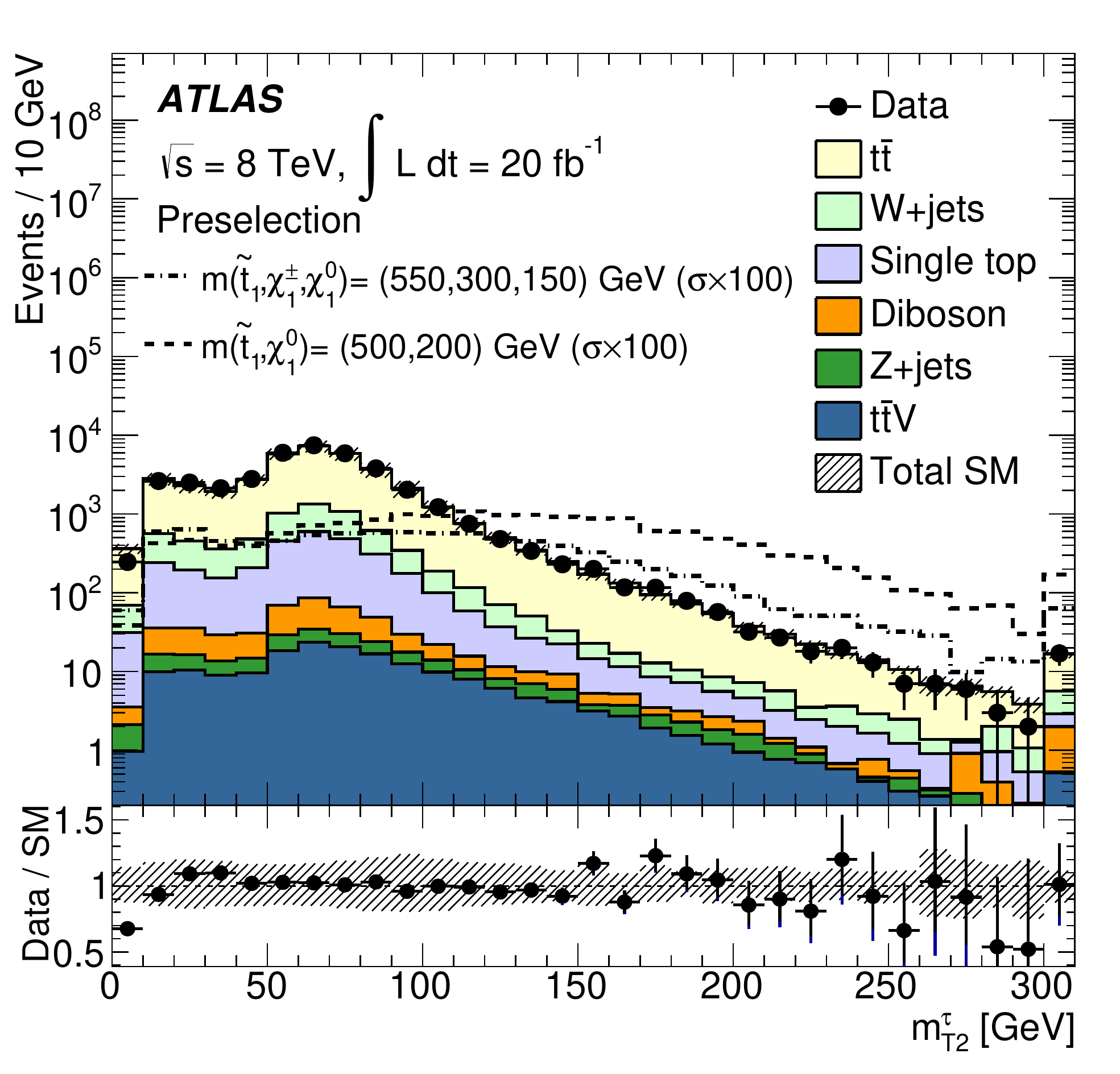} 
\caption{
Comparison of data with estimated backgrounds in the \met\ (top left), \topness\ (top right), \amtTwo\ (bottom left), and \mtTwoTau\ (bottom right)
distributions for the preselection defined in the text. 
The uncertainty band includes statistical and all experimental systematic uncertainties. The last bin includes overflows.
Benchmark signal models with cross-sections enhanced by a factor of $100$ are overlaid for comparison.
\label{fig:datamc_preselection}}
\end{center}
\end{figure}

\subsection{Selections for the \topLSP\ decay}\label{sec:topLSP_decay}

Stop pair production with subsequent \topLSP\ decays leads to
final state objects similar to that of \ttbar\ production augmented by two \ninoone.
Four SRs, labelled \SRtNonep, \SRtNtwox, \SRtNthreep, and \SRtNboost, 
target different regions in the $\tone - \ninoone$ mass plane and implement different analysis strategies.
\Tabref{tab:SRs_tN} details the event selections for these SRs.
Criteria based on a subset of the variables outlined in \secref{sec:variables}, as well as optimised jet thresholds, a more stringent \met\ requirement, and a requirement on the angular separation between the highest-\pt\ $b$-tagged jet and the lepton, $\Delta R(b \textnormal{\,-}\rm{jet}, \ell)$, are used to suppress \ttbar\ and \Wjets\ backgrounds as well as to reduce the multijet background to a negligible level.

\begin{table}
\begin{center}
\setlength{\tabcolsep}{0.36pc}
\renewcommand{\arraystretch}{1.5}
{\small
\begin{tabular}{| l | l | l | l | l |}
\hline
 & \SRtNonep & \SRtNtwox & \SRtNthreep &  \SRtNboost \\
\hline
\textbf{Preselection} & \multicolumn{4}{ |c| }{Default preselection criteria, cf. \tabref{tab:preselection}.}\\
\hline
\textbf{Lepton} & \multicolumn{4}{ |c| }{ $=1$ lepton } \\
\hline 
\textbf{Jets} & $\ge 4$ with $\pT>$ & $\ge 4$ with $\pT>$ &  $\ge 4$ with $\pT>$ & $\ge 4$ with $\pT>$ \\
                    & $60,60,40,25$\,\GeV & $80,60,40,25$\,\GeV &  $100,80,40,25$\,\GeV &  $75,65,40,25$\,\GeV \\
\hline
\textbf{$b$-tagging} & \multicolumn{4}{ |c| }{$\ge 1$ $b$-tag (70\% eff.) amongst four selected jets}\\
\hline
\textbf{\largeR jet} & \multicolumn{3}{ |c| }{--} & $\ge 1$, $\pT>270$\,\GeV\\
  & \multicolumn{3}{ |c| }{ } & and $m > 75$\,\GeV\\
\hline
$\bm{\Delta\phi(\text{jet}^{\text{\largeR}}_2, \vec{p}_\text{T}^\text{miss})}$ & \multicolumn{3}{ |c| }{--} & $> 0.85$ \\
\hline
$\bm{\met}$ & $>100$\,\GeV & $>200$\,\GeV & $>320$\,\GeV & $>315$\,\GeV \\
\hline
$\bm{\mt}$ & $>60$\,\GeV & $>140$\,\GeV &  $>200$\,\GeV &  $>175$\,\GeV \\
\hline
$\bm{\amtTwo}$ & -- & $>170$\,\GeV &  $>170$\,\GeV &  $>145$\,\GeV \\
\hline
$\bm{\mtTwoTau}$ & -- & -- &  $>120$\,\GeV & -- \\
\hline
$\bm{\topness}$ & -- & -- &  -- &  $>7$ \\
\hline
$\bm{\mtophad}$ & $\in$ [$130$, $205$]\,\GeV & $\in$ [$130$, $195$]\,\GeV &  $\in$ [$130$, $250$]\,\GeV &  \\
\hline
\textbf{$\tau$-veto} &  tight & -- &  -- &  modified, see text. \\
\hline
$\bm{\Delta R(b\text{-jet}, \ell)}$    & $< 2.5$ & -- & $< 3$ &$< 2.6$ \\
\hline
$\bm{\metsig}$ & $> 5\,\GeV^{1/2}$ & \multicolumn{3}{ |c| }{--} \\ 
\hline
$\bm{\HTmissSig}$ & -- & \multicolumn{2}{ |c| }{$>12.5$} & $>10$ \\
\hline
$\bm{\Delta\phi(\text{jet}_i, \vec{p}_\text{T}^\text{miss})}$ & $>0.8$ {\scriptsize ($i=1,2$)} & $>0.8$ {\scriptsize($i=2$)} & -- & $>0.5,0.3$ {\scriptsize($i=1,2$)} \\ 
\hline
\hline
\multicolumn{5}{ |l| }{\textbf{Model-dependent selection:}}  \\
\hline
 & shape-fit in \mt\ and& \multicolumn{3}{ |c| }{cut-and-count}   \\
 & \met, cf. \figref{fig:tN_3b_shapeFits}. & \multicolumn{3}{ |c| }{ }\\
\hline
\multicolumn{5}{ |l| }{\textbf{Model-independent selection:}}  \\
\hline
&  test 4 most & \multicolumn{3}{ |c| }{cut-and-count}   \\
 &  signal-sensitive& \multicolumn{3}{ |c| }{ } \\
 &  bins one-by-one.& \multicolumn{3}{ |c| }{ } \\
\hline
\end{tabular}
}
\end{center}

\caption{Selection criteria for SRs employed to search for \topLSP\ decays.
\label{tab:SRs_tN}
}
\end{table}

\begin{figure}
  \centering
\includegraphics[width=.49\textwidth]{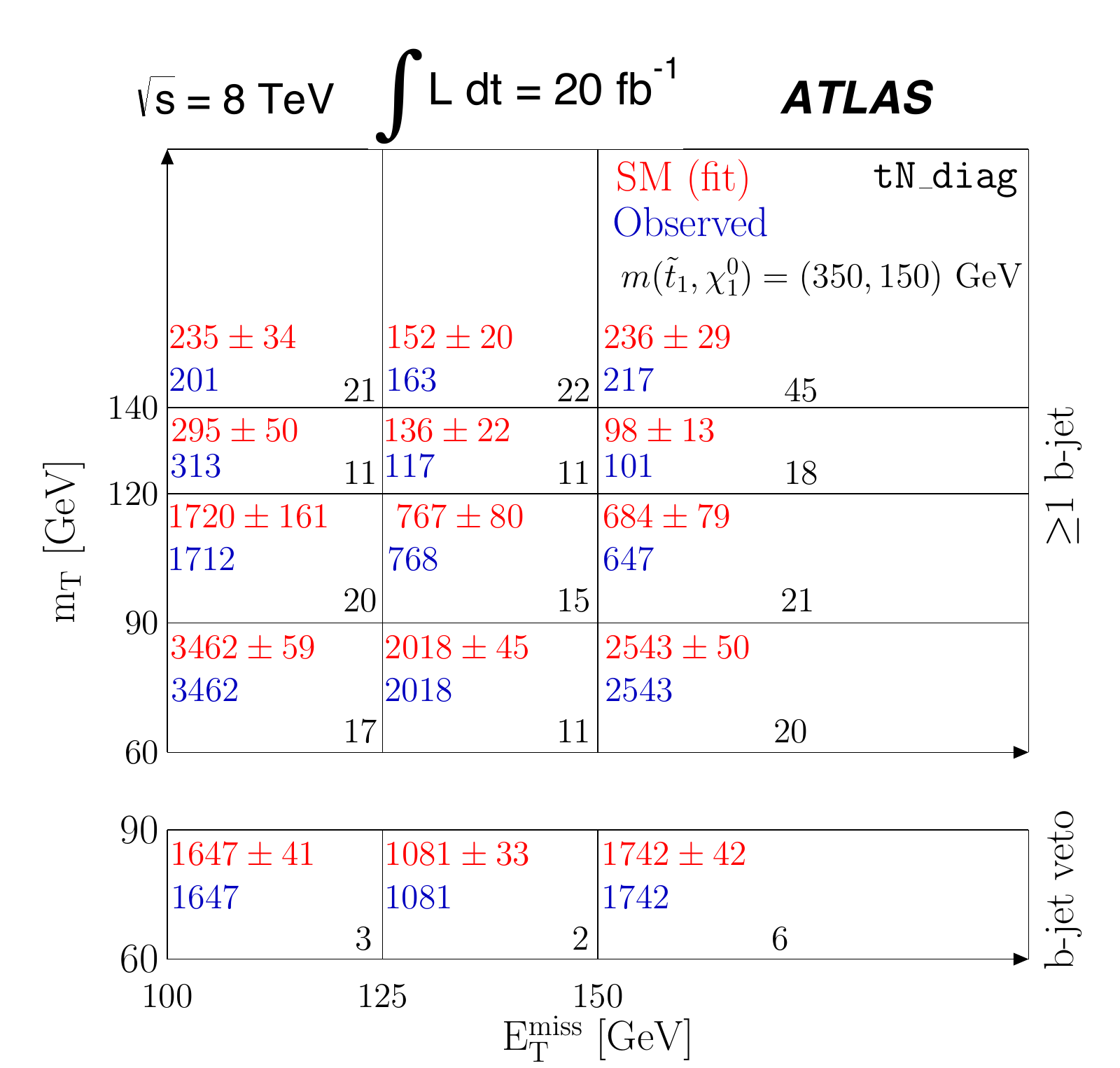}  
\includegraphics[width=.49\textwidth]{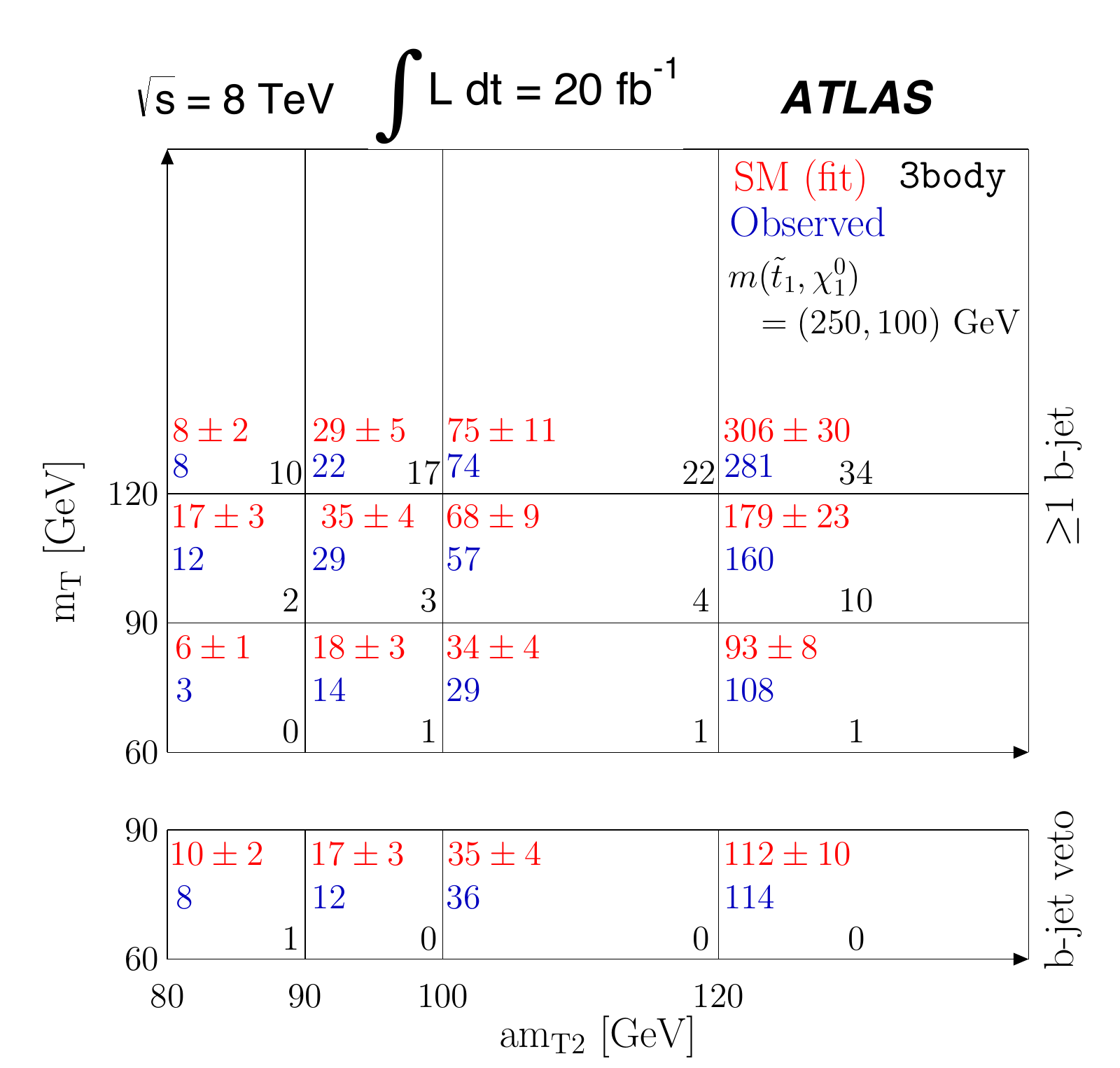}   
  \caption{
  Schematic illustration of the \SRtNonep\ (left) and \SRbWN\ (right) shape-fit binning. The \mt\ and \met\ (left) or \amtTwo\ (right) variables
  are used to define a matrix of $4 \times 3$ bins (left) or $3 \times 4$ (right) in the top part, which is sensitive to stop models, while also being enriched with \ttbar\ background.
  The bottom bins invert the $b$-tag requirement into a veto, and serve to normalise the \Wjets\ background.
  The numbers of observed events together with the estimated background,
  obtained using the background-only fit described in \secref{sec:backgrounds},
  are given for each bin. 
  The data and estimated background are in perfect agreement in the six bottom bins for the left plot because
  the fit is configured to use these six bins together with six free parameters; the fit used for the right plot employs
  the bottom eight bins and two free parameters.
  For comparison the expected numbers of events for one signal model are shown.
}
  \label{fig:tN_3b_shapeFits}
\end{figure}

The loosest selection, \SRtNonep, employs a multi-binned shape-fit that targets the challenging parameter
space where the stop and its decay products are nearly mass degenerate ($m_{\tone} \gtrsim m_{t} + m_{\ninoone}$), also referred to as
the `diagonal'.
The strategy of exploiting binned shape information significantly improves the sensitivity.
The two-dimensional shape-fit in the variables \mt\ and \met\ is illustrated in \figref{fig:tN_3b_shapeFits} (left plot). 
The top 12 bins serve both to probe a signal and to normalise the \ttbar\ background; 
a subset of the 12 bins has a high purity in \ttbar\ events.
Three additional bins with a $b$-tag veto, shown in the bottom part, are used to derive the 
normalisation of the \Wjets\ background.
The bins with $\met > 150\,\GeV$ or $\mt > 140\,\GeV$ are defined without upper boundaries.

The two SRs \SRtNtwox\ and \SRtNthreep\ target medium and high stop mass regions, respectively. 
Both SRs are based on a cut-and-count approach with relatively tight selections. 
The SR labelled \SRtNboost\ also targets models with a high stop mass and a nearly massless LSP, but takes advantage of the `boosted' topology. 
The selection assumes that either all decay products of the hadronically decaying top quark, or at least the decay products of the hadronically decaying $W$ boson, collimate into a jet with a radius of $\lesssim 1.0$.
\Figref{fig:datamc_preselection_boost} shows some of the relevant \largeR\ jet related distributions. The overlaid heavy stop benchmark model illustrates the separation power of the variables.
The \SRtNboost\ selection requires at least one \largeR\ jet with $\pt > 270$\,\GeV\ and an invariant mass above $75$\,\GeV. 
To further discriminate stop decays from the \ttbar\ background, events with a 
second (ordered by \pT) \largeR jet are required to have a minimum azimuthal distance between \Ptmiss\ and the second \largeR\ jet, 
$\Delta\phi(\text{jet}^{\text{\largeR}}_2, \vec{p}_\text{T}^\text{miss})$.
The extra-tight $\tau$-veto is applied to discard events with \tauh\ candidates well separated from \largeR\ jets, $\Delta R(\tauh,\text{\largeR-jet}) > 2.6$, that satisfy the above \pt\ and mass requirements. 

\begin{figure}
\begin{center}
\includegraphics[width=0.49\textwidth]{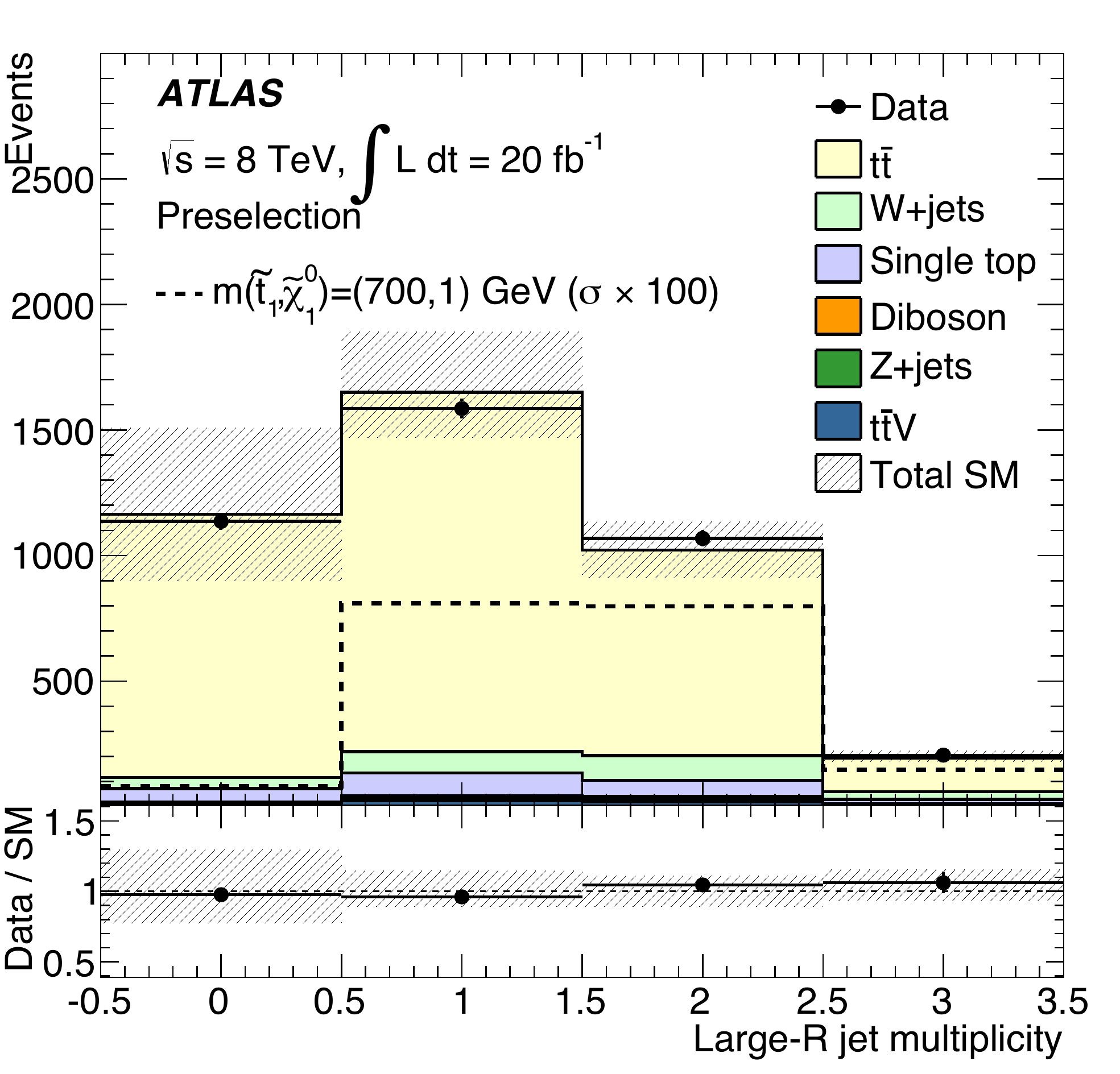} 
\includegraphics[width=0.49\textwidth]{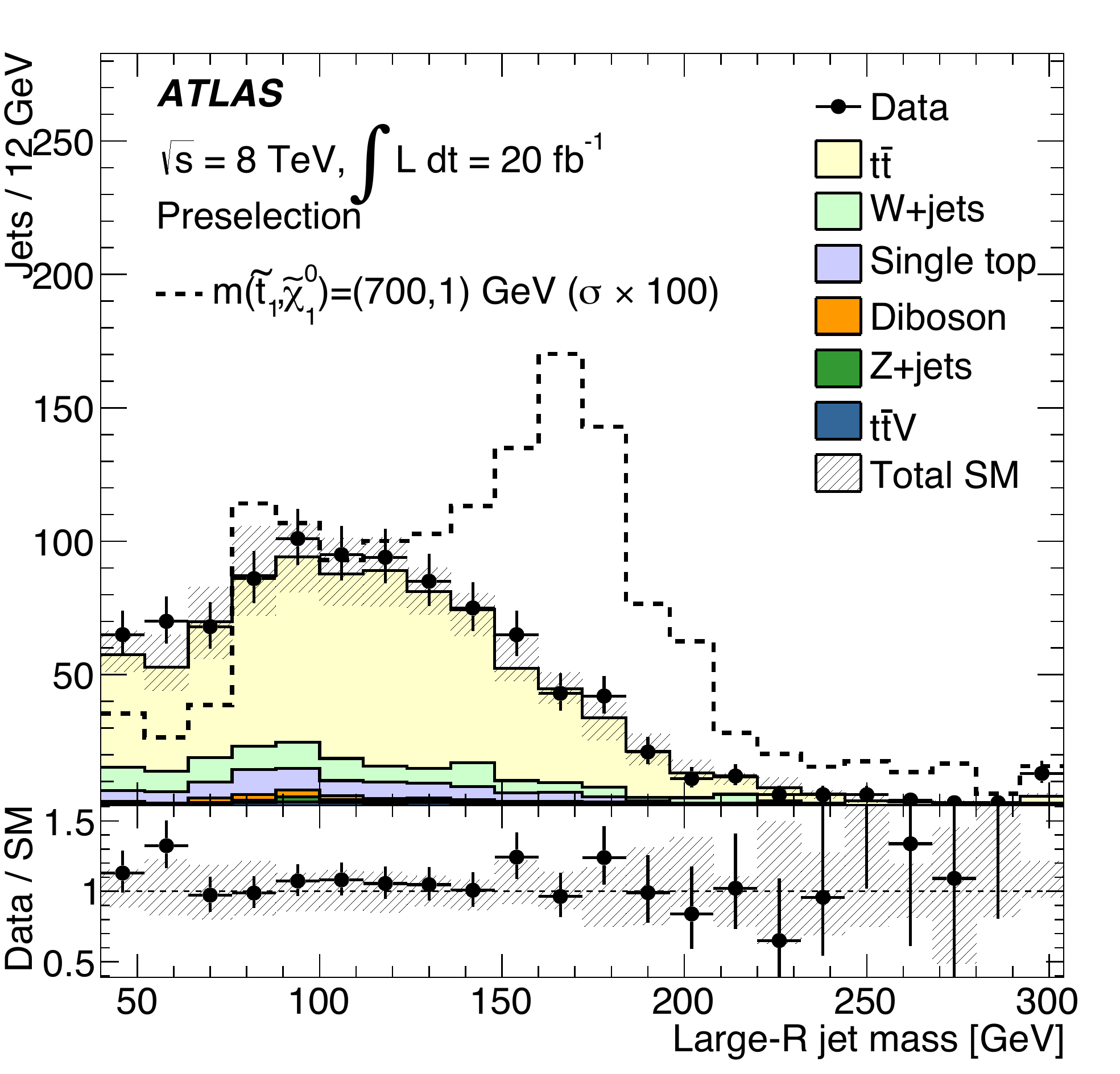} 
\includegraphics[width=0.49\textwidth]{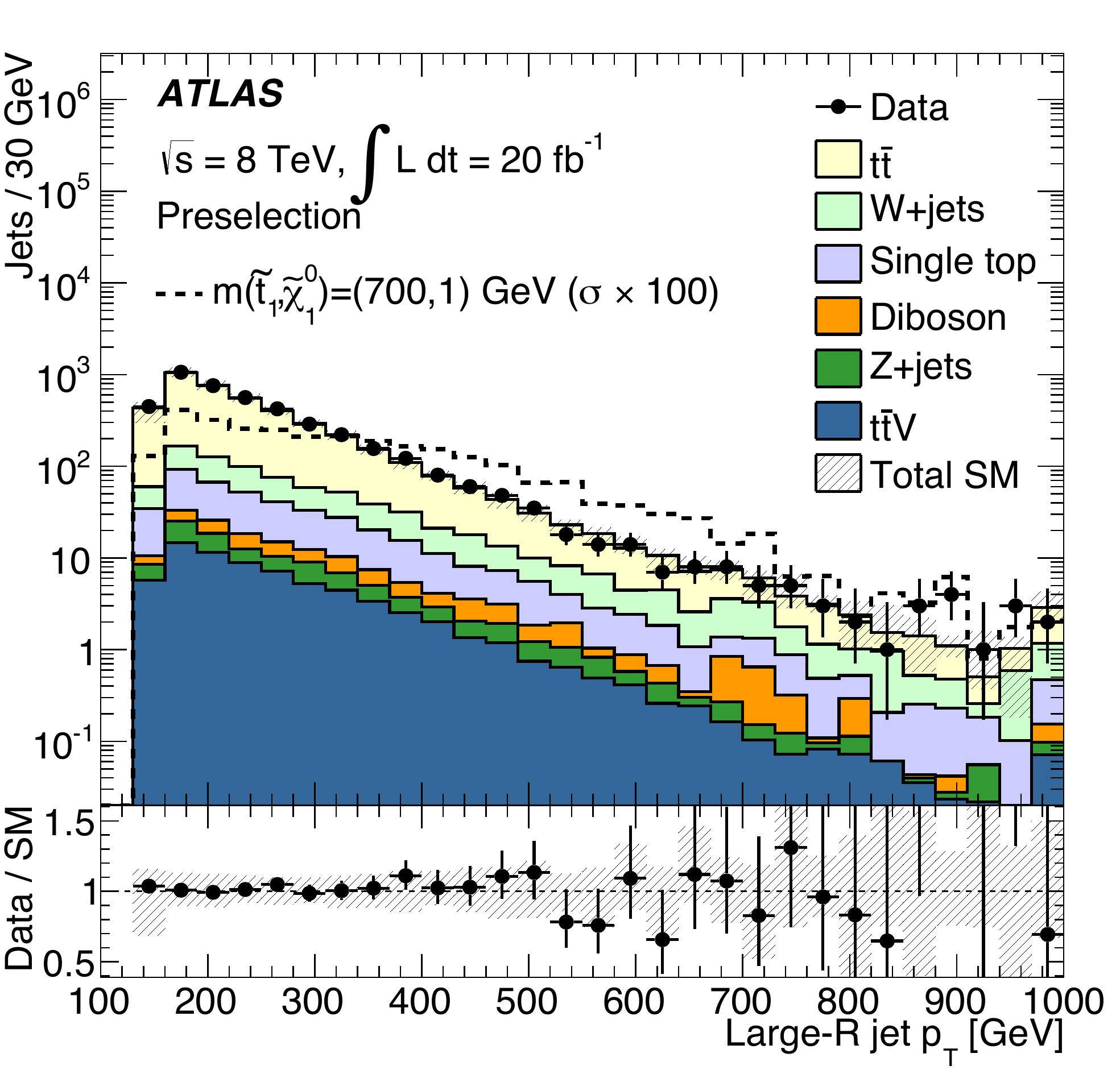} 
\includegraphics[width=0.49\textwidth]{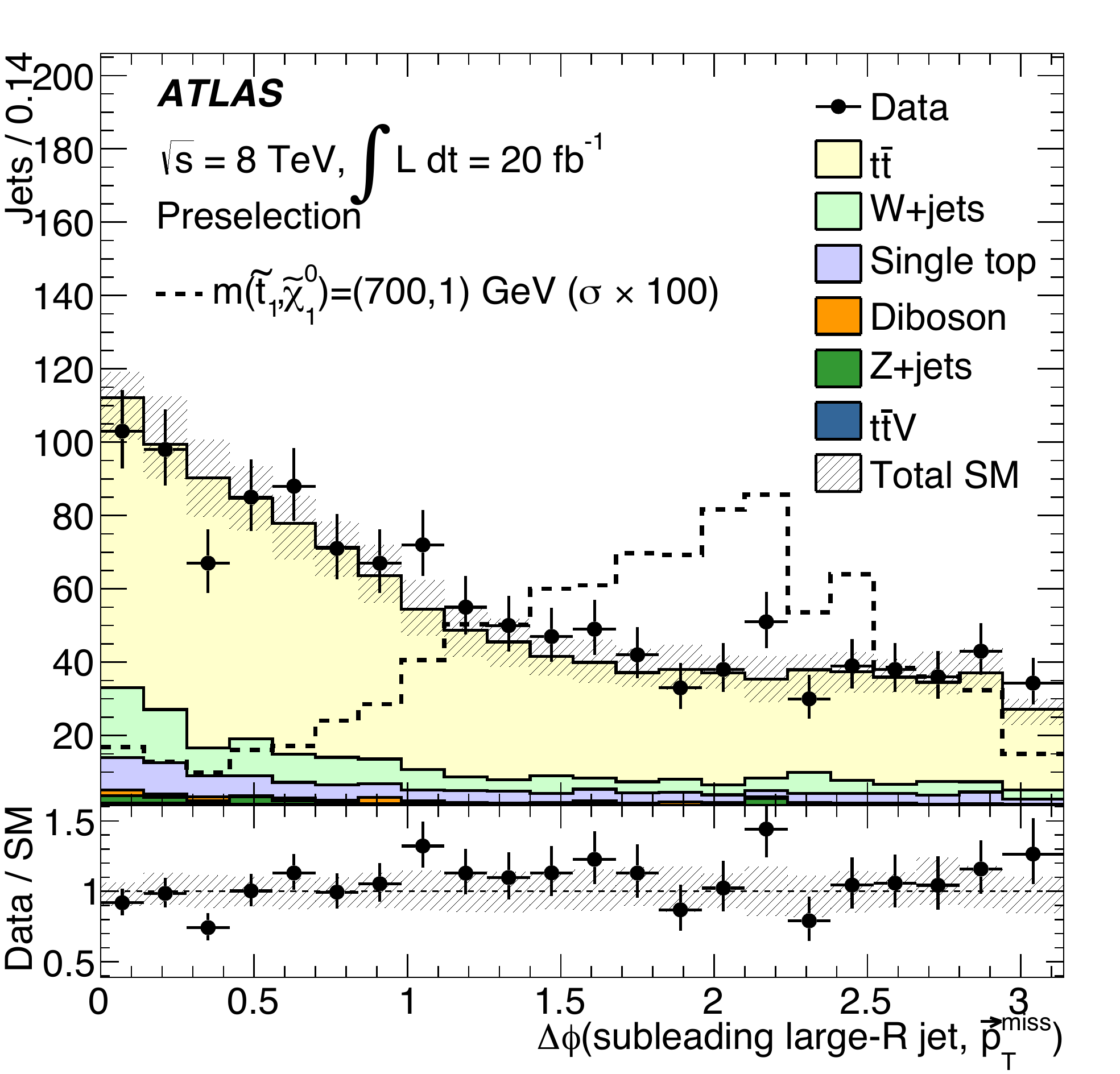} 
\caption{Comparison of data with background expectations of \largeR\ jet distributions: multiplicity (top left), invariant mass (top right), transverse momentum (bottom left), and distance in $\phi$ space between the second-highest-\pt\ \largeR\ jet and \Ptmiss. Events 
are required to pass the preselection defined in \secref{sec:signal:trigger_event_preselection}. In addition, the jet thresholds are tightened
($\pT > 75, 65, 40, 25$\,\GeV) and a requirement of $\mt > 120$\,\GeV\ is imposed. 
The uncertainty band includes statistical and all experimental systematic uncertainties. The last bin includes overflows. A benchmark signal model, with cross-section enhanced by a factor of $100$, is overlaid for comparison.
\label{fig:datamc_preselection_boost}}
\end{center}
\end{figure}

\subsection{Selections for the \bChargino\ decay}\label{sec:bChargino_decay}

Nine SRs target scenarios where both stops decay as \bChargino\ followed by subsequent 
$\chinoonepm \to W^{(*)} \ninoone$ decays.
The mass of the lightest chargino $m(\chinoonepm)$ relative to the \tone\ and \ninoone\ masses 
largely defines the kinematic properties. \Figref{fig:bChargino_mass_hierarchies} schematically illustrates the four distinct mass hierarchies, whose SRs are described below. 

\paragraph{Selections for mass hierarchy (a):\\}
The selection of signal events with a small overall mass splitting, $\Delta M=m(\tone)-m(\ninoone)$, relies on the presence of an initial-state radiation (ISR) jet, against which the stop decay products recoil. 
Consequently, events with a hard leading jet are selected together with a \softLepton\ and relatively soft sub-leading jets.
The leading jet must not satisfy the $b$-tagging criteria, while at least one $b$-tagged jet amongst the sub-leading jets is required.

Two SRs, labelled \SRoneLoneBc\ and \SRoneLoneBa, are defined to probe scenarios 
with a mass splitting $\Delta M \lesssim 50$\,\GeV, and 
$50\,\GeV \lesssim \Delta M \lesssim 80$\,\GeV, respectively. The SR event selections are listed in \tabref{tab:SRs_softLeptons}.
The requirement of $\ge 3$ jets suppresses the \Wjets\ background in \SRoneLoneBa. 
For \SRoneLoneBc, the jet multiplicity requirement is lowered to $\ge 2$ to avoid large signal acceptance losses, 
but tighter \met\ and $\met / \meff$ thresholds are applied 
to keep the \Wjets\ and multijet backgrounds suppressed.
\Figref{fig:datamc_preselection_softlep} compares data with estimated backgrounds in the lepton \pt\ and $\met/\meff$ distributions.
The overlaid stop benchmark model motivates the selection of low-\pt\ leptons, and the background estimates show the non-negligible contribution from multijet events (with mis-identified leptons).

\begin{table}
\begin{center}
\setlength{\tabcolsep}{0.36pc}
\renewcommand{\arraystretch}{1.5}
{\small
\begin{tabular}{| l | l | l | l | l |}
\hline
 & \SRoneLoneBc & \SRoneLoneBa & \SRoneLtwoBa & \SRoneLtwoBc \\
\hline
\textbf{Preselection} & \multicolumn{4}{ |c| }{ \softLeptonHyphen\ preselection, cf. \tabref{tab:preselection}.}\\
\hline
\textbf{Lepton} & \multicolumn{2}{ |c| }{ $=1$ \softLepton } & \multicolumn{2}{ |c| }{ $=1$ \softLepton\ with $\pT < 25$\,\GeV}\\
\hline 
\textbf{Jets} & $\ge 2$ with  & $\ge 3$ with  &  \multicolumn{2}{ |c| }{$\ge 2$ with}\\
                    & $\pT > 180,25$\,\GeV & $\pT > 180,25,25$\,\GeV &  \multicolumn{2}{ |c| }{$\pT > 60,60$\,\GeV}\\
\hline
\textbf{$\Delta\phi ({\mathrm jet}_{i}, \vec{p}_\text{T}^\text{miss})$} & \multicolumn{2}{ |c| }{--}  & \multicolumn{2}{ |c| }{$>0.4$ (i=1,2)} \\
\hline
\textbf{Jet veto} &  \multicolumn{2}{ |c| }{--}  & $\HTtwo < 50$\,\GeV  &    \multicolumn{1}{ |c| }{--} \\
\hline
\textbf{$b$-tagging} & \multicolumn{2}{ |c| }{$\ge 1$ sub-leading jet $b$-tagged (70\% eff.)} & \multicolumn{2}{ |c| }{Leading two jets $b$-tagged (60\% eff.)}\\
\hline
\textbf{$b$-veto} & \multicolumn{2}{ |c| }{$1^{\rm{st}}$ jet not $b$-tagged (70\% eff.)} & \multicolumn{2}{ |c| }{ -- }\\
\hline
$\bm{m_{bb}}$ & \multicolumn{2}{ |c| }{--} & \multicolumn{2}{ |c| }{$>150$\,\GeV}\\
\hline
$\bm{\met}$ & $>370$\,\GeV & $>300$\,\GeV & $>150$\,\GeV & $>250$\,\GeV \\
\hline
$\bm{\met / \meff}$ & $>0.35$ & $>0.3$ & \multicolumn{2}{ |c| }{--} \\
\hline
$\bm{\mt}$ & $>90$\,\GeV & $>100$\,\GeV &  \multicolumn{2}{ |c| }{--} \\
\hline
\hline
\multicolumn{5}{ |l| }{ \textbf{Model-dependent selection:} shape-fit}   \\
\hline
 & \multicolumn{2}{ |c| }{ 4 bins in lepton \pT\ range [$6 (7)$, $50$]\,\GeV} & \multicolumn{2}{ |c| }{ 6 bins in \amtTwo\ range [$0$, $500$]\GeV}\\
\hline
\multicolumn{5}{ |l| }{ \textbf{Model-independent selection:} 1 bin with}   \\
\hline
           & \multicolumn{2}{ |c| }{ lepton $\pT < 25$\,\GeV } & $\amtTwo >170$\,\GeV & $\amtTwo >200$\,\GeV \\
\hline
\end{tabular}
}
\end{center}

\caption{Selection criteria for \softLeptonHyphen\ SRs, employed to search for \bChargino\ decays.
The two leftmost/rightmost SRs target mass hierarchies (a)/(b), illustrated in \figref{fig:bChargino_mass_hierarchies}.
\label{tab:SRs_softLeptons}
}
\end{table}

\begin{figure}
\begin{center}
\includegraphics[width=0.49\textwidth]{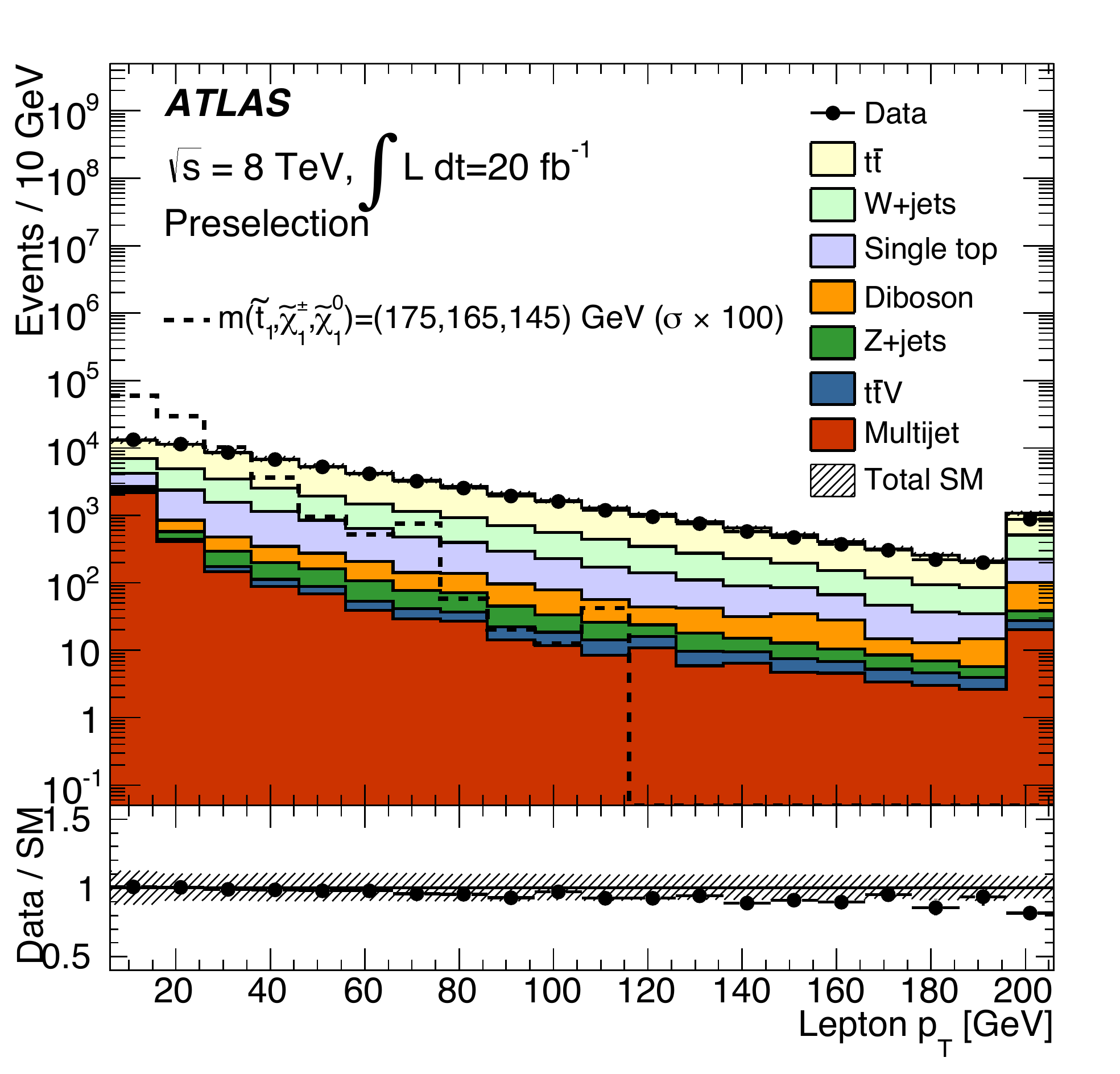} 
\includegraphics[width=0.49\textwidth]{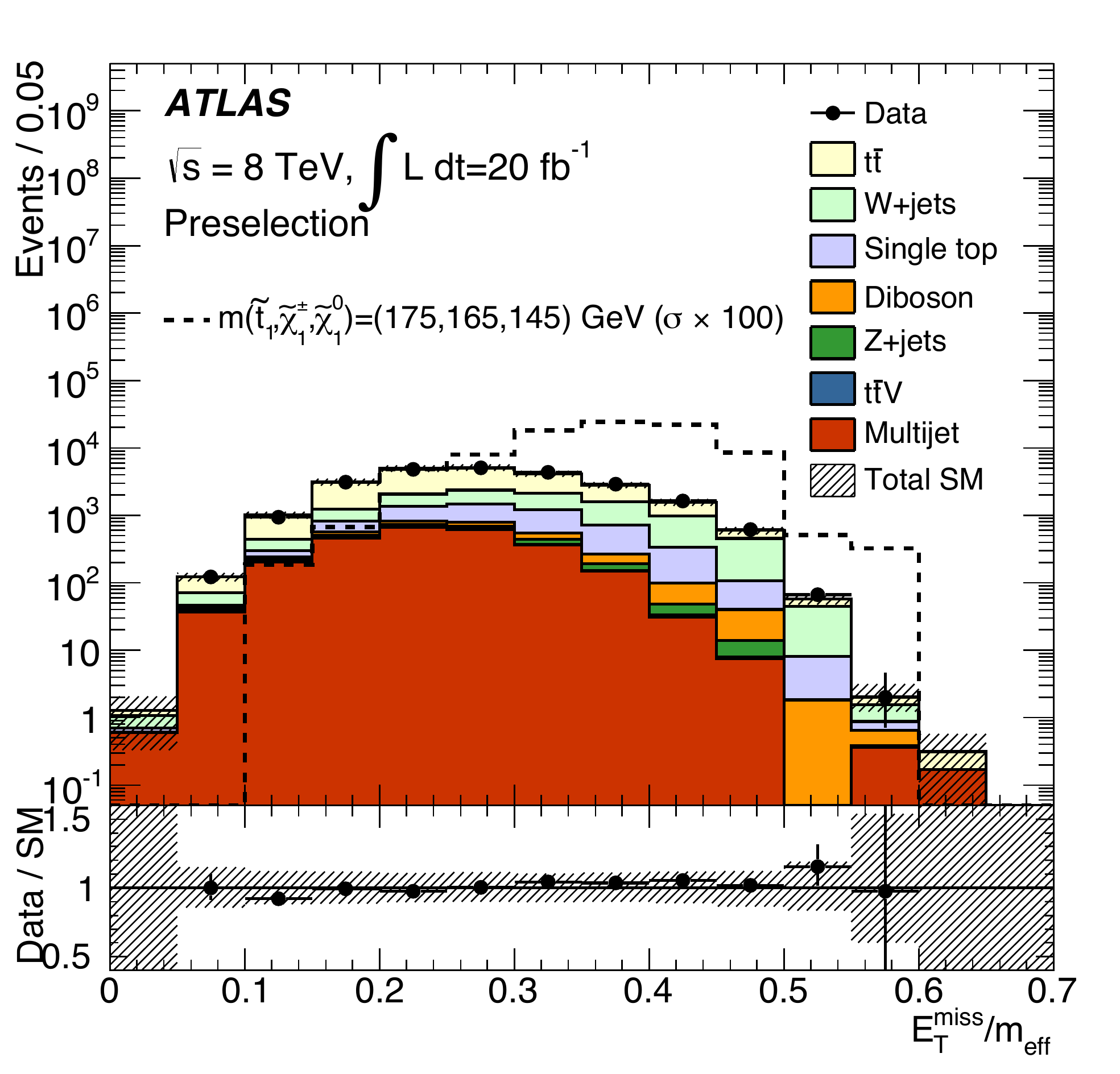} 
\caption{
Comparison of data with estimated backgrounds in the lepton \pt\ (left) and \met/\meff\ (right)
distributions. 
Events are required to satisfy the \softLeptonHyphen\ preselection criteria (cf. \tabref{tab:preselection}),
have $\mt > 40$\,\GeV, and contain two or more jets ($\pt > 130,25$\,\GeV) of which the leading one must not be 
$b$-tagged while the sub-leading one is required to be $b$-tagged.
The \ttbar\ and \Wjets\ backgrounds are normalised using control regions, 
the multijet background is estimated directly from data, and all other backgrounds are normalised to their theoretical predictions (as described in \secref{sec:backgrounds}).
The uncertainty band includes statistical and all experimental systematic uncertainties. The last bin includes overflows.
A benchmark signal model, with cross-section enhanced by a factor of $100$, is overlaid for comparison.
\label{fig:datamc_preselection_softlep}}
\end{center}
\end{figure}

Model-dependent exclusion results are obtained using a shape-fit in the lepton \pT\ variable with four bins of approximately uniform widths in the range [$6 (7)$, $50$]\,\GeV\ for muons (electrons).
For model-independent results, the cut-and-count approach is used with an additional lepton $\pT < 25$\,\GeV\ requirement.

\paragraph{Selections for mass hierarchy (b):\\}
Signal scenarios with a moderately large $\Delta M$ but a small $\Delta m = m(\chinoonepm)-m(\ninoone)$ feature 
two high-\pT\ $b$-jets and  low-momentum decay products from the two off-shell $W$ bosons.

Two SRs, labelled \SRoneLtwoBa\ and \SRoneLtwoBc, employ event selections based on the presence of one \softLepton\ and two $b$-tagged jets.
They target medium and high stop mass regions, respectively. 
The complete event selections are listed in \tabref{tab:SRs_softLeptons}.
The \SRoneLtwoBa\ SR employs an $\met > 150$\,\GeV\ requirement, the lowest possible to retain full \met\ trigger efficiency. 
For models with a heavier \tone, a higher \met\ threshold improves the sensitivity.
The dominant background stems from \ttbar\ production and is suppressed by vetoing additional hard jet activity. 
The variable \HTtwo\ is defined like \HT\ but without including the two leading jets.
The \SRoneLtwoBc\ SR omits the jet activity veto to compensate for the loss in signal acceptance associated with the more stringent \met\ requirement.
Beyond the kinematic \amtTwo\ bound, the dominant source of background arises from mis-tagged $c$-jets in semileptonic \ttbar\ events, and the production of a $W$ boson in association with heavy-flavour jets.
To minimise the mis-tagged background, the $b$-tagging algorithm is operated at the $60\%$ efficiency working point.
A minimum requirement on the invariant mass of the two $b$-tagged jets, $m_{bb}$, is imposed to reduce the contribution from $W + b\bar{b}$ events. 

Exclusion results are obtained using a shape-fit in the \amtTwo\ variable with six bins in the range [$0$, $500$]\,\GeV\ with a uniform bin width.
For all model-independent results, the cut-and-count approach is used but applying an $\amtTwo >170 (200)$\,\GeV\ requirement in 
\SRoneLtwoBa\ (\SRoneLtwoBc).

\begin{figure}
  \centering
  \includegraphics[width=0.49\textwidth]{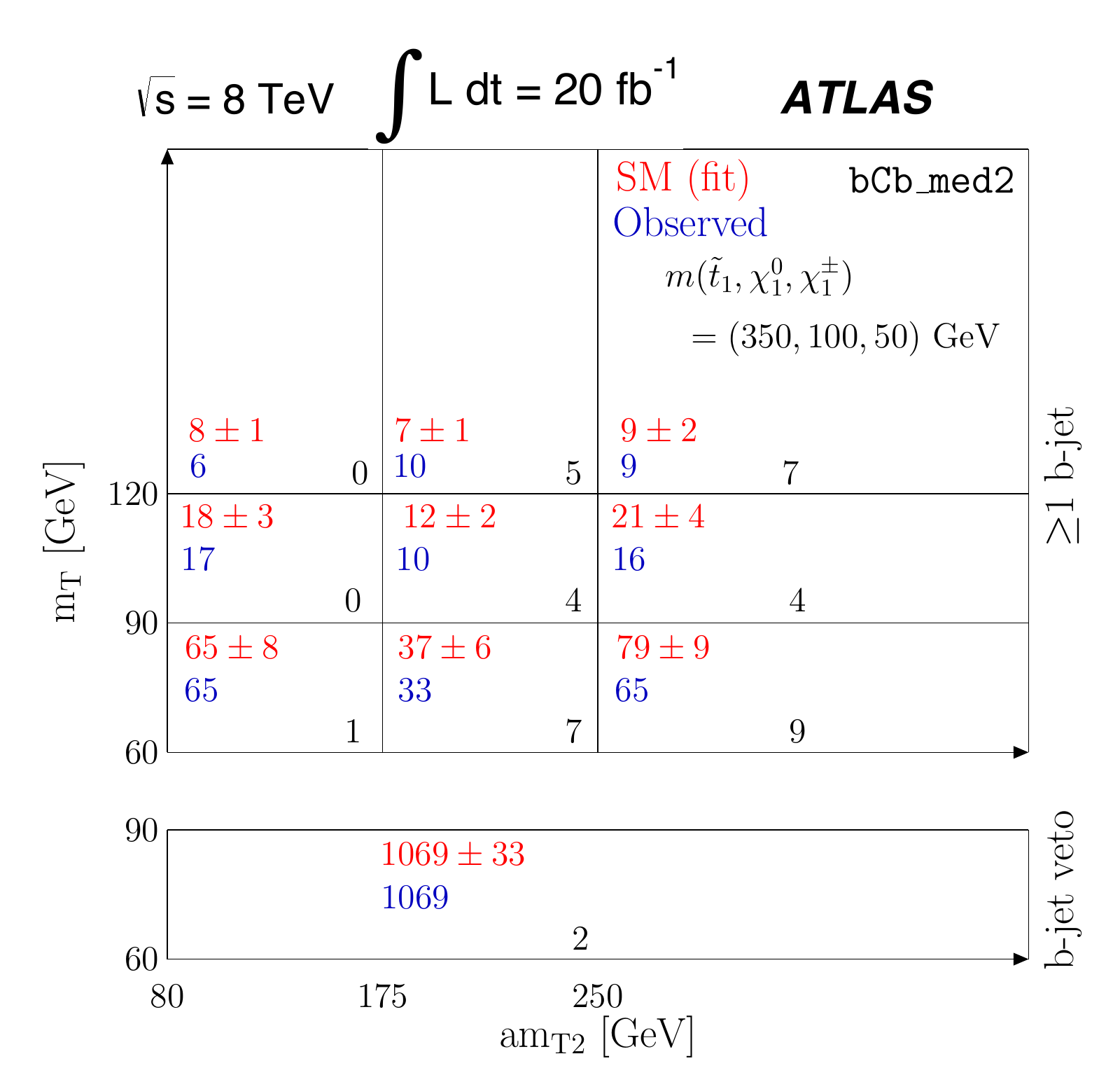}
  \includegraphics[width=0.49\textwidth]{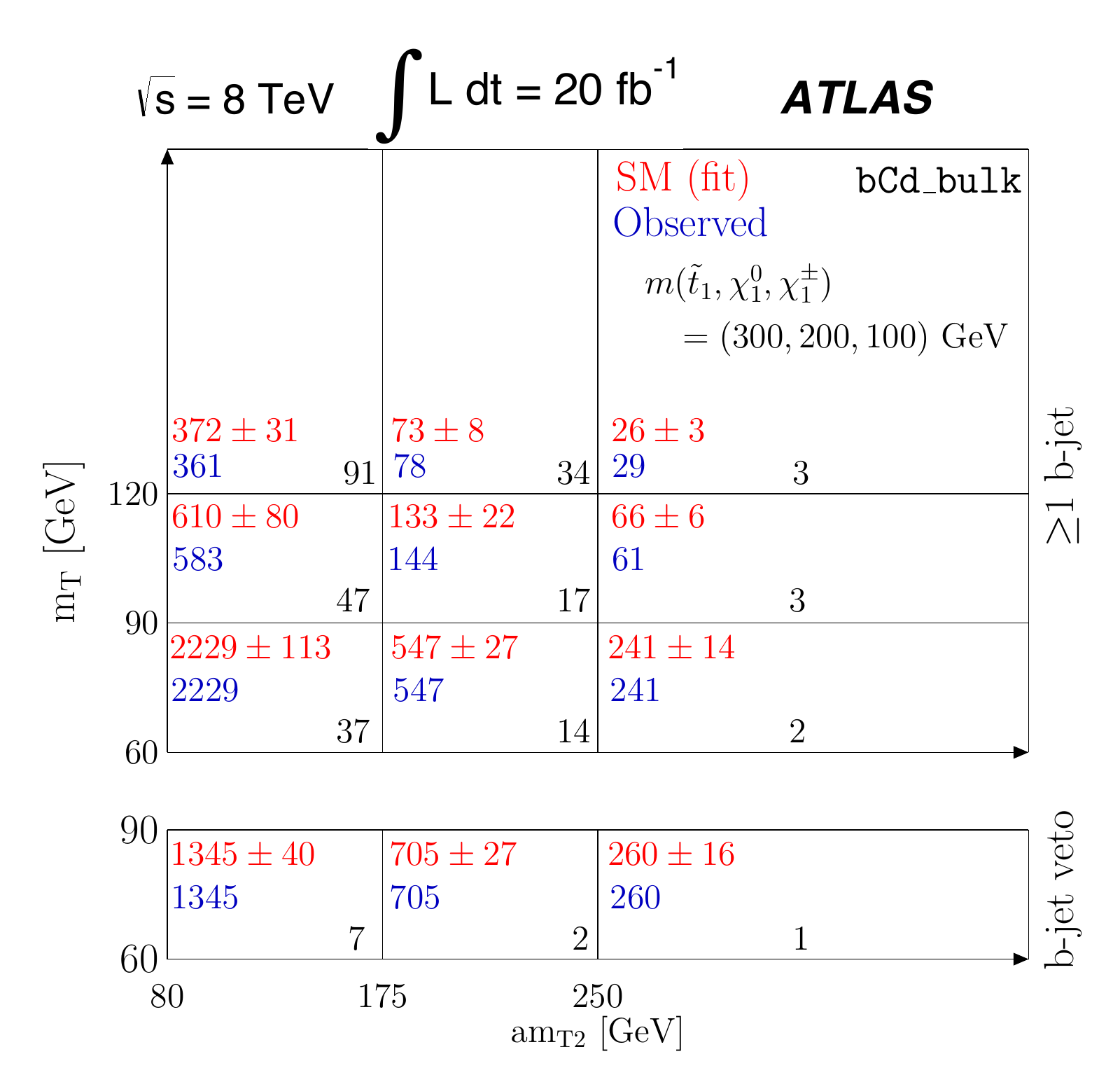}
  \caption{Schematic illustration of the \SRbCvW\ (left) and \SRbCone\ (right) shape-fit binning. More details are given in the caption of \figref{fig:tN_3b_shapeFits}.
    The data and estimated background are in perfect agreement in the six bottom bins of the right plot because
  the fit is configured to use these six bins together with six free parameters; the fit used for the left plot employs
  the bottom four bins and two free parameters.
  }
  \label{fig:bC_shapeFits}
\end{figure}

A third SR, labelled \SRbCvW, also targets intermediate stop masses. It is based on the default lepton selection ($\pT >25\,\GeV$), and requiring at least two high-\pT\ $b$-tagged jets
to exploit larger \tone--\chinoonepm\ mass splittings than \SRoneLtwoBa.
The full event selection is detailed in the leftmost column in \tabref{tab:overview_SRbC}.
The analysis employs a two-dimensional shape-fit technique similar to the one used for \SRtNonep\ but in the \mt\ and \amtTwo\ variables.
\Figref{fig:bC_shapeFits} 
(left plot) illustrates the configuration.  The highest signal sensitivity is obtained from bins in the top-right region.

\afterpage{
  \clearpage
\begin{landscape}
\begin{table}
{\centering
\setlength{\tabcolsep}{0.36pc}
\renewcommand{\arraystretch}{1.5}
{\small
\begin{tabular}{ | l | l | l | l | l | l | }
\hline
                                    & \SRbCvW &  \SRbCzero & \SRbCone & \SRbCfour & \SRbCfive  \\
\hline
\textbf{Preselection} & \multicolumn{5}{ |c| }{ Default preselection criteria, cf. \tabref{tab:preselection}.}\\
\hline
\textbf{Lepton}  & $=1$ lepton & $=1$ lepton with $|\eta( \ell)| < 1.2$ &  \multicolumn{3}{ |c| }{$=1$ lepton}  \\
\hline
\textbf{Jets}    & $\ge 4$ with & $\ge 3$ with & \multicolumn{2}{ |c| }{ $\ge 4$ with} & $\ge 4$ with  \\
& $\pT > 80,60,40,25$\,\GeV & $\pT > 80,40,30$\,\GeV & \multicolumn{2}{ |c| }{$\pT > 80,60,40,25$\,\GeV} & $\pT > 80,80,40,25$\,\GeV  \\
\hline
\textbf{$b$-tagging/veto} & $\ge 2$ (80\% eff.) with & $= 0$ (70\% eff.) with & $\ge 1$ (70\% eff.) with & $\ge 2$ (80\% eff.) with & $\ge 2$ (80\% eff.) with  \\
 & $\pT > 140,75$\,\GeV  & $\pT > 25$\,\GeV & $\pT > 25$\,\GeV  & $\pT > 75,75$\,\GeV & $\pT > 170,80$\,\GeV   \\
 \hline
$\bm{\met}$ & $>170$\,\GeV & $>140$\,\GeV & \multicolumn{2}{ |c| }{$>150$\,\GeV} & $>160$\,\GeV  \\
\hline
$\bm{\mt}$   & $>60$\,\GeV & $>120$\,\GeV & $>60$\,\GeV & \multicolumn{2}{ |c| }{$>120$\,\GeV}  \\
\hline
$\bm{\metsig}$   & $> 6\,\GeV^{1/2}$   & $> 5\,\GeV^{1/2}$ & $> 7\,\GeV^{1/2}$ &  $> 9\,\GeV^{1/2}$  &  $> 8\,\GeV^{1/2}$   \\
\hline
$\bm{\amtTwo}$       & \multicolumn{1}{ |c| }{$>80$\,\GeV} & --  & $>80$\,\GeV   & $>200$\,\GeV   & $>250$\,\GeV   \\
\hline
\textbf{Track, $\tau$-veto} & track \& loose $\tau$-veto & -- & \multicolumn{3}{ |c| }{track \& loose $\tau$-veto} \\
\hline
$\bm{\Delta R(j_1, \ell)}$        & --      & $\in$ [$0.8$, $2.4$]                        & \multicolumn{3}{ |c| }{--}  \\
\hline
$\bm{\Delta\phi(\text{jet}, \vec{p}_\text{T}^\text{miss})}$ &  $>0.8$ ($1^{\rm{st}}$ and $2^{\rm{nd}}$ jet)   & $>2.0$ ($1^{\rm{st}}$ jet), $>0.8$ ($2^{\rm{nd}}$ jet) &  \multicolumn{3}{ |c| }{$>0.8$ ($1^{\rm{st}}$ and $2^{\rm{nd}}$ jet)}   \\
\hline
\hline
\multicolumn{6}{ |l| }{ \textbf{Model-dependent selection:}}   \\
\hline
 & shape-fit in \mt\ and   & cut-and-count  & shape-fit in \mt\ and \amtTwo, & \multicolumn{2}{ |c| }{ cut-and-count }\\
 & \amtTwo, cf. \figref{fig:bC_shapeFits}. & & cf. \figref{fig:bC_shapeFits}. &  \multicolumn{2}{ |c| }{ } \\
\hline
\multicolumn{6}{ |l| }{ \textbf{Model-independent selection:}}   \\
\hline
 & test 4 most signal-sensitive & cut-and-count & test 4 most signal-sensitive & \multicolumn{2}{ |c| }{ cut-and-count }\\
 & bins one-by-one. & &  bins one-by-one. & \multicolumn{2}{ |c| }{ }\\
\hline

\end{tabular}
}
}
\caption{Selection criteria for SRs employed to search for \bChargino\ decays. 
The first SR targets mass hierarchy (b), the next SR is designed for mass hierarchy (c), and the last three SRs are optimised for mass hierarchy (d),
as illustrated in \figref{fig:bChargino_mass_hierarchies}.
\label{tab:overview_SRbC}
}
\end{table}
\end{landscape}
}

\paragraph{Selections for mass hierarchy (c):\\}

Models with $m_{\chinoonepm}$ just below $m_{\tone}$ yield two low-momentum $b$-jets.
The signal selection strategy is based on vetoing events with $b$-tagged jets, assuming both signal $b$-jets are below the jet \pT\ acceptance of the analysis, and therefore 
suppressing the \ttbar\ background.
The sensitivity for low-$m(\tone)$  models is improved by selecting events with ISR-like jet activity. 
One SR, labelled \SRbCzero, is employed and defined in \tabref{tab:overview_SRbC}.
The suffix `{\texttt{diag}}' refers to the diagonal region of the
$m_{\chinoonepm} = 2 m_{\ninoone}$ scenario, the benchmark region used to optimise this SR.
The event selection includes one central lepton ($|\eta| < 1.2$) to suppress the \Wjets\ background, and three or more jets, of which none must satisfy the $b$-tagging criteria. 
In signal events, two of the three required jets tend to originate from the hadronic $W$ boson decay, while
the highest-\pT\ jet typically arises from ISR. 
The $b$-veto strongly suppresses \ttbar\ events, leaving \Wjets\ as the dominant background. 
Requirements on  the $\Delta R(j_1, \ell)$ and \met\  variables further enhance the signal-to-background ratio, by selecting events where the two stops recoil from an ISR jet.

\paragraph{Selections for mass hierarchy (d):\\}

Signal models with relatively large mass splittings between the three mass states, \tone, \chinoonepm, and \ninoone, 
result in events where all particles from the two \tone\ decays are well above the identification \pt\ thresholds. 
Three SRs, labelled \SRbCone, \SRbCfour, and \SRbCfive, target specific
mass regions. 

The \SRbCone\ SR employs a two-dimensional shape-fit technique with bins in the $\mt - \amtTwo$ plane. 
\Figref{fig:bC_shapeFits} (right plot) illustrates the binning. 
Compared to \SRbCvW, which uses the same two variables in a shape-fit, a loose and more inclusive event selection is employed.
The \SRbCfour\ and \SRbCfive\ SRs are based on tight event selections, leading to low expected background yields.
\Tabref{tab:overview_SRbC} details the event selections. 

\subsection{Selections for the mixed, three- and four-body decays}\label{sec:other_decays}

Three additional stop decay modes are considered: 
Events where \topLSP\ and \bChargino\ decays are both allowed, with the branching ratios of the two decays summing to one; 
both stops decay via a three-body process (\threeBody); and both stops undergo a four-body decay (\fourBody).

In the mixed decay mode, models with a very large and a very small $\br(\topLSP)$ are well covered by the SRs targeting the pure \topLSP\ and \bChargino\ decays, respectively. A dedicated SR with the label \SRtNbC\ is optimised for models with $\br(\topLSP) \sim 0.5$.
It employs a requirement on the \topness\ variable, which was designed specifically for the mixed decay mode, to suppress the dominant dileptonic \ttbar\ background.
Diboson events that pass the selection tend to have a leptonic $W$ boson decay and a hadronic $W$ or $Z$ decay, accompanied by at least two additional jets. Large \met\ can be generated by the neutrino when the diboson system is sufficiently boosted; the two additional jets hence typically arise from ISR activity.  
The diboson background is suppressed by placing a loose upper requirement on the three-jet invariant mass, $m_{jjj}$. 
The jet-jet pair with an invariant mass above $60$\,\GeV\ that has the smallest $\mathrm{\Delta}R$ is selected to form the hadronic $V$ boson. The mass $m_{jjj}$ is reconstructed from the third jet closest in $\mathrm{\Delta}R$ to the hadronic $V$ boson momentum vector. 
\Tabref{tab:SRs_alternative} lists the entire event selection.

\begin{table}
\begin{center}
\setlength{\tabcolsep}{0.36pc}
\renewcommand{\arraystretch}{1.5}
{\small
\begin{tabular}{| l | l | l |}
\hline
 & \SRtNbC  & \SRbWN \\
\hline
\textbf{Preselection} & \multicolumn{2}{ |c| }{Default preselection criteria, cf. \tabref{tab:preselection}.}\\
\hline
\textbf{Lepton} & \multicolumn{2}{ |c| }{ $=1$ lepton } \\
\hline
\textbf{Jets}  & $\ge 4$ jets with $\pT > 80,70,50,25$\,\GeV & $\ge 4$ jets with $\pT > 80,25,25,25$\,\GeV \\
\hline
\textbf{$b$-tagging } &  $\ge 1$ $b$-tag ($70 \%$ eff.) with $\pt > 60$\,\GeV &  $\ge 1$ $b$-tag ($70 \%$ eff.) with $\pt > 25$\,\GeV \\
\hline
\met  & $> 270$\,\GeV & $> 150$\,\GeV \\
\hline
\mt  & $> 130$\,\GeV & $> 60$\,\GeV \\
\hline
\amtTwo  & $> 190$\,\GeV & $>80$\,\GeV \\
\hline
\topness & $> 2$ & $-$ \\
\hline
$\bm{m_{jjj}}$  & $ < 360$\,\GeV & $-$ \\
\hline
$\bm{\metsig}$  & $> 9\,\GeV^{1/2}$ & $> 5\,\GeV^{1/2}$ \\ 
\hline
\textbf{$\tau$-veto} & \multicolumn{2}{ |c| }{loose}  \\
\hline
$\bm{\Delta\phi(\text{jet}_i, \vec{p}_\text{T}^\text{miss})}$ & $>0.6$ {\scriptsize($i=1,2$)} & $>0.2$ {\scriptsize ($i=1,2$)} \\
\hline
$\bm{\Delta\phi(\ell, \vec{p}_\text{T}^\text{miss})}$  & $>0.6$ & $>1.2$ \\
\hline
$\bm{\Delta R(\ell, \text{jet}_i})$ & $<2.75$ {\scriptsize($i=1$)} & $>1.2$ {\scriptsize ($i=1$)}, $> 2.0$ {\scriptsize ($i=2$)} \\
\hline
$\bm{\Delta R(\ell, b\text{-jet})}$ & $<3.0$ & $-$ \\
\hline
\hline
\multicolumn{3}{ |l| }{\textbf{Model-dependent selection:}}  \\
\hline
 & cut-and-count  & shape-fit in \mt\ and \amtTwo, cf. \figref{fig:tN_3b_shapeFits}.  \\
\hline
\multicolumn{3}{ |l| }{\textbf{Model-independent selection:}}   \\
\hline
 & cut-and-count &test 4 most signal-sensitive   \\
 &  & bins one-by-one.   \\
\hline
\end{tabular}
}
\end{center}

\caption{Selection criteria for the two SRs employed to search for 
the mixed \topLSP\ and \bChargino\ decay mode (left), and
the three-body decay, \threeBody\ (right).
\label{tab:SRs_alternative}
}
\end{table}

A dedicated SR labelled \SRbWN\ is optimised for the three-body decay mode. 
Compared to the scenario with on-shell top quarks, three-body decays yield the same final state objects 
but with significantly lower momenta, although typically still above the reconstruction thresholds.
The dileptonic \ttbar\ background is separated from signal in the very low \amtTwo\ regime. The three-body signal peaks in \amtTwo\ below around $100$\,\GeV\ due to the kinematic construction of the variable and the fact that $m(\tone) - m(\ninoone)$ is below the top quark mass. 
A two-dimensional shape-fit technique using the \mt\ and \amtTwo\ variables
is employed, similar to that used in \SRbCone\ and \SRbCvW, but with different binning. The configuration is illustrated in \figref{fig:tN_3b_shapeFits} (right plot). 
Fine binning is used in the low \amtTwo\ region where the highest signal sensitivity is obtained.
The full \SRbWN\ event selection is detailed in \tabref{tab:SRs_alternative}.

The four-body decay mode is characterised by events with final state objects that tend to have even lower momenta than for three-body decays. The selections based on a \softLepton\ designed for the overall `compressed' mass hierarchy  (a) 
provide good search sensitivity for this scenario.

 \section{Background estimates}\label{sec:backgrounds}

The dominant sources of background are the production of \ttbar\ events and \Wjets\ where the $W$ decays leptonically. 
Other background processes considered are 
single top, dibosons, $Z$+jets, \ttbar\ produced with a vector boson ($\ttbar V$), and multijet events. 

The \ttbar\ and \Wjets\ backgrounds are estimated by isolating each of them in a dedicated control region, normalising simulation to match data in that control region, and then using simulation to extrapolate the background predictions into the SR. A detailed description of the method and its validation are given below. 

The multijet background is estimated from data using a matrix method described in refs.~\cite{Aad:2012ms,ATLAS:2011ad}. 
The contribution is found to be negligible for all but the \softLeptonHyphen\ selections. 
All other (small) backgrounds are determined entirely from simulation, normalised to the most accurate theoretical cross-sections available (cf. \secref{sec:samples}).

\subsection{Control regions}\label{sec:control_regions}

Each cut-and-count analysis has two orthogonal CRs which are enriched in either \ttbar\ events (TCR) or \Wjets\ events (WCR). These CRs are used to normalise the corresponding backgrounds in data, specifically for the associated SR.
For the two-dimensional shape-fits, bins enriched in \ttbar\ events are already part of the SR and act as a TCR, while additional bins are used to normalise the \Wjets\ background.
The one-dimensional shape-fits are set up similar to the cut-and-count analyses, with two additional bins acting as TCR and WCR.
The CRs are designed to select events as similar as possible to those selected by the corresponding SR
while keeping the contamination from other backgrounds and potential signal low. 
The CRs are also chosen to retain a sufficiently large number of events to not be limited by the 
statistical precision. 
This background estimation approach improves the robustness against potential mis-modelling effects in simulation since the dependence on simulation is reduced, and hence it reduces the uncertainties on the background estimates.

A likelihood fit is performed for each analysis, involving all associated bins:
SR, TCR, WCR for cut-and-count or all bins of a shape-fit.
Each bin is modelled by a separate probability density function based on a Poisson term, 
where the expected number of events is given by the sum over all background processes, and optionally a signal model.
The normalisation of the \ttbar\ and \Wjets\ backgrounds is controlled by two free parameters in the fit.\footnote{
For some of the two-dimensional shape-fits, the \ttbar\ and \Wjets\ backgrounds are controlled by more than two parameters, as discussed later in the text.} 
To obtain a set of background predictions that is independent of the observation in the SRs, the fit can be configured 
to use only the CRs to constrain the fit parameters: the SR bins (or signal-sensitive bins in shape-fits) are removed from the likelihood and any potential signal contribution is neglected everywhere. This fit configuration, referred to as the background-only fit, is used throughout this section.
The treatment of systematic uncertainties in the likelihood fits is discussed in \secref{sec:systematics}.
To quantify a potential excess, or to derive exclusion limits, the SR bins are included in the likelihood,
as further detailed in \secref{sec:results}. 

\begin{figure}
\begin{center}
\includegraphics[width=0.49\textwidth]{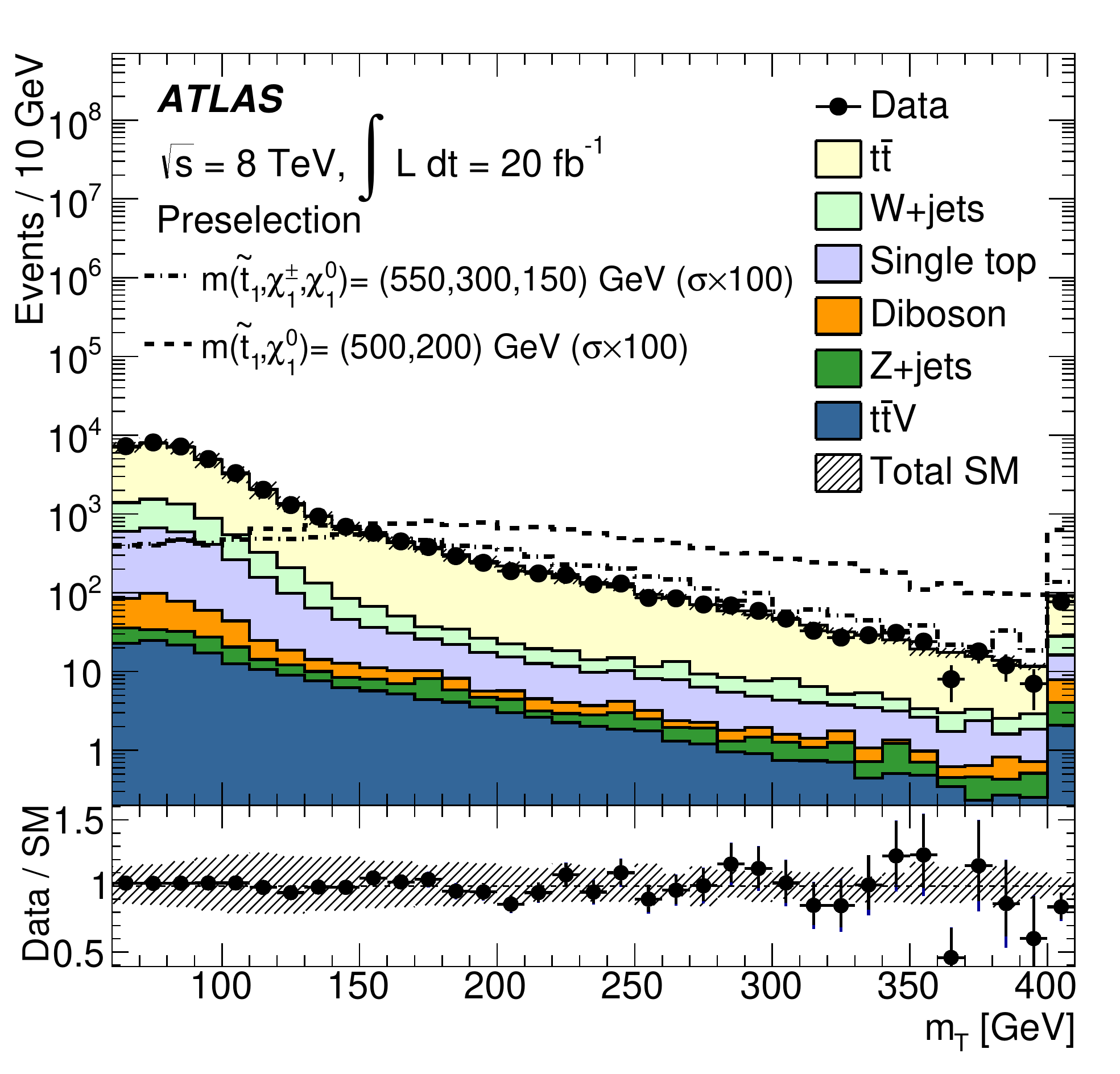} 
\includegraphics[width=0.49\textwidth]{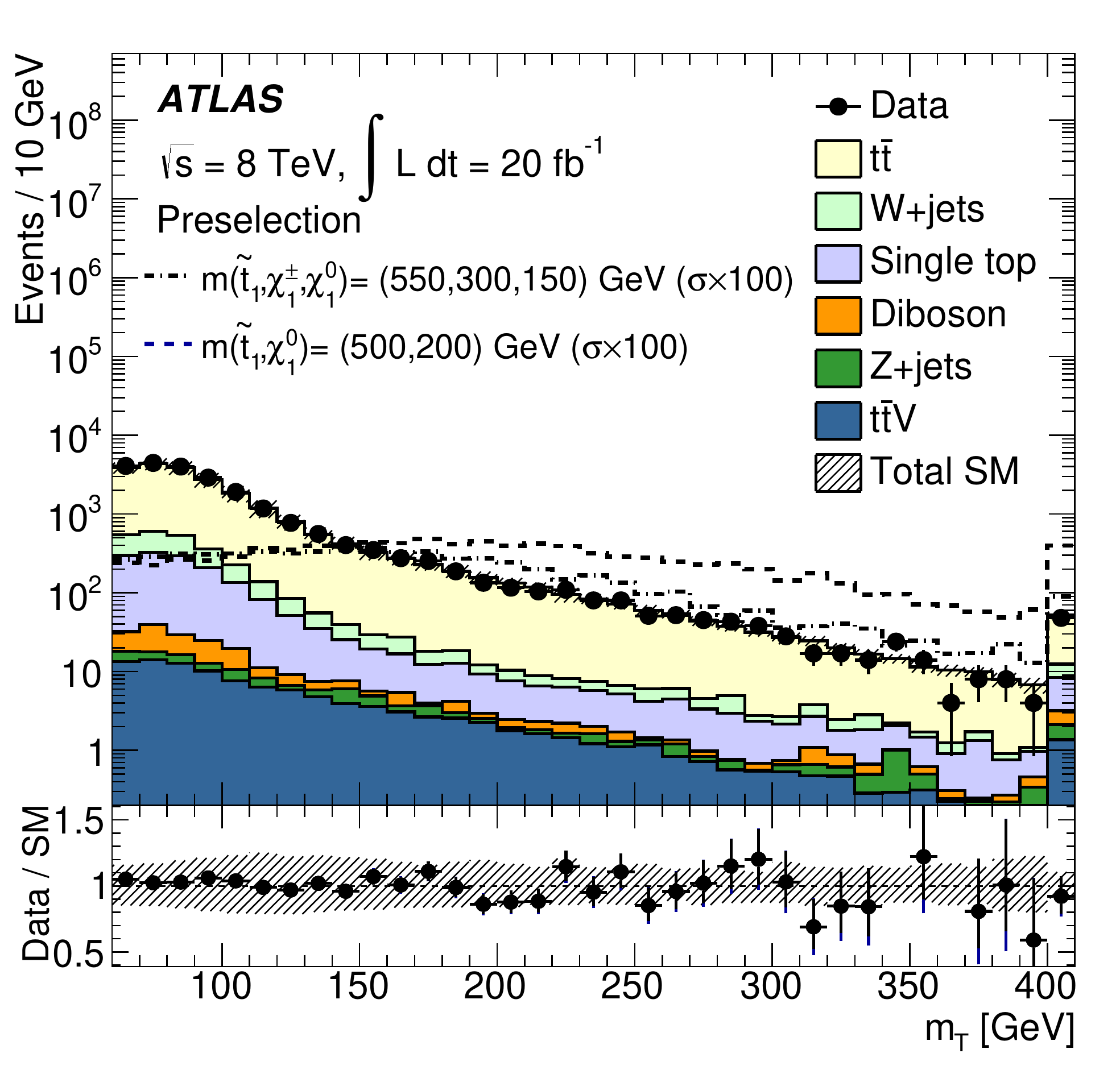} 

\caption{Distribution of the transverse mass, \mt, for events that pass the same preselection as used in \figref{fig:datamc_preselection} and with at least one (left) or at least two (right) $b$-tagged jets.
The uncertainty band includes statistical and all experimental systematic uncertainties. The last bins include the overflows. Benchmark signal models with cross-sections enhanced by a factor of $100$ are overlaid for comparison.
\label{fig:pre_mt}}
\end{center}
\end{figure}

The key observable used to define the TCR and WCR for most analyses is the transverse mass. 
\Figref{fig:pre_mt} shows the \mt\ distribution for events passing the preselection defined in \secref{sec:signal:trigger_event_preselection} and with $\ge 1$ (left) or $\ge 2$ $b$-tagged jets (right) using the 70\% and 80\% working points, respectively. The \ttbar\ and \Wjets\ backgrounds drop sharply beyond the $W$ boson mass, while signal events can exceed this kinematic endpoint due to the two additional LSPs in the event. 
In all cut-and-count analyses the CRs differ from the associated SR by the \mt\
requirement, which is set to $60$\,\GeV $< \mt < 90$\,\GeV\ for the CRs. 
For the four SRs based on a \softLepton,
which all employ one-dimensional shape-fits,  
the two CRs are defined by loosening the \met\ and \mt\ requirements (\SRoneLoneBa\ and \SRoneLoneBc), or
the \softLeptonHyphen\ selection of $6(7)\,\GeV < \pt(\ell) < 25$\,\GeV\ for muons (electrons) in the SR is changed to a $\pt(\ell) > 25$\,\GeV\ requirement in the CRs (\SRoneLtwoBa\ and \SRoneLtwoBc).
All WCRs have 
a $b$-tag veto instead of a $b$-tag requirement to reduce the \ttbar\ contamination;
consequently all requirements related to the presence of a $b$-jet are removed.
The $b$-tag requirement used in all but one SR enhances the heavy-flavour contribution of the \Wjets\ background, 
while the WCRs with the $b$-tag veto predominantly select light-flavour \Wjets\ events. 
A systematic uncertainty and a dedicated validation related to this effect are described in \secref{sec:systematics} and in the next subsection, respectively.
The TCRs employ the same $b$-tagging requirement as used in the associated SRs, except for \SRbCzero\ where the $b$-tag veto in the SR is turned into a $\ge 1$ $b$-tag requirement in the TCR (using the same $b$-tagging working point as for the $b$-tag veto in the SR).
Furthermore, some other kinematic requirements are slightly loosened or removed to increase the event yields in the CRs.
The event selections for all CRs associated with the cut-and-count or the one-dimensional shape-fit analyses are specified in \tabref{tab:CR_definitions}.
In each analysis the same two CRs are used for the model-dependent and the model-independent fit configurations.

\begin{table}
\small
\centering
\begin{tabular}{|l|l|l|l|l|l|l|}
\hline
Analysis & Variable &\multicolumn{2}{|c|}{Control regions} & \multicolumn{2}{|c|}{Validation regions} &  Signal reg.\\
 &   &\multicolumn{1}{|c|}{TCR} & \multicolumn{1}{|c|}{WCR} & TVR & WVR &  \multicolumn{1}{|c|}{SR} \\
\hline
\hline
All \texttt{tN\_*}  & \mt  & $[60, 90] $ & $[60,90] $ & $[90,120] $ & $[90,120] $ & $> [140,200] $ \\
			& $N_{b}$ & $\geq 1$ & $= 0$ & $\geq 1$ & $= 0$ & $\geq 1$ \\ 
\hline
\textbf{\SRtNtwox} & \amtTwo & $> 120 $ & $> 120 $ &$> 120 $ & $> 120 $ & $> 170 $ \\
\hline
\textbf{\SRtNthreep} & \amtTwo & - & - &  - & - & $ > 170 $\\
&\mtTwoTau & - & - & - & - & $> 120 $ \\
& \met & $> 225 $ & $ > 225 $  & $> 225 $ & $ > 225 $ & $> 320 $\\
& \HTsig & $ > 8.8$ & $> 8.8$ & $ > 8.8$ & $> 8.8$ & $> 12.5$ \\
\hline
\textbf{\SRtNboost} & \amtTwo & $> 130 $ & $ > 130 $ & $> 130 $ & $ > 130 $ &  $> 145 $  \\
& \met &$ > 260 $ & $> 260 $ & $ > 260 $ & $> 260 $ & $ > 315 $   \\
 & topness  & - & - & - & - &$> 7$  \\
\hline
\hline
All \texttt{bCa\_*}   
  & $\pt(\ell)$ & $>6(7) $ & $> 6(7) $ & [6(7),25] & [6(7),25]  & [6(7),25/50]  \\
  & $N_{b}$             & $\geq 1$ & $=0$ & $\geq 1$ & $\geq 1$ &  $\geq 1$ \\
\hline
\textbf{\SRoneLoneBc} & $\met$       & [200,250] & [200,250] & [250,370]  & [250,370] & $>370 $ \\
               & $\mt$        & [90,120]  & [90,120]  & $>90 $ & $>90 $ & $>90 $ \\
               & $\met/\meff$ & - & - & - & $>0.35$ & $>0.35$ \\
\hline
\textbf{\SRoneLoneBa} & $\met$       & [200,250]  & [200,250]  & [250,300]  & [250,300]  & $>300 $ \\
               & $\mt$        & [100,120] & [100,120]  & $>100 $ & $>100 $ & $>100 $ \\
               & $\met/\meff$ & - & - & - & $>0.3$ & $>0.3$ \\
\hline
All \texttt{bCb\_*} 
  & $\pt(\ell)$ & $>25 $ & $> 25 $ & \multicolumn{2}{|l|}{[6(7),25] } & [6(7),25]  \\
  & $N_{b}$             & $=2$ & $=0$ &  \multicolumn{2}{|l|}{ $=2$ } & $=2$ \\ 
  & $\mt$               & - & [40, 80]  &  \multicolumn{2}{|l|}{ -} & - \\
  & $m_{bb}$            & - & - & \multicolumn{2}{|l|}{$<150 $} & $>150 $ \\
\hline
\textbf{\SRoneLtwoBa} &  $\amtTwo$ & - & - & \multicolumn{2}{|l|}{-} & $> 170 $  \\
\hline
\textbf{\SRoneLtwoBc} &  $\amtTwo$ & - & - & \multicolumn{2}{|l|}{-} & $> 200 $  \\
             &  $\met$    & $>250 $ & $>150 $ &  \multicolumn{2}{|l|}{ $>250 $} & $>250 $ \\
\hline

All \texttt{bCc\_*}, \texttt{bCd\_*}
	& \mt  & $[60, 90] $ & $[60,90] $ & $[90,120] $ & $[90,120] $ & $> 120 $ \\
\hline
\textbf{\SRbCzero} & $N_{b}$ & $\geq 1$ & $= 0$ & $\geq 1$ & $= 0$ & $= 0$ \\ 
\hline
All \texttt{bCd\_*}
& $N_{b}$ & $\geq 2$ & $= 0$ & $\geq 2$ & $= 0$ & $\geq 2$ \\ 
\hline
\textbf{\SRbCfour} &  \amtTwo & $ > 120 $ & $> 200 $ &   $ > 120 $ & $> 200 $ & $> 200 $   \\
\hline
\textbf{\SRbCfive} &  \amtTwo & $ > 120 $ & $ > 250 $  & $ > 120 $ & $ > 250 $  & $> 250$\\
\hline 
\hline 
\textbf{\SRtNbC} & \mt  & $[60, 90] $ & $[60,90] $ & $[90,120] $ & $[90,120] $ & $> 130 $ \\
			& $N_{b}$ & $\geq 1$ & $= 0$ & $\geq 1$ & $= 0$ & $\geq 1$ \\ 
& \amtTwo & $> 120 $ & $> 120 $ & $> 120 $ & $> 120 $ & $> 190 $ \\
& \met & $> 170 $ & $> 170 $ & $> 170 $ & $> 170 $ & $> 270 $ \\
& \metsig & $> 5$ & $> 5$ & $> 5$ & $> 5$ & $> 9$ \\
\hline
\end{tabular}
\caption{
Event selections for control regions, validation regions, and the signal regions of the model-independent selection (defined in \tabsref{tab:SRs_tN}--\ref{tab:overview_SRbC}) associated with cut-and-count or one-dimensional shape-fit analyses.
The asterisk symbol `\texttt{*}' is used as a wildcard to describe variable requirements common to several regions. 
Only one validation region is defined for the \SRoneLtwoBa\ and \SRoneLtwoBc\ selections. 
Variables for which the requirements 
are the same between the regions are not listed. 
Requirements related to the presence of a $b$-tagged jet are removed in all selections with a $b$-tag veto (WCRs and WVRs). 
All units are in \GeV\ except for unitless quantities and \metsig\ which is quoted in \GeV$^{1/2}$.
\label{tab:CR_definitions}}
\end{table}

The four two-dimensional shape-fits (\SRtNonep, \SRbWN, \SRbCvW, and \SRbCone) have built-in bins enhanced in \ttbar\ events, which act as TCRs, while additional WCR bins are added with a $b$-tag veto and with a $60$\,\GeV$ < \mt < 90$\,\GeV\ requirement (as can be seen in \figsref{fig:tN_3b_shapeFits} and~\ref{fig:bC_shapeFits}).
For the background-only fits, the WCRs and the subset of the TCR shape-fit bins with $\mt < 90$\,\GeV\ are used to constrain the likelihood fit. The \ttbar\ and \Wjets\ backgrounds
are normalised separately in each \met\ or \amtTwo\ slice in the \SRtNonep\ and \SRbCone\ shape-fits, as these two have sufficiently large numbers of events in the low-\mt\ bins. Thus, there are three \ttbar\ and three \Wjets\ normalisation parameters, which are applied to all \mt\ bins in the given \met\ or \amtTwo\ range.
This approach increases the robustness of the fit against potential mis-modelling in the simulation at the expense of a reduced statistical precision.
For the other two shape-fits with lower event yields, \SRbWN\ and \SRbCvW, one \ttbar\ and one \Wjets\ normalisation parameter is applied across all bins. 
All shape-fit bins are used to extract model-dependent exclusion limits, while a subset is used for the model-independent results and for the background only fits. This subset includes all bins with $\mt < 90$\,\GeV, acting as CR bins, and in addition for the model-independent results one signal-sensitive bin is included.

Top pair and \Wjets\ production accounts for $70$--$80$\% of events in the TCRs and WCRs.
The signal contamination, for all signal models studied and all CRs, 
is typically at the percent level and never exceeds $10$\%. It is explicitly taken into account when setting 
model-dependent exclusion limits.

\Tabref{tab:background_breakdown} shows the background predictions in each SR. The number of \ttbar\ and \Wjets\ events are estimated using the background-only fit configuration. 
For the four two-dimensional shape-fits, the background predictions are given for the four bins with the highest signal-sensitivity. The quoted uncertainties include all statistical and systematic effects, described in \secref{sec:systematics}.
The numbers of \ttbar\ events normalised in the various TCRs are compatible with the predictions entirely based on simulation and the theoretical cross-section, while the \Wjets\ estimates are about $30$\% lower than, but nonetheless compatible with the predictions from simulation normalised to the theoretical cross-section.
Tables showing the estimated and fitted number of background events in the CRs, validation regions, and SRs of all analyses are shown in appendix~\ref{sec:appendixbkg}.

\afterpage{
\begin{landscape}
\begin{table}
{\centering
{\small
\setlength{\tabcolsep}{0.2pc}
\begin{tabular}{| rr | r | r | r | r | r | r | r | r |}
\hline
\multicolumn{2}{|l|}{Signal region} & \multicolumn{1}{|c|}{Total bkg.} & \multicolumn{1}{|c|}{\ttbar} &  \multicolumn{1}{|c|}{\Wjets} &  \multicolumn{1}{|c|}{Single top} &  \multicolumn{1}{|c|}{Diboson} &  \multicolumn{1}{|c|}{$Z$+jets} &  \multicolumn{1}{|c|}{$\ttbar V$} &  \multicolumn{1}{|c|}{Multijet} \\
\hline
\multicolumn{2}{|l|}{\SRtNtwox}        & $13.0 \pm 2.2$ & $6.5 \pm 1.7$ & $2.1 \pm 0.5$ & $1.1 \pm 0.5$ & $1.4 \pm 0.6$ & $0.009 \pm 0.005$ & $2.0 \pm 0.6$& --\\            
\multicolumn{2}{|l|}{\SRtNthreep}      & $5.0 \pm 1.0$ & $2.0 \pm 0.6$ & $0.87 \pm 0.26$ & $0.54 \pm 0.19$ & $0.86 \pm 0.31$ & $0.0030 \pm 0.0020$ & $0.75 \pm 0.25$& --\\ 
\multicolumn{2}{|l|}{\SRtNboost}       & $3.3 \pm 0.7$ & $1.1 \pm 0.4$ & $0.28 \pm 0.14$ & $0.39 \pm 0.15$ & $0.72 \pm 0.25$ & $0.0040 \pm 0.0020$ & $0.85 \pm 0.28$& --\\    
\hline
\multicolumn{2}{|l|}{\SRoneLoneBc}     & $6.5 \pm 1.4$ & $2.8 \pm 0.9$ & $1.1 \pm 0.5$ & $0.9 \pm 0.4$ & $1.0 \pm 0.6$ & $0.018 \pm 0.017$ & $0.23 \pm 0.09$ & $0.6^{+0.8}_{-0.6}$\\    
\multicolumn{2}{|l|}{\SRoneLoneBa}     & $17 \pm 4$ & $10.5 \pm 3.1$ & $1.2 \pm 0.5$ & $1.7 \pm 0.9$ & $1.1 \pm 0.6$ & $0.022^{+0.134}_{-0.022}$ & $0.47 \pm 0.17$ & $2.5 \pm 2.0$\\    
\multicolumn{2}{|l|}{\SRoneLtwoBa}     & $32 \pm 5$ & $15 \pm 4$ & $9.5 \pm 2.8$ & $6.2 \pm 1.2$ & $0.060 \pm 0.018$ & $0.23 \pm 0.13$ & $0.13 \pm 0.04$ & $0.7^{+2.1}_{-0.7}$\\        
\multicolumn{2}{|l|}{\SRoneLtwoBc}     & $9.8 \pm 1.6$ & $3.8 \pm 0.8$ & $2.5 \pm 0.9$ & $2.9 \pm 0.8$ & $0.075 \pm 0.024$ & $0.10 \pm 0.06$ & $0.27 \pm 0.10$ & $0.18^{+0.52}_{-0.18}$\\
\multicolumn{2}{|l|}{\SRbCzero}        & $470 \pm 50$ & $140 \pm 40$ & $248 \pm 27$ & $12.1 \pm 3.0$ & $59 \pm 17$ & $5.2 \pm 2.6$ & $3.3 \pm 1.1$& --\\                                
\multicolumn{2}{|l|}{\SRbCfour}        & $11.0 \pm 1.5$ & $5.2 \pm 1.0$ & $1.9 \pm 0.5$ & $1.9 \pm 0.8$ & $0.38 \pm 0.18$ & $0.047 \pm 0.024$ & $1.5 \pm 0.5$& --\\                   
\multicolumn{2}{|l|}{\SRbCfive}        & $4.4 \pm 0.8$ & $1.8 \pm 0.4$ & $0.71 \pm 0.33$ & $1.1 \pm 0.5$ & $0.25 \pm 0.19$ & $0.047 \pm 0.024$ & $0.48 \pm 0.16$& --\\                  
\hline
\multicolumn{2}{|l|}{\SRtNbC}         & $7.2 \pm 1.0$ & $2.9 \pm 0.6$ & $1.30 \pm 0.30$ & $0.70 \pm 0.30$ & $0.8 \pm 0.4$ & $<0.001$ & $1.4 \pm 0.4$& --\\

\hline
\multicolumn{2}{|l|}{{\bf \SRtNonep}} &   & & & & & & & \\
~$125 < \met  < 150$\,\GeV, & $120 < \mt < 140$\, \GeV     & $136 \pm 22$ & $123 \pm 22$ & $6.5 \pm 2.5$ & $5.3 \pm 2.2$ & $0.29 \pm 0.18$ & $0.17 \pm 0.08$ & $1.5 \pm 0.5$& --\\ 
~$125 < \met  < 150$\,\GeV, & $\mt > 140$\, \GeV           & $152 \pm 20$ & $137 \pm 20$ & $5.8 \pm 2.6$ & $5.8 \pm 2.1$ & $0.8 \pm 0.5$ & $0.24 \pm 0.12$ & $2.9 \pm 0.9$& --\\   
~$\met > 150$\,\GeV, & $120 < \mt < 140$\, \GeV            & $98 \pm 13$ & $85 \pm 12$ & $4.6 \pm 1.5$ & $5.8 \pm 2.1$ & $0.34 \pm 0.32$ & $0.30 \pm 0.15$ & $2.5 \pm 0.8$& --\\   
~$\met > 150$\,\GeV, & $\mt > 140$\, \GeV                  & $236 \pm 29$ & $202 \pm 27$ & $10 \pm 4$ & $9.0 \pm 3.5$ & $4.0 \pm 1.8$ & $0.53 \pm 0.26$ & $10.8 \pm 3.3$& --\\     
\hline
\multicolumn{2}{|l|}{{\bf \SRbCvW}} &   & & & & & & & \\
$175 < \amtTwo < 250$\,\GeV, & $90 < \mt < 120$\,\GeV      & $12.1 \pm 2.0$ & $8.5 \pm 1.8$ & $1.8 \pm 0.8$ & $1.6 \pm 0.7$ & $0.018 \pm 0.007$ & $<0.001$ & $0.26 \pm 0.10$& --\\     
$175 < \amtTwo < 250$\,\GeV, & $\mt > 120$\,\GeV           & $7.4 \pm 1.4$ & $4.8 \pm 1.2$ & $0.47 \pm 0.19$ & $1.2 \pm 0.6$ & $0.01 \pm 0.10$ & $<0.001$ & $0.85 \pm 0.27$& --\\    
$\amtTwo > 250$\,\GeV, & $90 < \mt < 120$\,\GeV            & $21 \pm 4$ & $12.1 \pm 3.1$ & $3.9 \pm 1.4$ & $4.5 \pm 2.0$ & $0.32 \pm 0.12$ & $<0.001$ & $0.25 \pm 0.08$& --\\          
$\amtTwo > 250$\,\GeV, & $\mt > 120$\,\GeV                 & $9.1 \pm 1.6$ & $4.0 \pm 1.1$ & $2.2 \pm 0.9$ & $1.8 \pm 0.8$ & $0.29 \pm 0.13$ & $0.047 \pm 0.024$ & $0.74 \pm 0.27$& --\\
\hline
\multicolumn{2}{|l|}{{\bf \SRbCone}} &   & & & & & & & \\
$175 < \amtTwo < 250$\,\GeV, & $90 < \mt < 120$\,\GeV       & $133 \pm 22$ & $87 \pm 16$ & $29 \pm 7$ & $13 \pm 5$ & $2.2 \pm 1.0$ & $0.019 \pm 0.010$ & $1.5 \pm 0.5$& --\\         
$175 < \amtTwo < 250$\,\GeV, & $\mt > 120$\,\GeV            & $73 \pm 8$ & $46 \pm 7$ & $12.2 \pm 3.5$ & $6.9 \pm 2.5$ & $3.0 \pm 1.4$ & $0.29 \pm 0.15$ & $4.8 \pm 1.4$& --\\       
$\amtTwo > 250$\,\GeV, & $90 < \mt < 120$\,\GeV             & $66 \pm 6$ & $33 \pm 7$ & $20 \pm 4$ & $11 \pm 4$ & $1.8 \pm 1.0$ & $0.11 \pm 0.06$ & $0.56 \pm 0.18$& --\\            
$\amtTwo > 250$\,\GeV, & $\mt > 120$\,\GeV                  & $26.5 \pm 2.6$ & $10.8 \pm 2.5$ & $6.9 \pm 1.6$ & $4.7 \pm 1.6$ & $1.9 \pm 0.9$ & $0.22 \pm 0.11$ & $2.0 \pm 0.6$& --\\
\hline
\multicolumn{2}{|l|}{{\bf \SRbWN}} &   & & & & & & & \\
$100 < \amtTwo < 120$\,\GeV, &  $90 < \mt < 120$\,\GeV      & $68 \pm 9$ & $60 \pm 9$ & $3.9 \pm 2.1$ & $3.0 \pm 2.8$ & $0.38 \pm 0.20$ & $0.08 \pm 0.04$ & $0.28 \pm 0.12$& --\\   
$100 < \amtTwo < 120$\,\GeV, & $\mt > 120$\,\GeV            & $75 \pm 11$ & $67 \pm 12$ & $4.1 \pm 1.8$ & $2.3 \pm 1.4$ & $0.55 \pm 0.18$ & $0.16 \pm 0.08$ & $0.73 \pm 0.23$& --\\ 
$\amtTwo > 120$\,\GeV, & $90 < \mt < 120$\,\GeV             & $179 \pm 23$ & $145 \pm 21$ & $22 \pm 6$ & $9 \pm 5$ & $1.4 \pm 0.7$ & $0.20 \pm 0.10$ & $1.1 \pm 0.4$& --\\        
$\amtTwo > 120$\,\GeV, & $\mt > 120$\,\GeV                  & $306 \pm 30$ & $239 \pm 32$ & $35 \pm 10$ & $18 \pm 7$ & $5.8 \pm 2.2$ & $1.6 \pm 0.8$ & $6.5 \pm 2.1$& --\\       
\hline
\end{tabular}
}
\caption{Background estimates in the SRs (model-independent selection) of the 15 analyses obtained from CRs for \ttbar\ and \Wjets, from data for multijet events, and from simulation normalised to theoretical cross-sections for all other backgrounds. The quoted uncertainties include all statistical and systematic effects. 
The sum in quadrature of the uncertainties of all backgrounds may not add up to the total uncertainty due to correlations.
}
\label{tab:background_breakdown}
}
\end{table}
\end{landscape}
}

\subsection{Validation}\label{sec:validation_regions}

The background fit predictions are validated using dedicated event samples.
For each cut-and-count and one-dimensional shape-fit analysis one or more dedicated validation regions (VRs) are defined for the \ttbar\ and \Wjets\ backgrounds. 
The VRs are designed to be kinematically close to the associated SRs to test the background estimates in regions of phase space 
as similar as possible to the SRs.
For most analyses the associated VRs are defined following a similar strategy as used for the CRs but with 
a $90\,\GeV < \mt < 120\,\GeV$ requirement, 
which leads to a set of events orthogonal to both the associated CRs and the SR.
The event selections for the \ttbar\ and \Wjets\ VRs, TVR and WVR respectively, are given in \tabref{tab:CR_definitions}. 
Another set of \ttbar\ validation regions, referred to as TVR2 but not shown in the table, is defined where applicable by inverting the SR requirement on \amtTwo\ while keeping all other requirements the same as in the SR. 
For the four two-dimensional shape-fits, a subset of the shape-fit bins in the region falling in between that dominated by \ttbar\ and the region enhanced by a potential signal is used for the \ttbar\ background validation. These signal-depleted shape-fit bins are referred to as validation bins.

For the cut-and-count and one-dimensional shape-fit analyses, the VRs are not used in any fit configuration to constrain the fit parameters. 
The validation bins of the two-dimensional shape-fits, on the other hand, are not used in the background-only fit configuration but are included in the model-dependent fit configuration. 
The number of background events in each VR or validation bin is predicted by the background-only fit 
(using simulation for the extrapolation) and compared to the data, as shown in the upper panel of \figref{fig:pulls}. 
The lower panel shows the pull for each bin, where the pull is defined as the difference between the predicted background and the observed number of events divided by the total uncertainty. The latter is given by the full uncertainty of the prediction (described in \secref{sec:systematics}) added in quadrature with the statistical uncertainty of the observed number of events.
No indication of background mis-modelling is found. VRs or validation bins belonging to different analyses can share events, and the systematic uncertainties are correlated across different regions and bins.

\begin{figure}
\begin{center}
\includegraphics[width=1.0\textwidth]{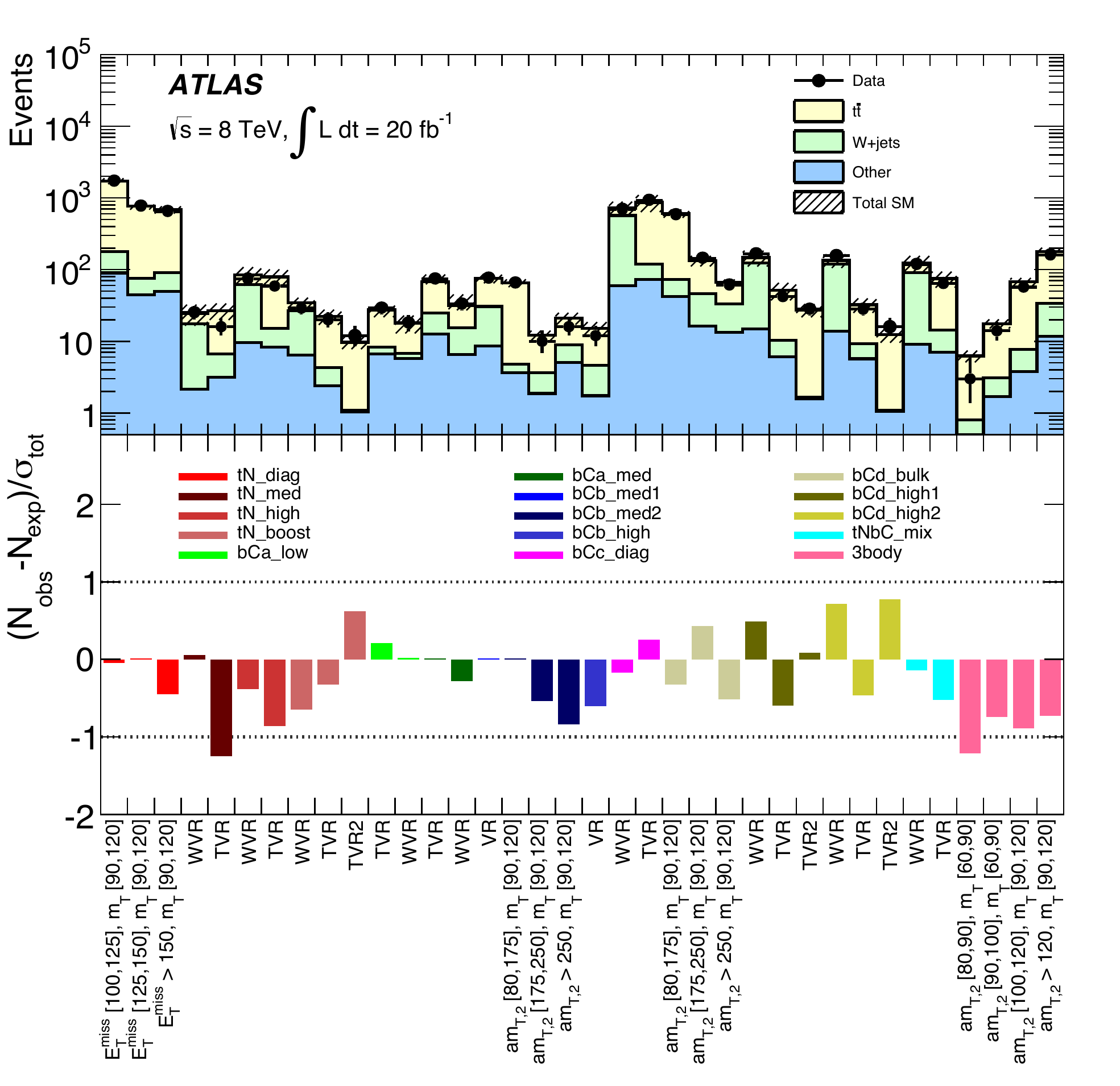} 
\caption{The upper panel compares data with background predictions in 
the VRs of the cut-and-count and one-dimensional shape-fit analyses as well as the validation bins of the two-dimensional shape-fit analyses.
The lower panel shows the pull of the same bins. The \ttbar\ and \Wjets\ background estimates are obtained using the background-only fit to the CRs (described in the text). All statistical and systematic uncertainties are included. 
\label{fig:pulls}}
\end{center}
\end{figure}

\begin{figure}
\begin{center}
\includegraphics[width=0.49\textwidth]{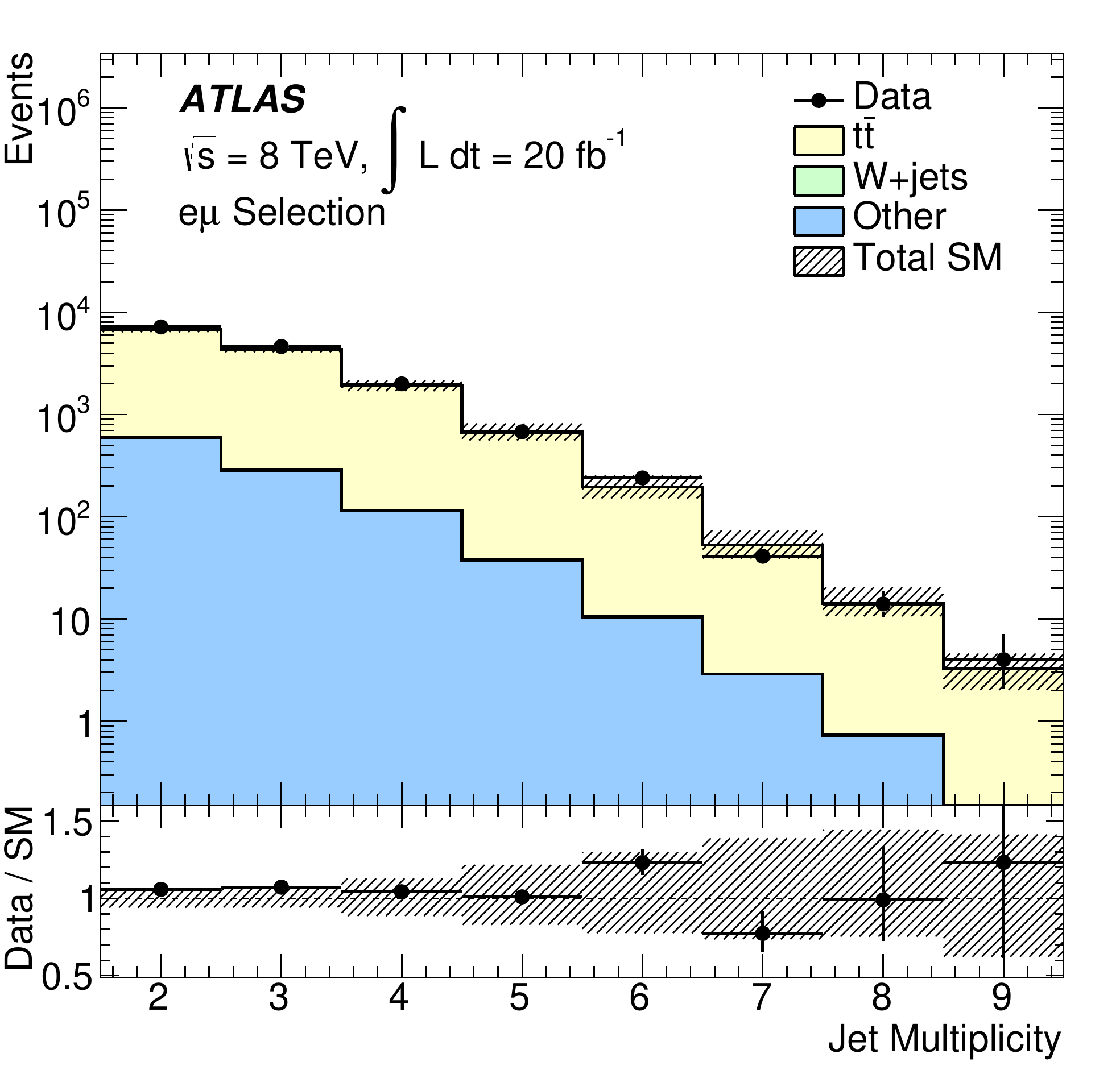} 
\includegraphics[width=0.49\textwidth]{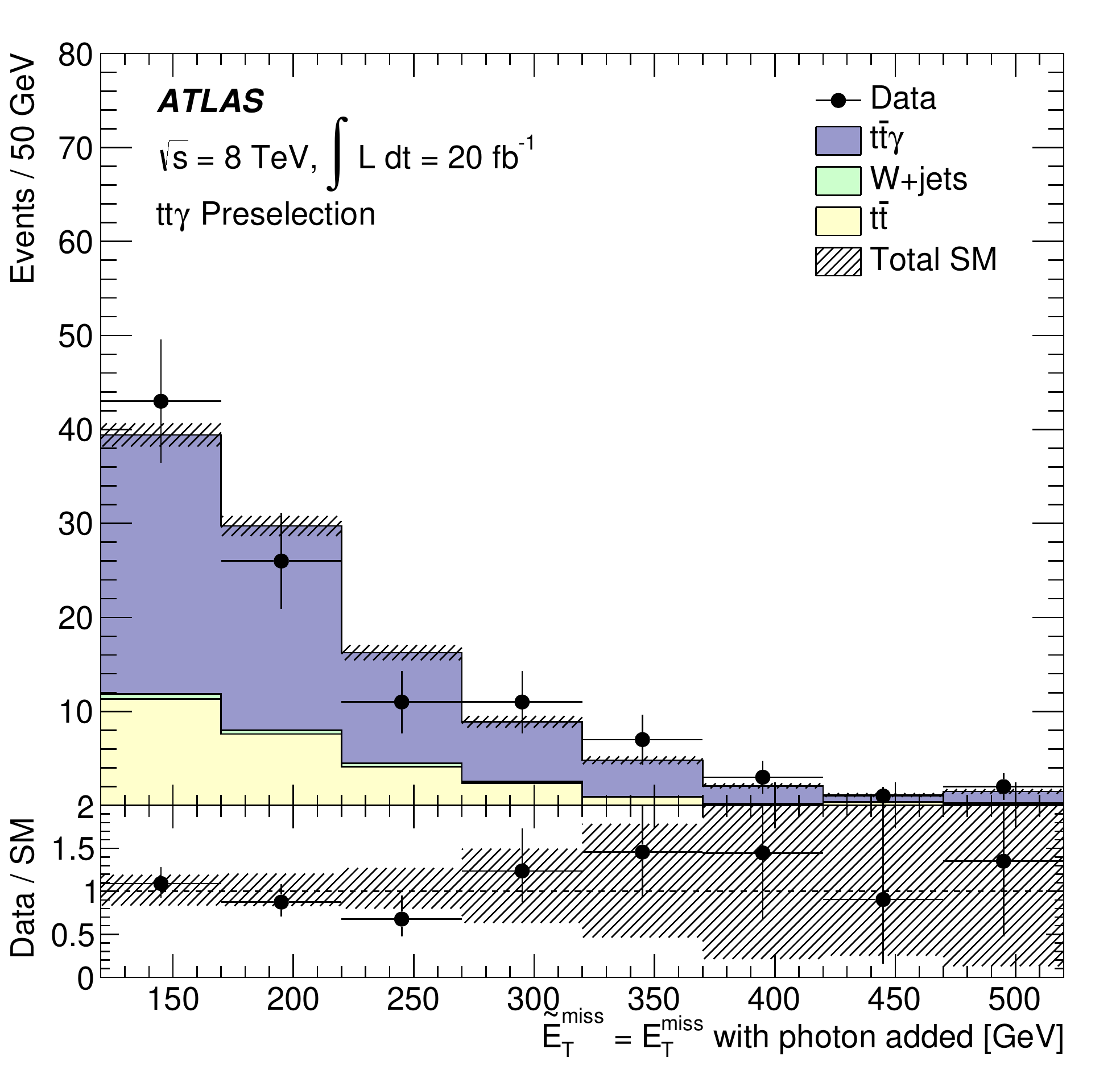} 
\caption{
Left: Jet multiplicity distribution for events with one opposite charged electron-muon pair and at least two jets of which one or more is $b$-tagged. Other processes include single top quark production, \ttbar\ production in association with a vector boson, $Z$+jets, and diboson production.
Right: Missing transverse momentum where photons are treated as invisible particles ($\tilde E_{\mathrm{T}}^{\mathrm{miss}}$) for an event selection of $\ttbar\ +$ photon (described in the text). 
Both plots: The uncertainty band includes all statistical and experimental uncertainties, and 
the last bins include overflows.
\label{fig:tt2L_VR}}
\end{center}
\end{figure}

Several other cross checks are performed to further validate the background estimations.
For the SRs requiring more than two jets, dileptonic \ttbar\ events can pass the event selection only if they contain additional jets beyond the two $b$-jets from the leading-order description of the decay.
The modelling of these additional jets, which in the simulation arise from radiation or higher-order-corrections, 
and which is relevant for the background estimation, is validated using a dedicated sample. The event selection is based on requiring one isolated electron and one 
 oppositely-charged, isolated muon, as well as two or more jets of which at least one is $b$-tagged using the 70\% working point. This selects a clean sample of \ttbar\ events, 
which is used in \figref{fig:tt2L_VR} (left) to compare the jet multiplicity distributions between data and simulation.  
Data is modelled sufficiently well within the systematic uncertainties.
Further dedicated validation samples are used to test the modelling of the \ttbar\ background with a \tauh\ or isolated track. 
These samples are based on the common event preselection and inverting either the track- or $\tau$-veto. 
The simulation is found to model data well within uncertainties.

The \Wjets\ light- vs heavy-flavour composition in the WCR can be different from that in the SR.
A dedicated validation is performed by selecting a sample enriched with $W$+heavy-flavour jets events. 
The event selection is based on exactly one isolated lepton, and exactly three jets (the fourth jet veto reduces \ttbar\ events), of which at least one is $b$-tagged.
Furthermore, events are required to have $60$\,\GeV $< \mt < 90$\,\GeV,  $\met > 150$\,\GeV, and the two jets with the highest $b$-tagging weights are required to yield an invariant mass below $80$\,\GeV\ 
and to have a limited separation in $\eta$--$\phi$ space to increase the sensitivity to pair-produced heavy-flavour jets in association with a $W$ boson.
The selected sample of $166$ events has a predicted $W$+heavy-flavour jets component 
of about $40$\%; data are found to be in good agreement with simulation, predicting $171$ events, when the overall \Wjets\ background is normalised to
match data in a $b$-veto control region.\footnote{The \Wjets\ background is normalised using the WCR associated with \SRbCzero, which 
requires three or more jets with a jet \pt\ selection similar to that used in the validation sample.}

Another dedicated validation sample is constructed to test the background prediction for \ttbar\ produced with a $Z$ boson that decays to two neutrinos, $\ttbar Z(\to\nu\bar{\nu})$. This process represents an irreducible background that becomes important for SRs with stringent requirements on kinematic variables, such as \SRtNthreep\ or \SRtNboost.
The validation strategy is to select \ttbar\ events produced in association with a photon, $\ttbar \gamma$. This process closely resembles $\ttbar Z(\to\nu\bar{\nu})$ 
in terms of Feynman diagrams and kinematic properties when the vector boson \pt\ is well above $m_Z$.
The event selection is based on one isolated lepton, four or more jets with at least one $b$-tag, one high-\pT\ photon, as well as requirements on modified versions of \mt\ and \met\ where  photons are treated as invisible particles. 
\Figref{fig:tt2L_VR} (right) compares data and background predictions, illustrating the accuracy of data modelling. 
The sample of $104$ events has a purity in $\ttbar \gamma$ of more than $70$\%.
The production of $\ttbar \gamma$ events is estimated using simulation, 
based on the same generator (\madgraph) as used for the $\ttbar Z$ process, and normalised to the NLO theoretical cross-section~\cite{PhysRevD.83.074013}.

 \section{Systematic uncertainties}\label{sec:systematics}

The systematic uncertainties affecting the results can be divided into two classes: 
uncertainties due to theoretical predictions and modelling, and uncertainties stemming 
from experimental effects. 
The impact of both types of uncertainty is evaluated for all background and signal samples. 
Since the yields for the dominant background sources, \ttbar\ and \Wjets, are obtained in 
dedicated control regions, the modelling uncertainties for these processes affect only the 
extrapolation from the CRs into the signal regions (and between TCR  and WCR), but not the overall normalisation. 
The systematic uncertainties are included as nuisance parameters and profiled in the likelihood fits.
The nuisance parameters are constrained by Gaussian
terms with widths corresponding to the sizes of the systematic uncertainties.
The same set of nuisance parameters is used across all bins, with the exception of the two shape-fits that have three \ttbar\ and three \Wjets\ normalisation parameters and hence also have three sets of nuisance parameters.
The effects of the sources of uncertainties
discussed in this section are quantified in terms of the corresponding relative uncertainty 
on the estimated number of background events in the various signal regions, this is referred to as the `impact on the background estimate'.

The dominant experimental uncertainties arise from imperfect knowledge of the 
jet energy scale (JES) and jet energy resolution (JER) as well as from the modelling of the $b$-tagging 
efficiency.
The JES uncertainty is derived from a combination of simulation and data samples~\cite{Aad:2011he,Aad:2014bia}
taking into account the dependence on the \pt, $\eta$ and flavour of the jet as well as the amount of pileup.
The impact of JES on the background estimate varies from $1$\% to $13$\%.
The JER uncertainties are determined with in-situ measurements of
the jet response balance in dijet events~\cite{Aad:2012ag}, 
and the impact on the background estimate is $1$\%--$21$\%.
The JES, JER, and jet mass scale and resolution uncertainties for \largeR\ jets are derived from a combination of data and simulation samples~\cite{Aad:2013gja,JetMassAndSubstructure}, 
and their combined impact on the background estimate amounts to 3\%.
The $b$-tagging uncertainty is estimated by varying the $b$-tagging efficiency and mis-tag rate correction factors, 
obtained from data-driven measurements of these quantities in \ttbar\ and dijet events~\cite{ATLAS-CONF-2014-004,ATLAS-CONF-2012-043,ATLAS-CONF-2012-039,ATLAS-CONF-2012-040}, 
within their uncertainties.
The impact of these uncertainties on the background estimate
ranges from $1$\% to $8$\%,
and is dominated by the uncertainty on the $b$-tagging efficiency.
Other sources of experimental uncertainty are 
the modelling of the average number of $pp$ interactions per bunch crossing,
the modelling of the contribution to the \met\ from energy deposits not associated with any reconstructed objects 
and from pileup, the modelling of lepton-related quantities (trigger and identification efficiency, 
energy and momentum scale and resolution, isolation and $\tau$-veto) as well as imperfect knowledge of the 
integrated luminosity. 
The combined impact of these sources on the background estimate 
is between $1$\% and $5$\%.

Uncertainties related to theoretical predictions and MC modelling are evaluated for all signal and background processes
obtained entirely or partly from simulated events. 
The sources of uncertainty considered for both the \ttbar\ and \Wjets\ background processes are the variations 
of the renormalisation and factorisation scales by factors of $0.5$ and $2.0$ 
as well as PDF variations, which are studied 
following the PDF4LHC recommendations~\cite{Botje:2011sn} comparing CT10 NLO, 
MSTW2008 NNLO
and NNPDF21\_100~\cite{Ball:2012cx} PDF error sets.
For the \ttbar\ background, the uncertainty on the hadronisation modelling is derived from a 
comparison between events generated with \powheg\ and interfaced with \pythia\ for the shower model 
and those generated in the same way, but interfaced with \herwig+\jimmy~\cite{Butterworth:1996zw}. 
Furthermore, the effect of the modelling of 
ISR and final-state radiation (FSR)
is studied using samples of \ttbar\ events 
generated with \acermc\ with reduced and increased amounts of additional radiation 
(constrained by the measurement of ref.~\cite{ATLAS:2012al}).
The impact of the \ttbar\ modelling on the background estimate 
is $2$\%--$6$\%.
For the \Wjets\ background, the effect of varying the number of partons used in the hard-scatter process is 
estimated by comparing samples generated with up to four extra partons to samples generated with up to five extra partons.
The impact of merging matrix elements and parton showers is studied by varying the \sherpa\ scales
related to the matching scheme.
As the \Wjets\ background is normalised in a region with a $b$-tag veto, additional
uncertainties on the flavour composition of the \Wjets\ events in the signal region,
based on the uncertainties on the measurement of ref.~\cite{Aad:2013vka} extrapolated to
higher jet multiplicities, are applied in all regions requiring at least one $b$-tagged jet.
The impact of the \Wjets\ modelling on the background estimate 
is $1$\%--$7$\%.

Background sources other than \ttbar\ and \Wjets\ are estimated from simulated events and are normalised to the most accurate cross-section predictions available. 
The cross-section uncertainty for the single-top process is $7$\%~\cite{Kidonakis:2011wy,Kidonakis:2010ux,Kidonakis:2010tc}, while it is $22$\% for $\ttbar V$~\cite{Campbell:2012dh,Garzelli:2012bn}.
The $ZZ$ and $WZ$ cross-section uncertainties are  $5$\% and $7$\%, respectively~\cite{Campbell:1999ah,Campbell:2011bn}.
Other sources of systematic uncertainty considered depend on the physics process, 
but include 
the choice of renormalisation and factorisation scales, PDF variations,
hadronisation modelling, 
choice of MC generator, 
modelling of ISR and FSR, 
variations of the matrix element to parton shower matching scales, 
the generation of a finite number of partons, 
and the interference between single-top and \ttbar\ production at NLO. 
The uncertainty on the interference treatment is estimated using inclusive $WWbb$ samples at LO generated 
with \acermc\ (which includes both the \ttbar\ and $Wt$ processes).
The total impact of the modelling of the smaller backgrounds on the background estimate 
ranges from $1$\% to $11$\%.

The theoretical uncertainties affecting the signal yields originate from the uncertainty on the production cross-section~\cite{Kramer:2012bx}, and from the uncertainty on the acceptance.
The latter includes PDF variations assessed using the PDF4LHC prescription~\cite{Botje:2011sn}, as well as modelling uncertainties of ISR and FSR and variations of the renormalisation and factorisation scales,  evaluated by varying the relevant parameters in \madgraph.
The total uncertainty on the production cross-section varies as a function of the stop mass; it amounts to about 15\% for $m_{\tone} = 200$\,\GeV\ and increases to $18$\% for $m_{\tone} = 700$\,\GeV. The impact of the ISR/FSR modelling uncertainty on the signal acceptance ranges from
$10$\% to $20$\% for signal regions that select events with ISR activity, such as \SRbCzero, and for signal models of mass hierarchy (c). It is negligible 
for the other signal regions. 

The search sensitivity is directly connected to the fitted uncertainty of the signal strength parameter, where the signal strength is 
a fit parameter that scales the signal yield predicted by the model in question; a signal strength of one corresponds to the nominal signal yield. 
The impact of the various sources of uncertainty, including the statistical precision, on the signal strength uncertainty is quantified in \tabref{tab:signal_uncertainty}
for selected signal regions and signal benchmark models.
The breakdown of the size of the systematic uncertainties is evaluated by re-running the fit, fixing the relevant nuisance parameter in question to its value from the nominal fit, and taking the difference in quadrature between the signal strength uncertainty of this fit and the nominal fit.
The statistical uncertainty is obtained from re-running the fit without any systematic uncertainties, again fixing the nuisance parameters to their values from the nominal fit.
The tightest signal regions, such as \SRtNboost, are statistically limited. Systematic uncertainties dominate the looser signal regions.
Overall, the largest contributions to the systematic uncertainty on the signal strength come from 
JER and \ttbar\ modelling. 
The energy scale and energy resolution of \largeR\ jets is relevant in the \SRtNboost\ signal region.

\begin{table}
\begin{center}
{\small
\begin{tabular}{|l|c|c|c|c|}
\hline
\textbf{Uncertainty on signal strength} & \SRtNboost\ & \SRtNonep\ & \SRbCzero\ & \SRbCone\ \\
\hline
Total &0.37 & 0.19 & 0.11 & 0.16 \\
\hline
Statistical & 0.36 & 0.05 & 0.07 & 0.09 \\
\hline
Systematic & 0.09 & 0.18 & 0.09 & 0.13 \\
\hline
\hline
\multicolumn{5}{| l |}{\textbf{Contribution of systematic uncertainty components}}\\
\hline
Jet energy scale & 0.02 & 0.03 & 0.02 & 0.03 \\
\hline
Jet energy resolution & 0.06 & 0.11 & 0.06 & 0.07 \\
\hline
\LargeR-jet related & 0.03 & - & - & -  \\
\hline
\met\ (non-associated energy and pileup)  & 0.01& 0.06 & 0.01 & 0.03  \\
\hline
Pileup  & $< 0.01$ & 0.03 & 0.01 & 0.02 \\
\hline
$b$-tagging & 0.03 & 0.01 & 0.04 & 0.01\\
\hline
\ttbar\ modelling & 0.01 & 0.15 & 0.04 & 0.08 \\
\hline 
\Wjets\ modelling & $< 0.01$ & 0.01 & 0.02 & 0.03 \\
\hline
Other backgrounds modelling & 0.04 & 0.03 & 0.02 & 0.02 \\
\hline
Signal acceptance modelling & 0.02 & 0.01  & 0.02 & 0.01 \\
\hline

\end{tabular}
}
\caption{Breakdown of the size of uncertainties on the signal strength parameter of the likelihood fit. 
The central values of the signal strength parameters (not shown) are close to zero because the data are compatible with the predicted
backgrounds.
The uncertainty components are obtained from the difference in quadrature between the signal strength uncertainty of the nominal fit and a fit where the systematic uncertainty in question is disabled by fixing the corresponding nuisance parameter(s) to the value(s) from the nominal fit. 
Some systematic uncertainty components, such as the jet energy scale or the modelling of backgrounds, are displayed as single entries while the likelihood fit employs a more detailed description.
The sum in quadrature of the systematic uncertainty components may not add up to the total systematic uncertainty due to correlations. The following benchmark signal models are used: $m_{\tone} = 700$\,\GeV\ and $m_{\ninoone} = 1$\,\GeV\ for \SRtNboost, $m_{\tone} = 350$\,\GeV\ and $m_{\ninoone} = 150$\,\GeV\ for \SRtNonep\, $m_{\tone} = 180$\,\GeV, $m_{\chinoonepm} = 174$\,\GeV\ and $m_{\ninoone} = 87$\,\GeV\ for \SRbCzero\, $m_{\tone} = 300$\,\GeV, $m_{\chinoonepm} = 200$\,\GeV\ and $m_{\ninoone} = 100$\,\GeV\ for \SRbCone.
\label{tab:signal_uncertainty}
}
\end{center}
\end{table}

 \section{Results}\label{sec:results}

\begin{figure}
\begin{center}
\includegraphics[width=0.49\textwidth]{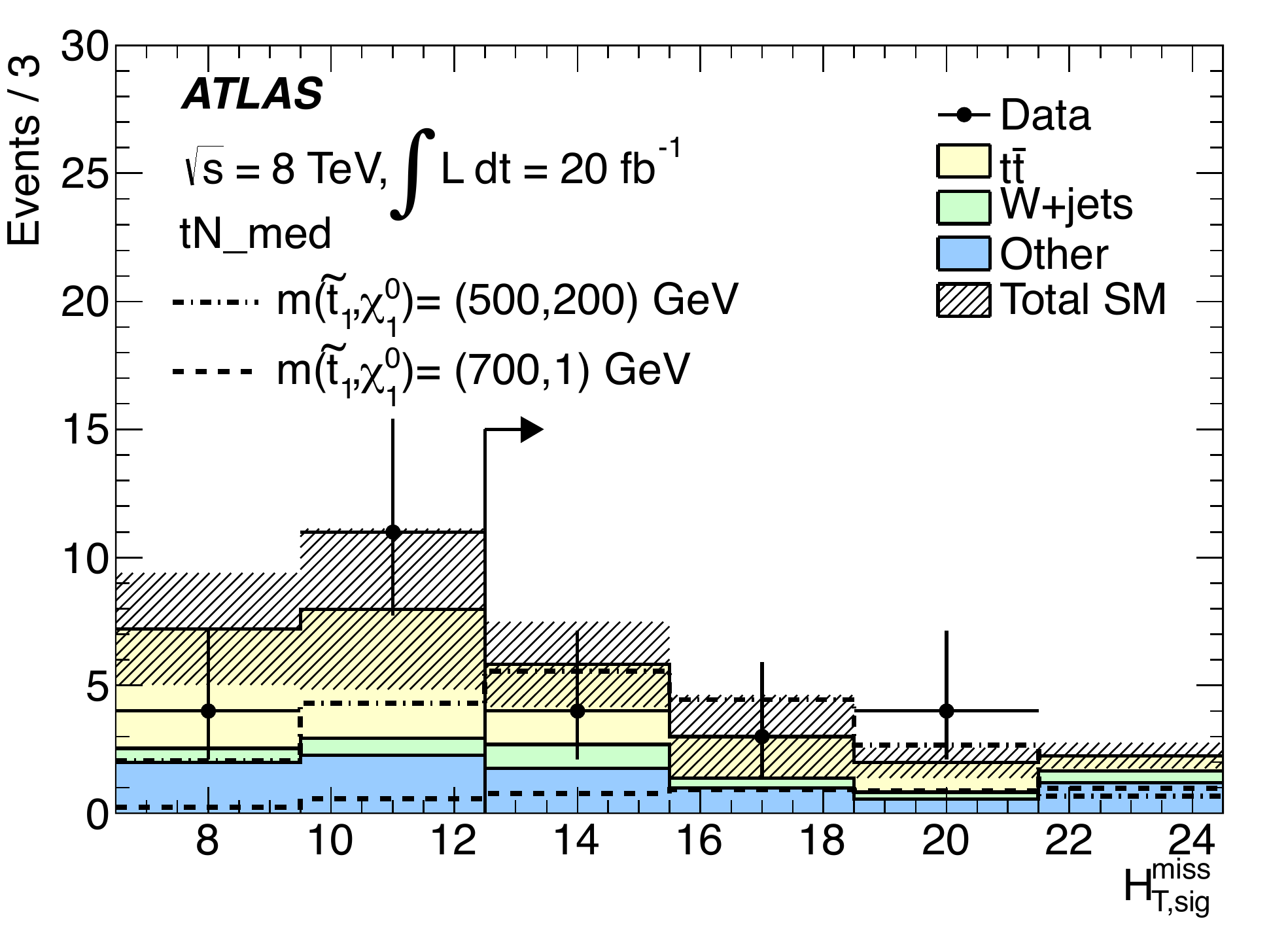}
\includegraphics[width=0.49\textwidth]{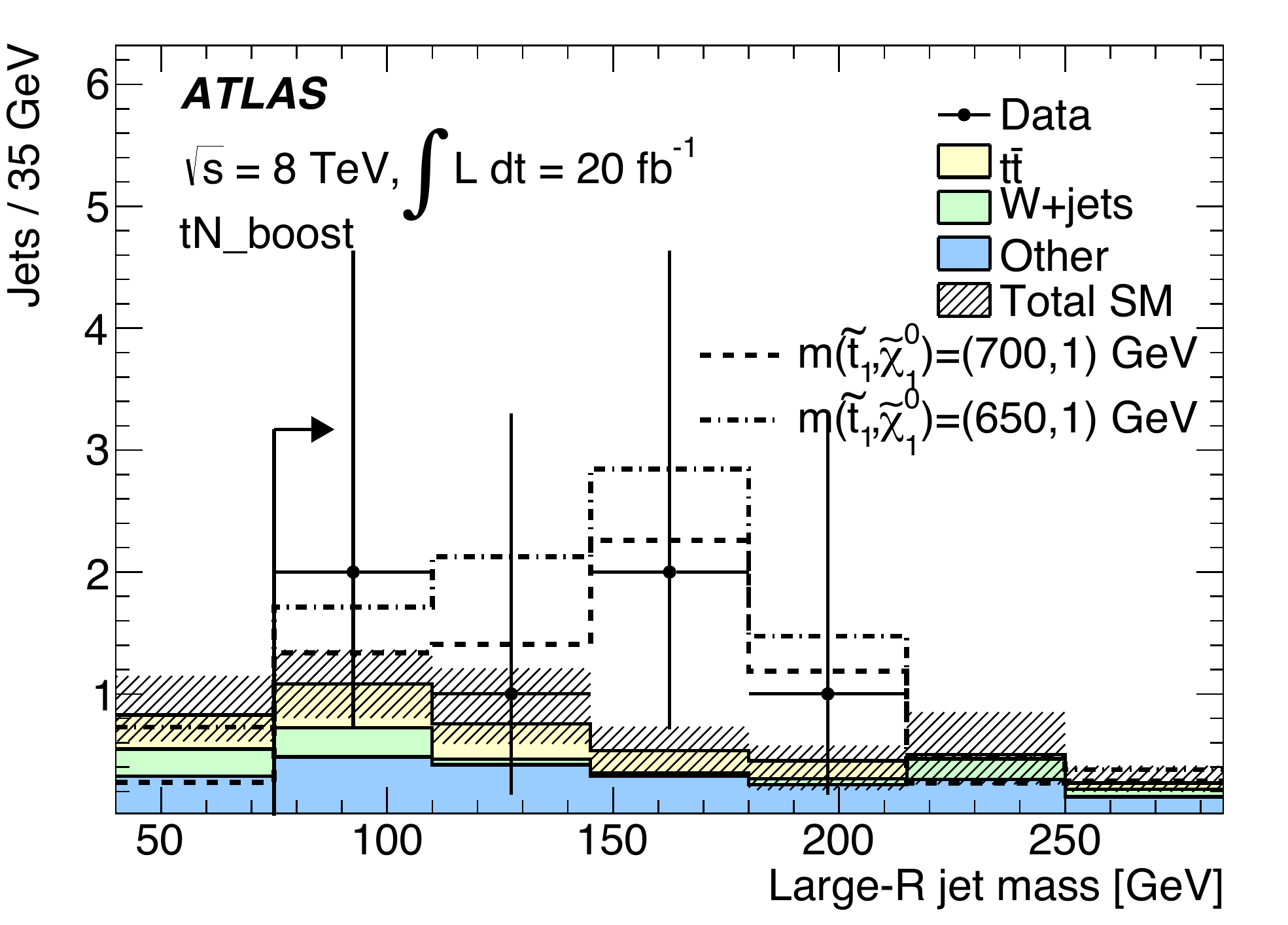}
\includegraphics[width=0.49\textwidth]{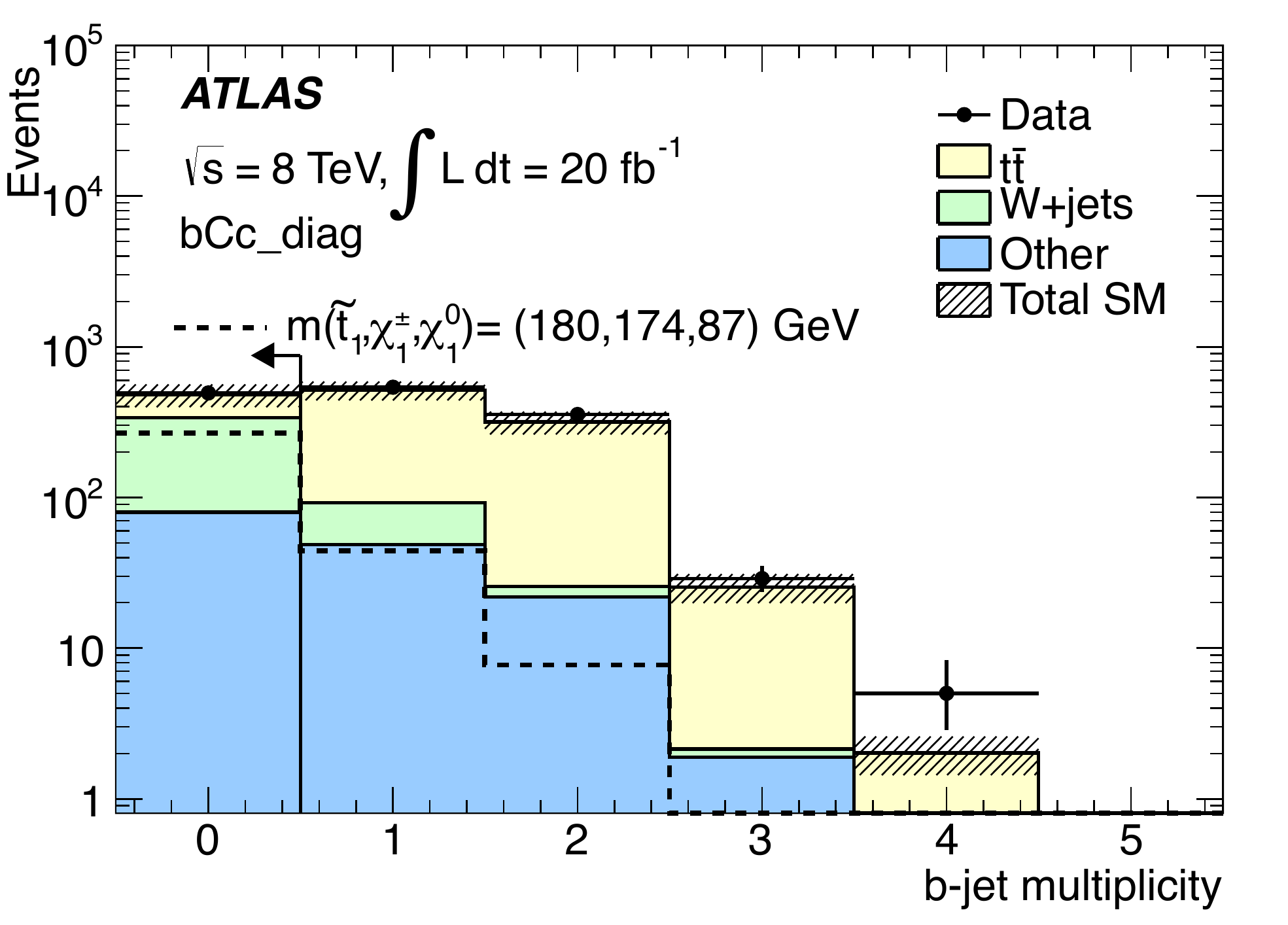}
\includegraphics[width=0.49\textwidth]{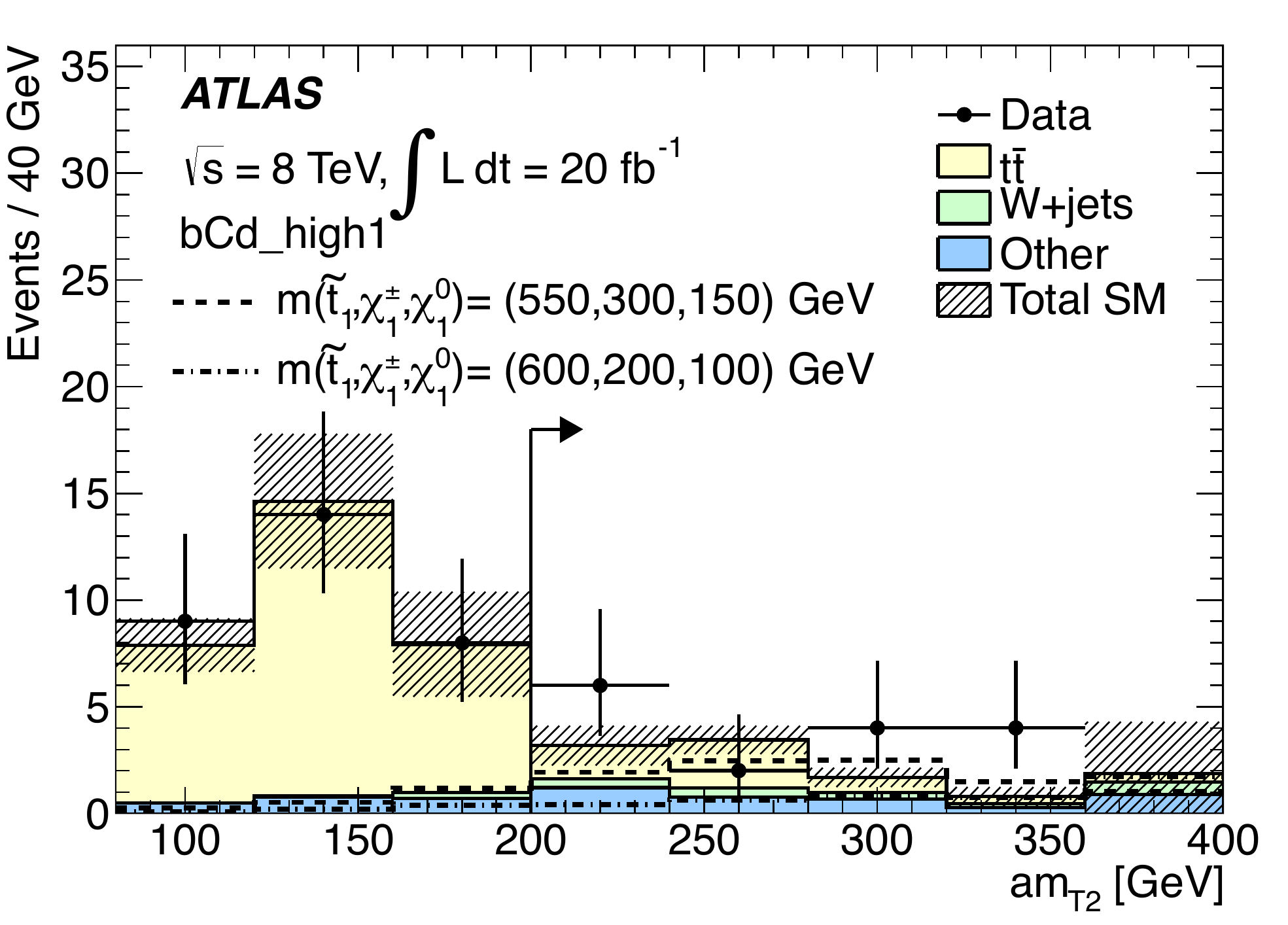}
\includegraphics[width=0.49\textwidth]{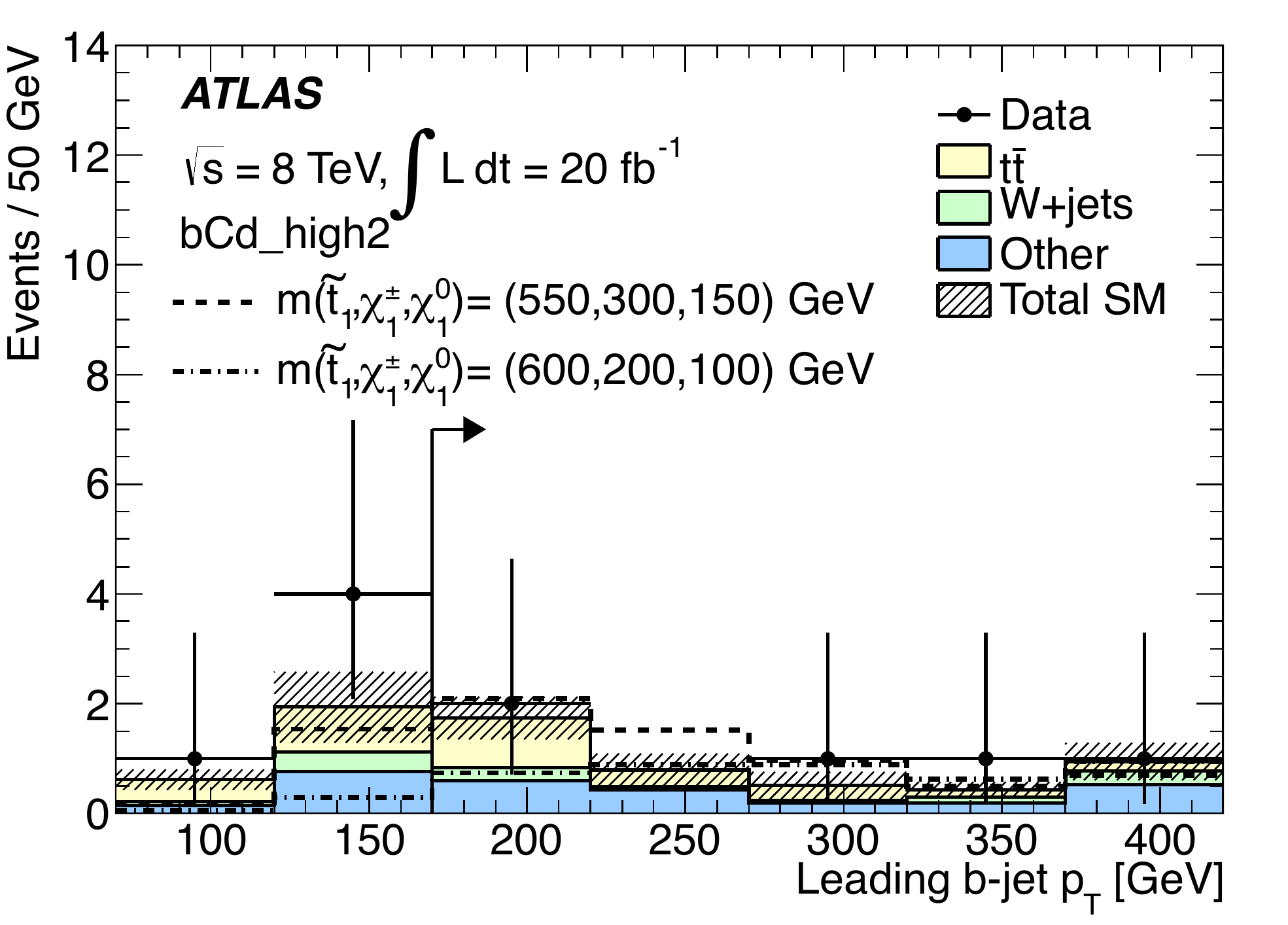}
\includegraphics[width=0.49\textwidth]{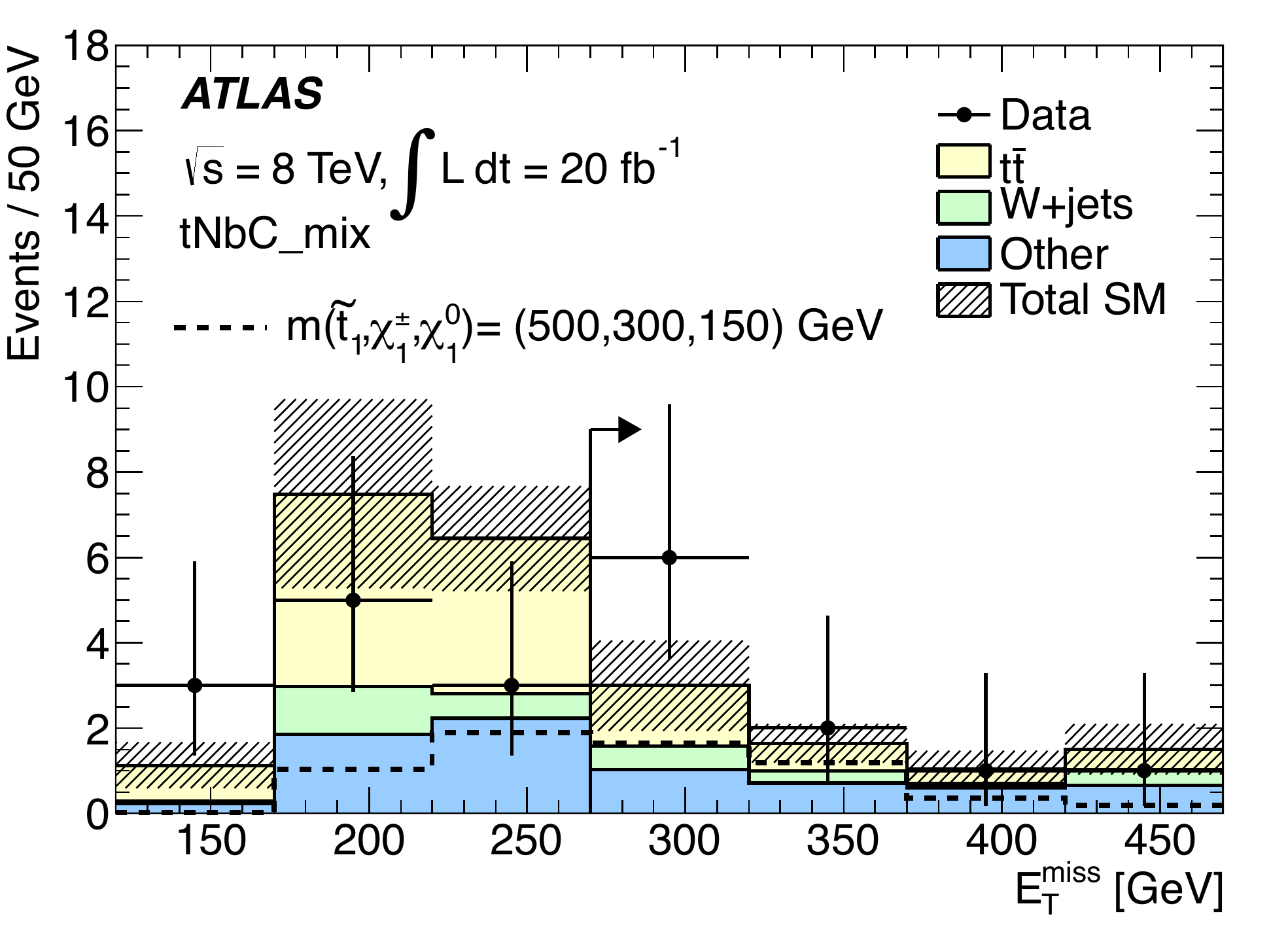}
\caption{For each signal region one characteristic distribution is shown, 
with the full event selection of the signal region applied, except for the 
requirement (indicated by an arrow) on the shown quantity. 
The uncertainty band includes statistical and all experimental systematic uncertainties. The last bin includes overflows.
Benchmark signal models are overlaid for comparison.
\label{fig:n-1_cutbased}}
\end{center}
\end{figure}

\begin{figure}
\begin{center}
\includegraphics[width=0.49\textwidth]{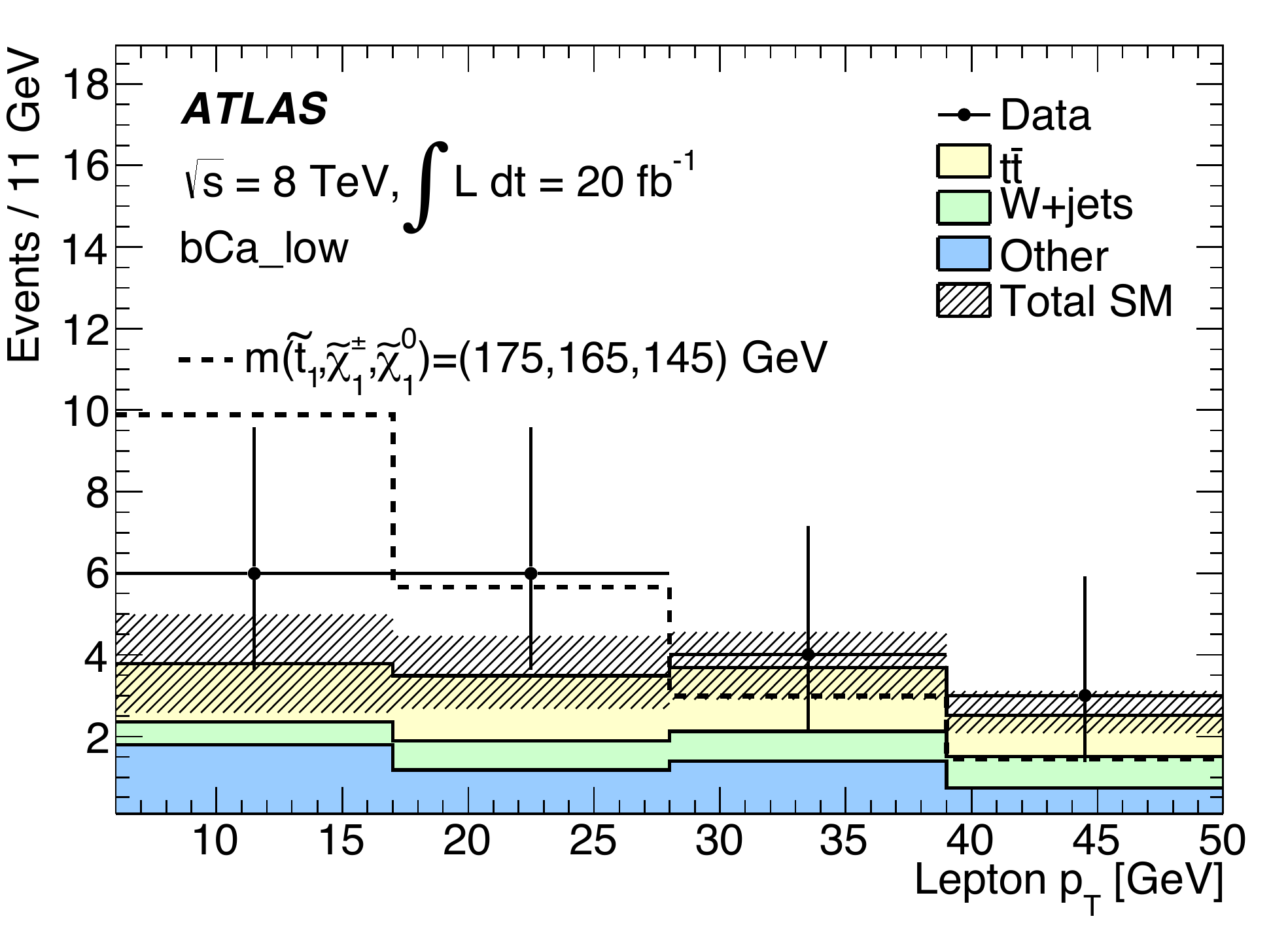}
\includegraphics[width=0.49\textwidth]{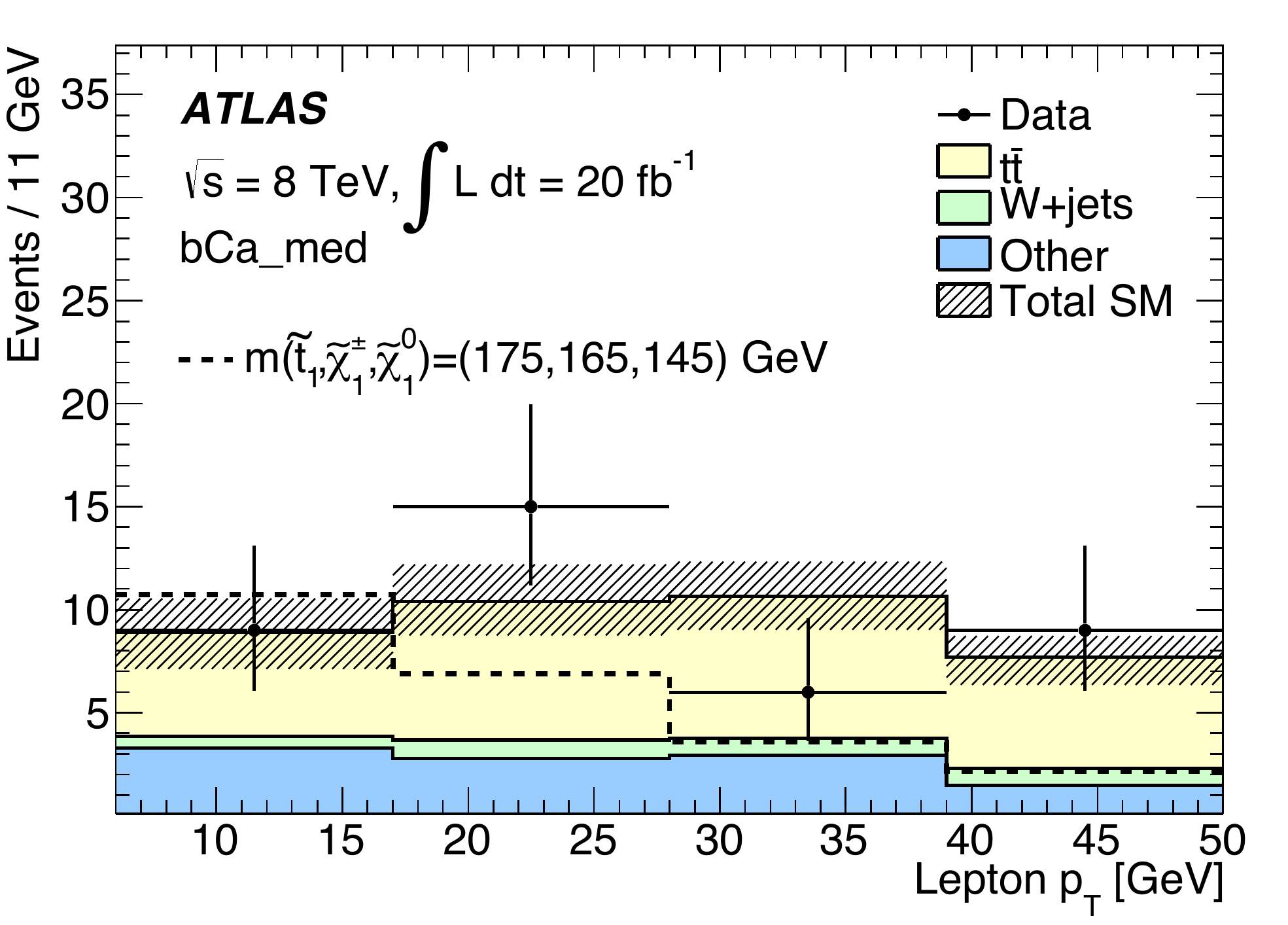}
\includegraphics[width=0.49\textwidth]{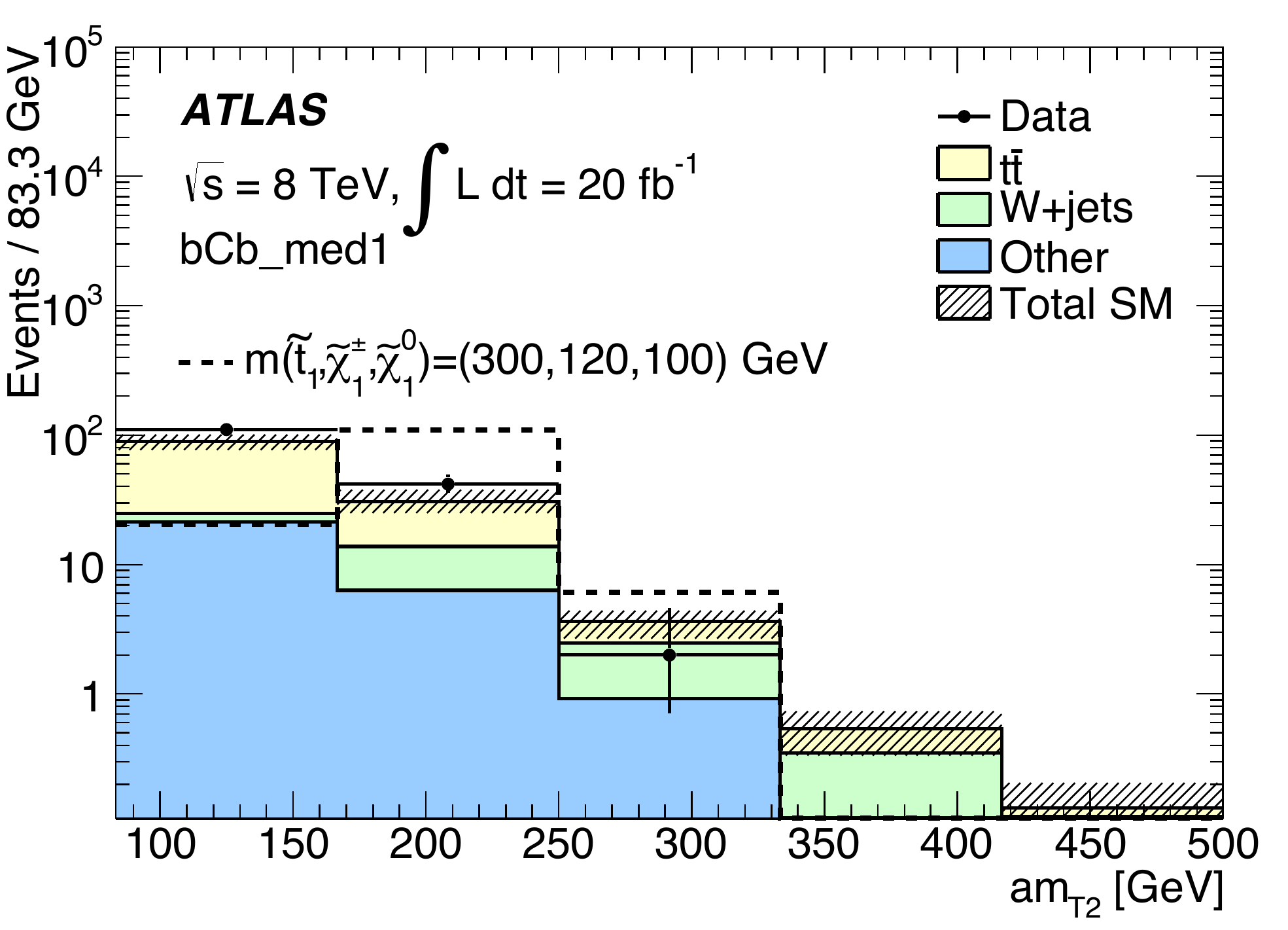}
\includegraphics[width=0.49\textwidth]{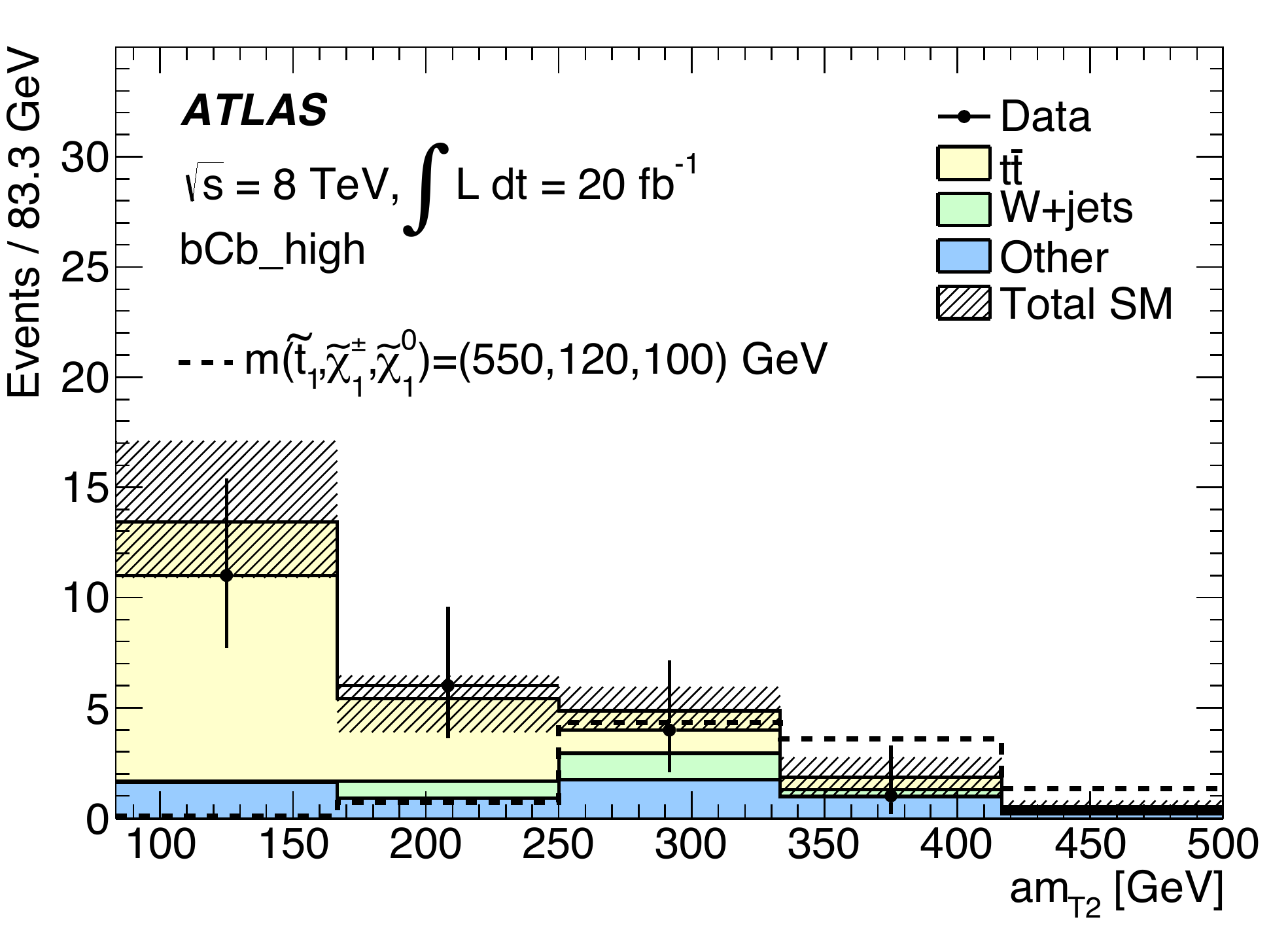}
\caption{For each signal region one characteristic distribution is shown, 
with the full event selection of the signal region applied, except for the 
requirement on the shown quantity.
The binning is similar to that used for the one-dimensional shape-fits of the corresponding analyses.
The uncertainty band includes statistical and all experimental systematic uncertainties. The last bin includes overflows.
Benchmark signal models are overlaid for comparison.
\label{fig:n-1_sl}}
\end{center}
\end{figure}

\Figsref{fig:n-1_cutbased} and~\ref{fig:n-1_sl} 
show comparisons between the observed data and the SM background prediction from the background-only fit
with all selections applied except the requirement on the plotted variable. 
In all SRs, the plots indicate good compatibility between the data and the SM background. 
The expected distributions from representative signal benchmark models are overlaid.

\Tabref{tab:obs_exp_p0_limits} shows the number of observed events together with 
the predicted number of background events in the SRs using the model-independent selection of the 15 analyses.
The predicted numbers of background events are obtained using the background-only fits to 
the number of observed events in the CRs as described in 
\secref{sec:backgrounds}. These fitted background estimates in the CRs 
are then used to obtain the fitted numbers of background events in the SRs by extrapolations that use transfer factors obtained with simulated events.
The observed numbers of events are found to agree well with the fitted numbers of background events in all SRs.

\begin{table}[ht]
\begin{center}
{\small
\setlength{\tabcolsep}{0.2pc}
\begin{tabular}{|rr|c|c|c|c|c|c|c|}
\hline
 \multicolumn{2}{|l|}{\multirow{2}{*}{Signal region}} & \multirow{2}{*}{Obs.} & \multirow{2}{*}{Exp. bkg.} & \multirow{2}{*}{$p_0$} & \multicolumn{2}{c|}{$N_{\mathrm{non-SM}}$} & \multicolumn{2}{c|}{$\sigma_{\mathrm{vis}}$ [fb]} \\
\multicolumn{2}{|l|}{} & & & & Obs.   & Exp.   & Obs.  & Exp.  \\
\hline
\multicolumn{2}{|l|}{\SRtNtwox} & 12 & $13.0 \pm 2.2$ & $\ge 0.5$ & $8.5$ & $9.2$ & $0.4$ & $0.5$ \\
\multicolumn{2}{|l|}{\SRtNthreep}& 5 & $5.0 \pm 1.0$ & $\ge 0.5$ & $6.0$ & $6.0$ & $0.3$ & $0.3$ \\
\multicolumn{2}{|l|}{\SRtNboost}& 5 & $3.3 \pm 0.7$ & 0.17 & $7.0$ & $5.3$ & $0.3$ & $0.3$ \\
\hline
\multicolumn{2}{|l|}{\SRoneLoneBc}& 11 & $6.5 \pm 1.4$ & 0.08 & 12.2 & 7.8 & 0.61 & 0.92\\
\multicolumn{2}{|l|}{\SRoneLoneBa}& 20 & $17 \pm 4$ & 0.33 & 14.4 & 12.3 & 0.72 & 0.68 \\
\multicolumn{2}{|l|}{\SRoneLtwoBa}& 41 & $32 \pm 5$ & 0.12 & 23.5 & 16.0 & 1.17 & 0.88 \\
\multicolumn{2}{|l|}{\SRoneLtwoBc}& 7 & $9.8 \pm 1.6$ & $\geq 0.5$ & 6.5 & 7.9 & 0.32 & 0.22 \\
\multicolumn{2}{|l|}{\SRbCzero}& 493 & $470 \pm 50$ & 0.27 & $110.6$ & $95.1$ & $5.4$ & $4.7$ \\
\multicolumn{2}{|l|}{\SRbCfour}& 16 & $11.0 \pm 1.5$ & 0.09 & $13.2$ & $8.5$ & $0.7$ & $0.4$ \\
\multicolumn{2}{|l|}{\SRbCfive}& 5 & $4.4 \pm 0.8$ & 0.36 & $6.3$ & $5.7$ & $0.3$ & $0.3$ \\
\hline
\multicolumn{2}{|l|}{\SRtNbC}& 10 & $7.2 \pm 1.0$ & $0.13$ & $9.7$ & $7.0$ & $0.5$ & $0.3$ \\
\hline
\multicolumn{2}{|l|}{{\bf \SRtNonep}} & & & & & & & \\
$125 < \met  < 150$\,\GeV, & $120 < \mt < 140$\, \GeV  & 117 & $136 \pm 22$& $\geq 0.5$ & $42.1$ & $55.7$ & $2.1$ & $2.7$ \\
$125 < \met  < 150$\,\GeV, & $\mt > 140$\, \GeV  & 163 & $152 \pm 20$& 0.35 & $55.4$ & $47.8$ & $2.7$ & $2.4$ \\
$\met > 150$\,\GeV, & $120 < \mt < 140$\, \GeV  & 101 & $98 \pm 13$& 0.43 & $36.1$ & $33.9$ & $1.8$ & $1.7$ \\
$\met > 150$\,\GeV, & $\mt > 140$\, \GeV  & 217 & $236 \pm 29$& $\geq 0.5$ & $58.7$ & $71.4$ & $2.9$ & $3.5$ \\
\hline
\multicolumn{2}{|l|}{{\bf \SRbCvW}} & & & & & & & \\
$175 < \amtTwo < 250$\,\GeV, & $90 < \mt < 120$\,\GeV & 10 & $12.1 \pm 2.0$& $\geq 0.5$ & $7.3$ & $8.8$ & $0.4$ & $0.4$ \\
$175 < \amtTwo < 250$\,\GeV, & $\mt > 120$\,\GeV  & 10 & $7.4 \pm 1.4$& 0.10 & $9.7$ & $7.3$ & $0.5$ & $0.4$ \\
$\amtTwo > 250$\,\GeV, & $90 < \mt < 120$\,\GeV  & 16 & $21 \pm 4$& $\geq 0.5$ & $9.3$ & $12.3$ & $0.5$ & $0.6$ \\
$\amtTwo > 250$\,\GeV, & $\mt > 120$\,\GeV  & 9 & $9.1 \pm 1.6$& $\geq 0.5$ & $7.7$ & $7.8$ & $0.4$ & $0.4$ \\
\hline
\multicolumn{2}{|l|}{{\bf \SRbCone}} & & & & & & & \\
$175 < \amtTwo < 250$\,\GeV, & $90 < \mt < 120$\,\GeV & 144 & $133 \pm 22$& 0.29 & $36.1$ & $33.9$ & $1.8$ & $1.7$ \\
$175 < \amtTwo < 250$\,\GeV, & $\mt > 120$\,\GeV   & 78 & $73 \pm 8$& 0.34 & $58.7$ & $71.4$ & $2.9$ & $3.5$ \\
$\amtTwo > 250$\,\GeV, & $90 < \mt < 120$\,\GeV   & 61 & $66 \pm 6$& $\geq 0.5$ & $17.5$ & $20.9$ & $0.9$ & $1.0$ \\
$\amtTwo > 250$\,\GeV, & $\mt > 120$\,\GeV  & 29 & $26.5 \pm 2.6$& 0.34 & $14.8$ & $12.6$ & $0.7$ & $0.6$ \\
\hline
\multicolumn{2}{|l|}{{\bf \SRbWN}} & & & & & & & \\
$80 < \amtTwo < 90$\,\GeV, &  $90 < \mt < 120$\,\GeV & 12 & $16.9 \pm 2.8$ &$\geq 0.5$ & $7.3$ & $9.9$ & $0.4$ & $0.5$ \\
$80 < \amtTwo < 90$\,\GeV, & $\mt > 120$\,\GeV    & 8 & $8.4 \pm 2.2$ & $\geq 0.5$ & $7.9$ & $7.8$ & $0.4$ & $0.4$ \\
$90 < \amtTwo < 100$\,\GeV, & $90 < \mt < 120$\,\GeV   & 29 & $35 \pm 4$ & $\geq 0.5$ & $11.7$ & $14.7$ & $0.6$ & $0.7$ \\
$90 < \amtTwo < 100$\,\GeV, & $\mt > 120$\,\GeV  & 22 & $29 \pm 5$ & $\geq 0.5$ & $55.4$ & $47.8$ & $2.7$ & $2.4$ \\
\hline
\end{tabular}
}
\caption{Columns two to four show the numbers of observed events in the SRs (model-independent selection) of the 15 analyses together 
with the expected numbers of background events (as predicted by the background-only fits) and the probabilities, represented by the 
$p_0$ values, that the observed numbers of events
are compatible with the background-only hypothesis.
The $p_0$ values are obtained with pseudo-experiments with the exception of the shape-fit bins where
only the smallest $p_0$ is derived with pseudo-experiments while the others are calculated from asymptotic 
formulae~\cite{Cowan:2010js}.
The $p_0$ value is set to 0.5 whenever the number of observed events is below
the number of expected events.
Columns five to eight show the 95\% CL upper limits on the number of
beyond-SM events ($N_{\mathrm{non-SM}}$) and on the visible signal cross-section
($\sigma_{\mathrm{vis}} = \sigma_{\mathrm{prod}} \times A \times \epsilon$).
The observed and (median) expected limits are given for a generic model without uncertainties other than on the luminosity.
}
\label{tab:obs_exp_p0_limits}
\end{center}
\end{table}

To assess the compatibility of the SM background-only hypothesis with the observations in 
the SRs, a profile likelihood ratio test is performed implementing the methodology described in ref.~\cite{Read:2002hq}. The model-independent selection is 
used, and the likelihood for a given test includes one SR and all its associated CRs.
Each SR, and each signal-sensitive bin in the two-dimensional shape-fits, is probed separately. 
\Tabref{tab:obs_exp_p0_limits} shows the $p_0$ values obtained using these fits, 
indicating that the data in all SRs are compatible with the background-only hypothesis.
Good agreement is found when comparing the results obtained using pseudo-experiments to those calculated from
asymptotic formulae~\cite{Cowan:2010js}; the latter is used as the default for all exclusion results presented below. 

As no significant excess over the expected background from SM processes is observed,
the data are used to derive one-sided limits at 95\% CL.
The results are obtained from a profile likelihood ratio test following the \CLs\ prescription~\cite{Read:2002hq}.
Model-independent upper limits on beyond-SM contributions are derived
separately for each analysis, and in case of the two-dimensional shape-fits for each signal-sensitive bin.
The model-independent selection is used, and the likelihood of the fit is configured to include one SR or shape-fit bin and all its associated CRs.
A generic signal model, which contributes only to the SR, is assumed and no experimental or theoretical signal systematic 
uncertainties are assigned other than the luminosity uncertainty.
The resulting limits on the number of beyond-SM events and on the visible signal cross-section 
are shown in the rightmost columns of \tabref{tab:obs_exp_p0_limits}. 
The visible signal cross-section ($\sigma_{\mathrm{vis}}$) is defined as the product of acceptance ($A$), reconstruction efficiency ($\epsilon$) and production cross-section ($\sigma_{\mathrm{prod}}$);
it is obtained by dividing the upper limit on the number of beyond-SM events by the integrated luminosity.

Exclusion limits are also derived in various SUSY scenarios.
The results are obtained using the same \CLs\ prescription as used for the model-independent limits,
but with the model-dependent selection. The likelihood for each analysis includes 
the full set of bins: SR, TCR, WCR for cut-and-count and the full set of SR and CR bins for shape-fit analyses. 
The signal uncertainties and potential signal contributions to all bins are taken into account.
All uncertainties except on the theoretical signal cross-section are included in the fit.
Combined exclusion limits are obtained by selecting a priori the signal region with the lowest
expected \CLs\ value for each signal grid point.

The expected and observed exclusion contours for the \topLSP\ decay mode are shown in 
\figref{fig:exclusion_combined_tN} overlaying the results for the signal regions targeting 
two-, three- and four-body decays.
The $\pm 1\,\sigma_{\rm exp}$ 
uncertainty band indicates the impact on the expected limit of all uncertainties included in the fit.  The $\pm 1\,\sigma_{\mathrm{theory}}^{\mathrm{SUSY}}$ 
uncertainty lines around the observed limit illustrate the change in the
observed limit as the nominal signal cross-section is scaled up and
down by the theoretical cross-section uncertainty.  
Quoted limits are derived from the $-1\,\sigma_{\mathrm{theory}}^{\mathrm{SUSY}}$  
observed limit contours.
In the four-body scenario stop masses are excluded between $100$ and $170$\,\GeV, for an LSP mass of about $75$\,\GeV.
Stop masses between $100$ and nearly $175$\,\GeV\ in the three-body scenario, and between $210$ and $640$\,\GeV\ in the two-body scenario are excluded for a massless LSP, while for a stop mass 
around $550$\,\GeV\ the exclusion reaches up to an LSP mass of $230$\,\GeV.
The non-excluded area between the four- and three-body decay regions is 
due to a reduction in search sensitivity as the kinematic properties of the signal change significantly when transitioning from a four-body to a three-body decay. In particular, approaching this boundary from the three-body side, the momenta of the two $b$-jets decrease to zero and
hence the acceptance of the \pt\ requirement on the $b$-tagged jet in the \SRbWN\ signal region drops quickly.
The kinematic properties change again at the other diagonal, between the three-body and on-shell top quark decay modes.
When approaching this diagonal from the on-shell top quark side the search sensitivity 
is limited by the difficulty to disentangle the signal from the \ttbar\ background, as the two processes begin to closely resemble each other in kinematic properties. In the limit of reaching the diagonal from the righthand side, the two LSPs have no phase space, thus carrying away no momentum, leading to a stop signature similar to that of \ttbar\ except for small deviations induced by the difference in spin.
This region is also referred to as `stealth stop'. 
The \SRtNonep\ signal region has the best expected sensitivity for stop masses up to $400$\,\GeV\ and close to the 
$m_{\tone} \gtrsim m_{t} + m_{\ninoone}$ kinematic boundary,
while the \SRtNboost\ signal region has the best expected sensitivity for stop masses above $600$\,\GeV. In the
intermediate mass region the best expected sensitivity comes from the \SRtNtwox\ signal region.
The use of large-R jets in the \SRtNboost\ signal region extends the reach for a heavy stop by about $30$\,\GeV, as obtained from a comparison with the \SRtNthreep\ signal region.

\Figsref{fig:exclusion_bC_x2} to~\ref{fig:exclusion_bC_stch10} show the expected and observed exclusion contours for the \bChargino\ decay mode
with different \chinoonepm\ mass hypotheses. 
If the mass of the \chinoonepm\ is twice that of the LSP (\figref{fig:exclusion_bC_x2}),
stop masses up to $500$\,\GeV\ are excluded for an LSP mass in the range of $100$ to $150$\,\GeV. 
The various regions in the exclusion area can be mapped to the mass hierarchies illustrated in 
\figref{fig:bChargino_mass_hierarchies}:  Models in  
the bulk region correspond to mass hierarchy (d), 
those in the (bottom) region with a low \ninoone\ mass are mass hierarchy (b), 
and the ones in the (diagonal) region close to the kinematic boundary are mass hierarchy (c). 
The strongest exclusion sensitivity is provided by the signal regions designed for the given mass hierarchy, 
for example \SRbCone\ (bulk), \SRoneLtwoBa\ and \SRoneLtwoBc\ (bottom), and \SRbCzero\ (diagonal).
The small region around $m_{\tone} \sim 175$\,\GeV\ and $m_{\ninoone} \lesssim 70$\,\GeV\ is not excluded because 
signal events cannot be sufficiently well distinguished from \ttbar\ background events; for larger $m_{\ninoone}$ values, the $b$-tag veto
in the \SRbCzero\ signal region becomes effective.

If the \chinoonepm\ mass is set to $150$\,\GeV\ (\figref{fig:exclusion_bC_150}), 
stop masses below $490$\,\GeV\ are excluded for LSP masses up to $80$\,\GeV.  
Models in the top, left, and bulk regions within the exclusion contour correspond to the mass hierarchies
 (a-b), (c), and (d) respectively; the exclusion power is mostly provided by
  \SRoneLtwoBa\ and \SRoneLtwoBc\ (top), \SRbCzero\ (left), \SRbCone\ (bulk), and \SRbCvW\ (bulk with high $m_{\tone}$).
The vertical drop in search sensitivity for $m_{\tone} \lesssim  260$\,\GeV\ in the top part of the plane is caused by the $b$-tagged jet \pt\ and $m_{bb}$ requirements in the two \texttt{bCb\_*} signal regions, which effectively imposes a minimum mass splitting between the \tone\ and \chinoonepm\ states. The other two signal regions that are based on soft-lepton selections (\texttt{bCa\_*}) require ISR activity and are hence 
limited to the low stop mass region where the cross-section is large.
The sharp horizontal contour line at $m_{\ninoone} = 145$\,\GeV\ is an artifact caused by the fact that signal grid points were generated only up to this LSP mass. However, the upper region is excluded by the search described in ref.~\cite{sbottom0L_8TeV}.
If the mass of the \chinoonepm\ is set to $106$\,\GeV\ (\figref{fig:exclusion_bC_106}), 
stop masses between
$240$ and $550$\,\GeV, together with a small region around a stop mass of $150$\,\GeV, are excluded for an LSP mass of $70$\,\GeV. 
The identification of regions with mass hierarchies, and the best signal regions are similar to the previous scenario.
The vertical exclusion gap around $m_{\tone} \sim 175$\,\GeV\ is due to the same effect seen in \figref{fig:exclusion_bC_x2} and described above, while the 
reduction in exclusion power for a decreasing LSP mass is due to the transition of the \chinoonepm\ decay from a three-body to a two-body process, which happens at an LSP mass of $\sim 26$\,\GeV; the two-body decay produces less \met\ on average, amongst other changes of the kinematic properties.

If the \chinoonepm\ mass is only $5$\,\GeV\ ($20$\,\GeV) above the LSP mass (\figsref{fig:exclusion_bC_plus5} 
and~\ref{fig:exclusion_bC_plus20}, respectively), 
stop masses between $265$ and $600$\,\GeV\ ($240$ and $600$\,\GeV) are excluded for an LSP mass of $100$\,\GeV.
 The exclusion in these `compressed' scenarios, corresponding to the mass hierarchies (a) and (b), is achieved using the 
 \softLeptonHyphen\ selections of the \SRoneLoneBc\ and \SRoneLoneBa\ 
 signal regions. 
The sensitivity decreases for smaller mass splittings because of the lepton \pt\ threshold.
The diagonal exclusion gap in \figref{fig:exclusion_bC_plus20} is caused by the effects described above in the discussion of \figref{fig:exclusion_bC_150} together with the impact of the $b$-veto on the leading jet in the \texttt{bCa\_*} signal regions.
If the \chinoonepm\ mass is only $10$\,\GeV\ below the stop mass (\figref{fig:exclusion_bC_stch10}),
stop masses below $390$\,\GeV\ are excluded for a massless LSP. 
The models in this scenario correspond to mass hierarchy (c). The bulk exclusion power comes from
the \SRbCzero\ signal region with a $b$-veto, while using soft leptons in the \SRoneLoneBc\ signal region extends the sensitivity in the top left region. 

The complementarity of the signal regions to various mass splittings of \chinoonepm\ and \ninoone\ is illustrated by fixing the stop mass to $300$\,\GeV\
and presenting the exclusion limit as a function of the \chinoonepm\ and \ninoone\ masses (\figref{fig:exclusion_stop300}).
LSP masses up to about $100$\,\GeV\ are excluded for all possible \chinoonepm\ masses, with one small exception in the bottom left corner.
The exclusion power close to the kinematic boundary
comes from the signal regions designed for mass hierarchy (a) and (b), while for larger mass splittings the sensitivity is provided by the selections for mass hierarchy (d), and (c) in case of a large \chinoonepm\ mass.
The non-excluded region around a \chinoonepm\ mass of $270$\,\GeV\ and a \ninoone\ mass of $175$\,\GeV\ is caused by the signal region transition.

In scenarios where both the \topLSP\ and \bChargino\ decay modes are allowed and where $m_{\chinoonepm} = 2 m_{\ninoone}$
(\figref{fig:exclusion_asym}),
the largest excluded stop mass for an
LSP mass of 100\,\GeV\ gradually increases from $530$\,\GeV\ to $660$\,\GeV\ as the branching ratio for \topLSP\ is 
increased from $0$\% to $100$\%. Here, the quoted limits correspond to the central observed limit contour.  
The signal regions providing the best expected sensitivity for models with
mixed decays are mainly \SRbCone\ and \SRtNbC, with some contribution from \SRtNonep, \SRtNtwox\ and \SRtNboost\ for models
with a large fraction of \topLSP\ decays, as well as some contributions from \SRoneLtwoBa\ and \SRoneLtwoBc\ for models with a low LSP mass and a small fraction of 
\topLSP\ decays. 

The upper limits on the signal cross-section for models where
the \tone\ is a pure \tleft\ and models where it is predominantly a \tright\ are compared in \figref{fig:leftright_limits}
for the \topLSP\ decay mode with two assumptions for the lightest neutralino, $m_{\ninoone} = 50$\,\GeV\ and $m_{\ninoone} = 150$\,\GeV.
The predominantly \tright\ mixing composition is the default setting used for all simplified-model \topLSP\ scenarios.
The weaker \tleft\ model exclusion is mainly the result of a reduced lepton and \mt\ acceptance. 
The excluded \tone\ mass reach of the \tleft\ model is reduced by about $50$\,\GeV\ for the two considered LSP masses.

The change in sensitivity when varying parameters other than the stop and \ninoone\ masses is studied using  
27 pMSSM samples, which can be classified into three groups of similar stop and \ninoone\ masses; a detailed description is given in \secref{sec:samples_signal}. The expected and observed \CLs\ significance values for the 27 pMSSM models 
and two simplified models are shown in \figref{fig:pMSSM_CLs}. The strongest impact on the \CLs\ significance is 
found to be from the sum of the branching ratios for \topLSP\ and \bChargino, where the
\CLs\ significance is smaller for
models where stop decays other than \topLSP\ and \bChargino\ are kinematically allowed. 
This is a consequence of the signal selections being optimised using only simplified models. 
In addition to the branching ratio dependence, the sensitivity also depends on the kinematic properties of the events, which are affected, e.g., by the stop mixing matrix and 
by the masses and field content of other SUSY
particles. These additional dependencies explain the large spread in \CLs\ significance for the models where the stop decays only to $t \ninoone$ and $b \chinoonepm$.
For $m_{\tone} \sim 400$\,\GeV\ and 
$m_{\ninoone} \sim 50$\,\GeV, 11 of the 14 models are excluded.
No models with $m_{\tone} \sim 550$\,\GeV\ are excluded.

\begin{figure}
\begin{center}
\includegraphics[width=\textwidth]{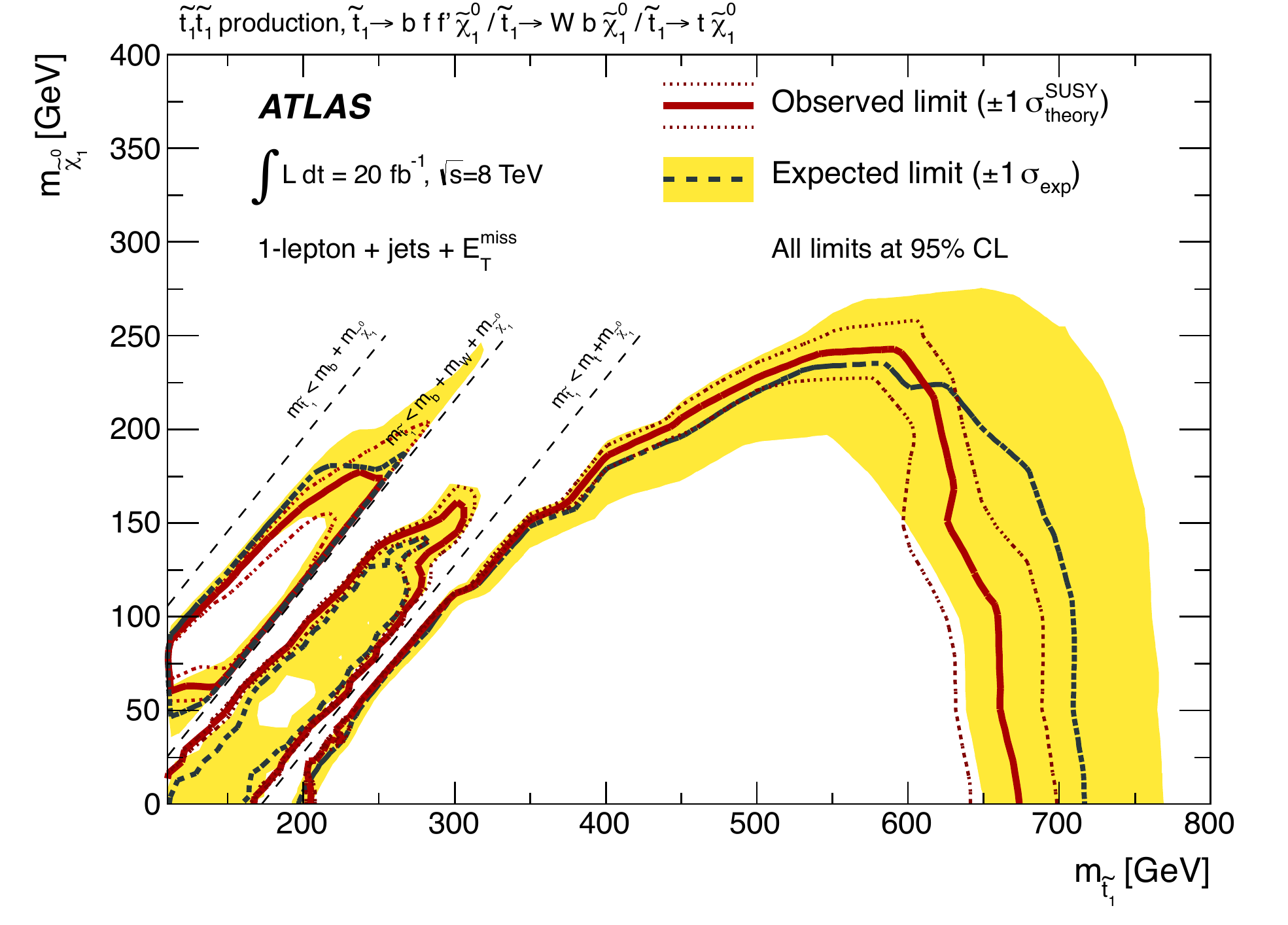}
\caption{Expected (black dashed) and observed (red solid) 95\% 
CL excluded region in the plane of $m_{\ninoone}$ vs. 
$m_{\stopone}$ , assuming $\mathcal{BR}(\stopone \to t \ninoone) = 100$\%.
In the region $m_b + m_W + m_{\ninoone} < m_{\stopone} < m_t + m_{\ninoone}$ the decay of the
$\stopone$ involves a virtual top quark (three-body decay), while in the region
$m_{\stopone} < m_b + m_W + m_{\ninoone}$ it involves both 
a virtual top quark and a virtual $W$ boson (four-body decay). 
The $m_{\tone} < 100$\,\GeV\ region for the three-body decay mode is excluded by the search described in ref.~\cite{stop2L_8TeV}. 
Furthermore, the $m_{\tone} < 78$\,\GeV\ region in the four-body scenario is excluded by the search in ref.~\cite{Heister:2002hp}.
\label{fig:exclusion_combined_tN}
}
\end{center}
\end{figure}

\begin{figure}
\begin{center}
\includegraphics[width=0.8\textwidth]{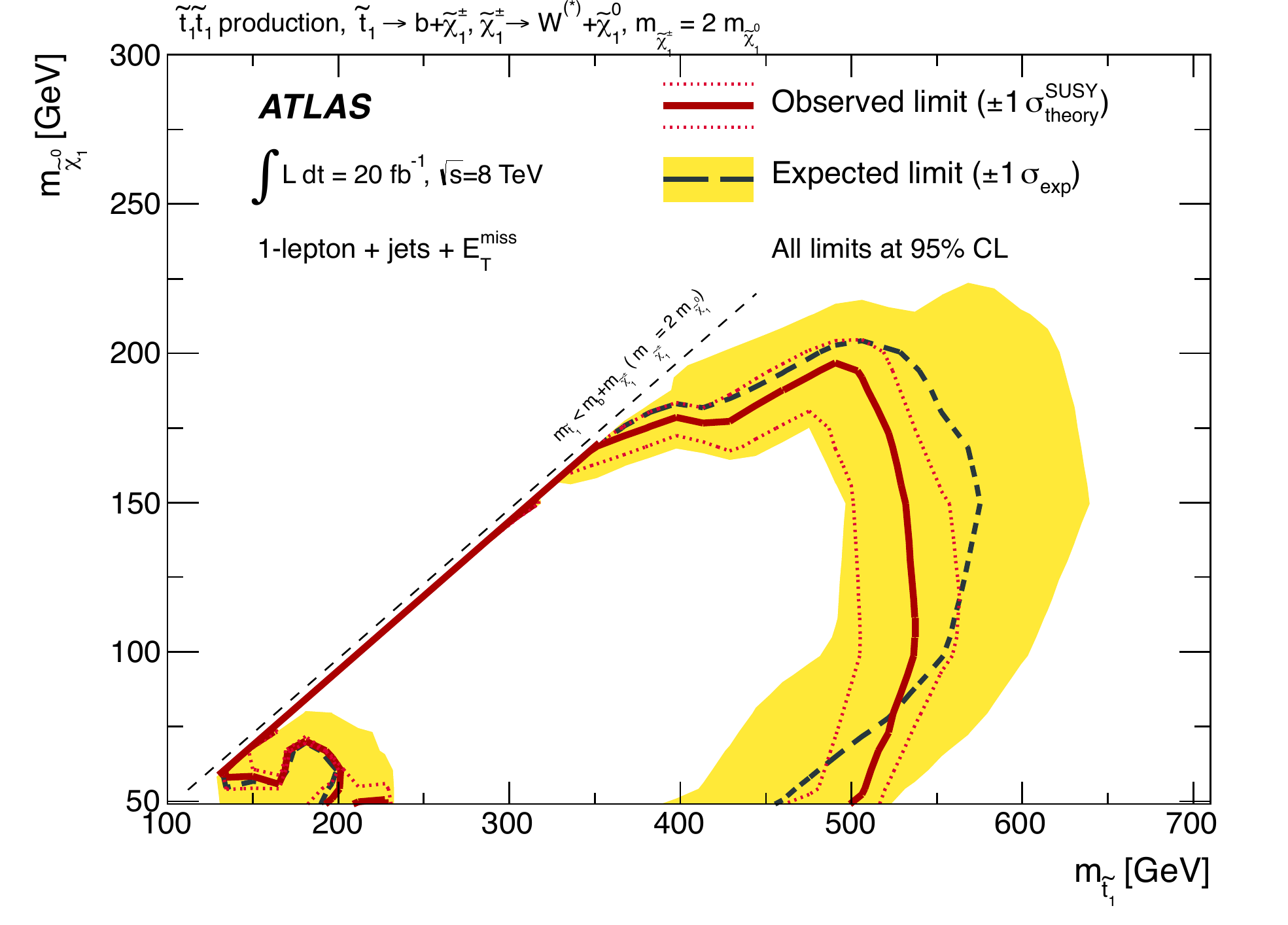}
\caption{Expected (black dashed) and observed (red solid) 95\% 
CL excluded region in the plane of $m_{\ninoone}$ vs. 
$m_{\stopone}$ , assuming $\mathcal{BR}(\bChargino) = 100$\%,
$\mathcal{BR}(\chinoonepm \to W^{(\star)} \ninoone) = 100$\% and
$m_{\chinoonepm} = 2 m_{\ninoone}$.
\label{fig:exclusion_bC_x2}}
\end{center}
\end{figure}

\begin{figure}
\begin{center}
\includegraphics[width=0.8\textwidth]{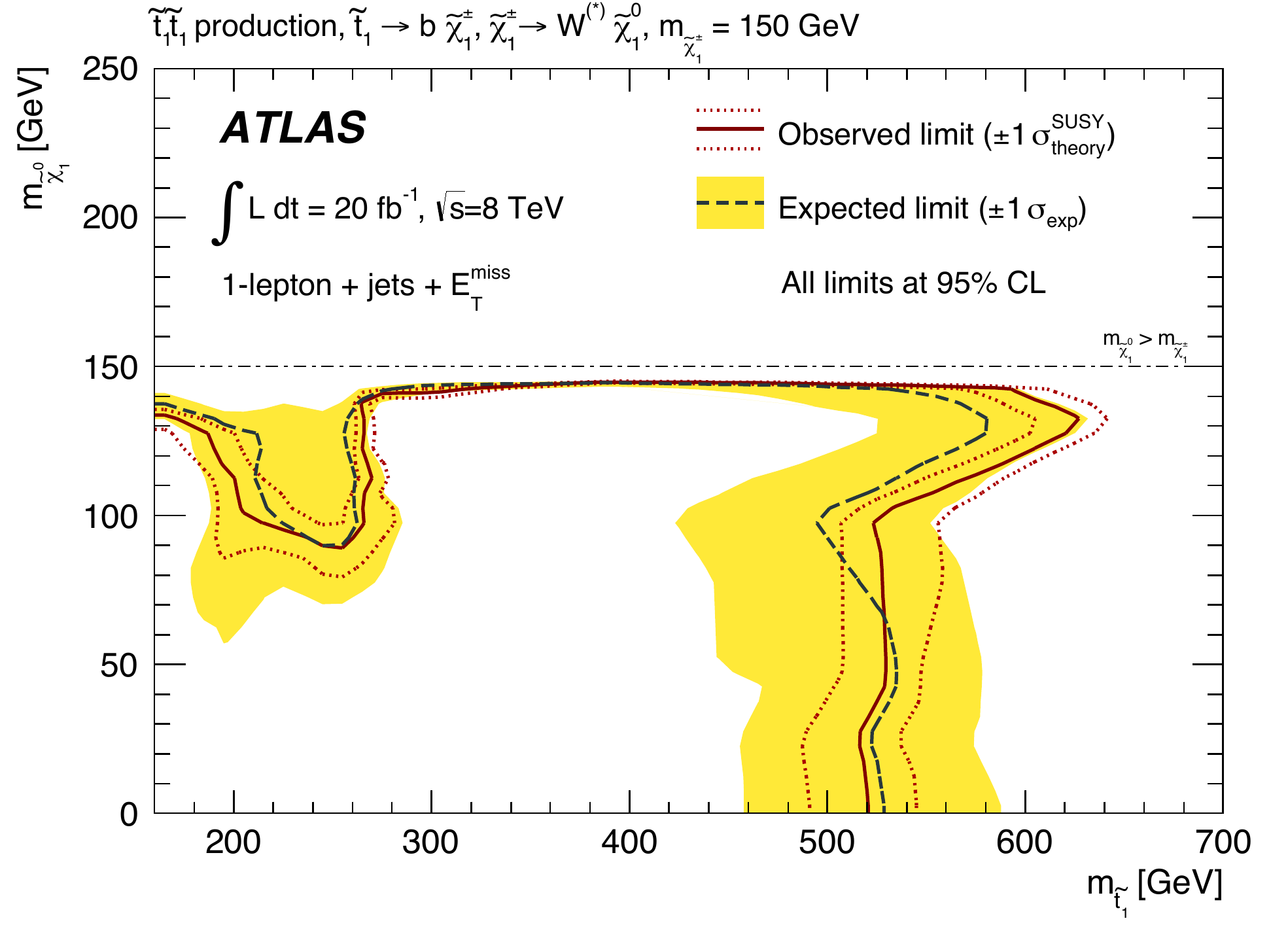}
\caption{Expected (black dashed) and observed (red solid) 95\% 
CL excluded region in the plane of $m_{\ninoone}$ vs. 
$m_{\stopone}$ , assuming $\mathcal{BR}(\bChargino) = 100$\%, 
$\mathcal{BR}(\chinoonepm \to W^{(\star)} \ninoone) = 100$\% and
$m_{\chinoonepm} = 150$\,\GeV.
\label{fig:exclusion_bC_150}}
\end{center}
\end{figure}

\begin{figure}
\begin{center}
\includegraphics[width=0.8\textwidth]{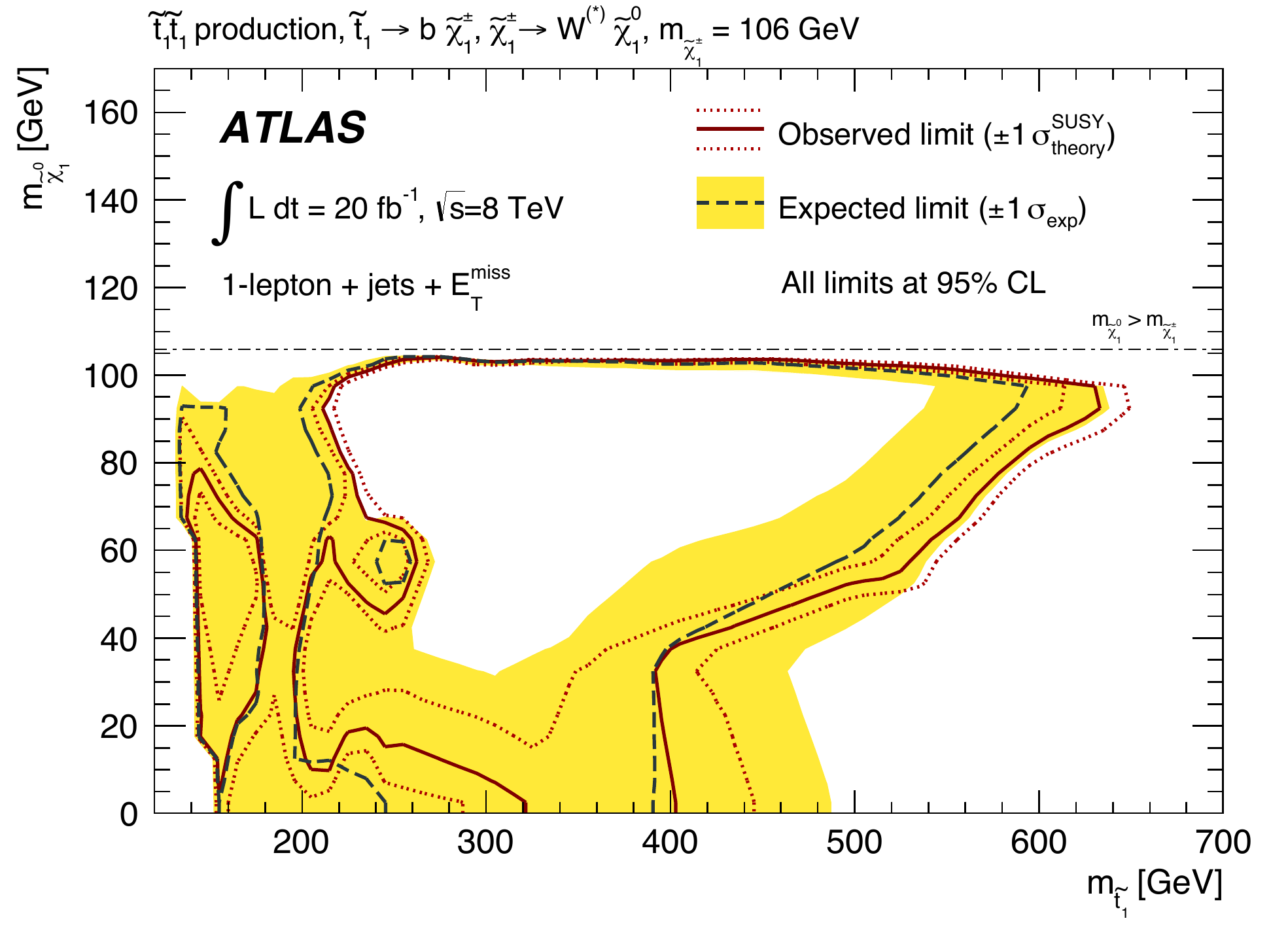}
\caption{Expected (black dashed) and observed (red solid) 95\% 
CL excluded region in the plane of $m_{\ninoone}$ vs. 
$m_{\stopone}$ , assuming $\mathcal{BR}(\bChargino) = 100$\%, 
$\mathcal{BR}(\chinoonepm \to W^{(\star)} \ninoone) = 100$\% and
$m_{\chinoonepm} = 106$\,\GeV.
\label{fig:exclusion_bC_106}}
\end{center}
\end{figure}

\begin{figure}
\begin{center}
\includegraphics[width=0.8\textwidth]{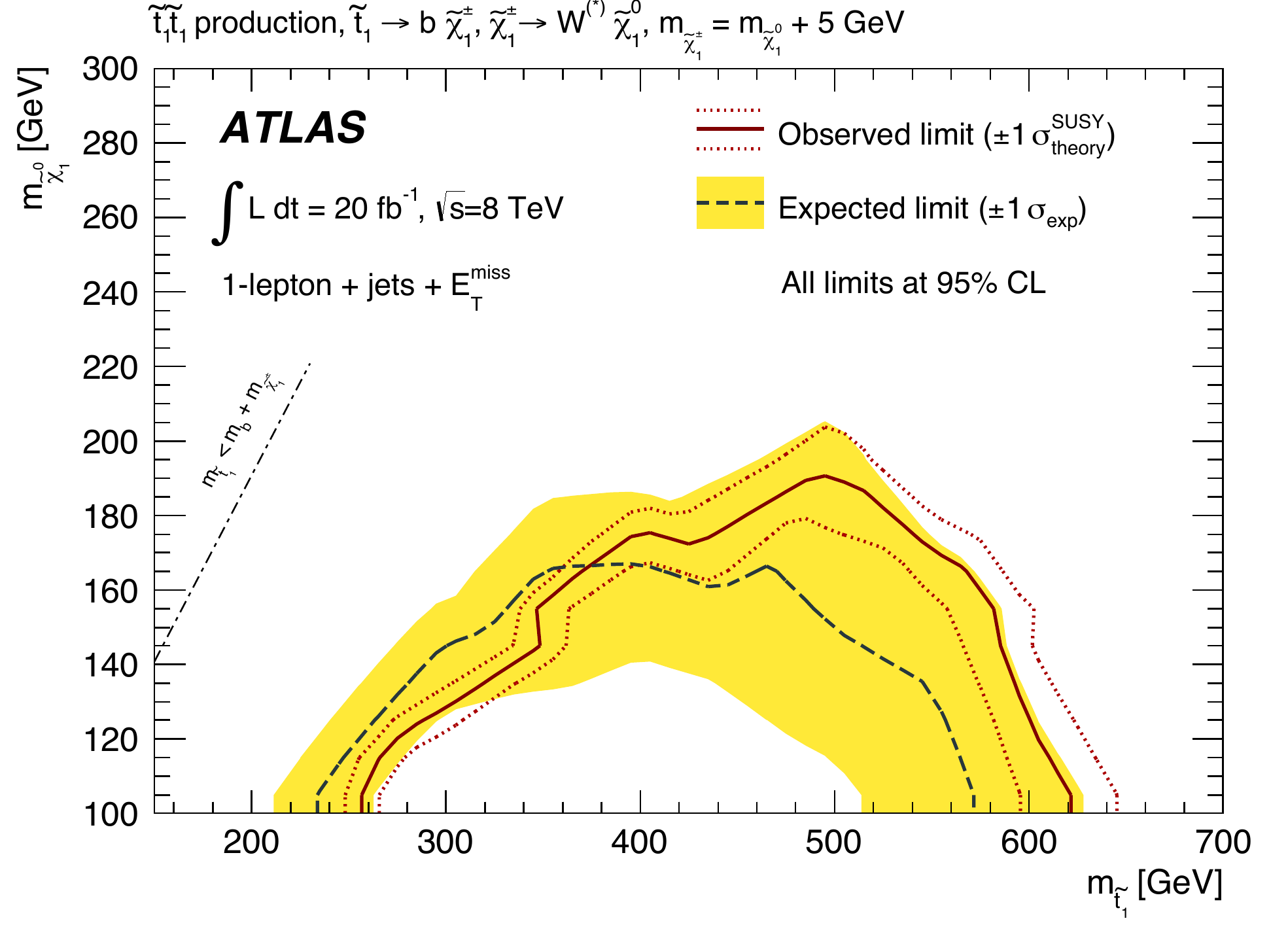}
\caption{Expected (black dashed) and observed (red solid) 95\% 
CL excluded region in the plane of $m_{\ninoone}$ vs. 
$m_{\stopone}$ , assuming $\mathcal{BR}(\bChargino) = 100$\%,
$\mathcal{BR}(\chinoonepm \to W^{\star} \ninoone) = 100$\% and
$m_{\chinoonepm} = m_{\ninoone} + 5$\,\GeV.
\label{fig:exclusion_bC_plus5}}
\end{center}
\end{figure}

\begin{figure}
\begin{center}
\includegraphics[width=0.8\textwidth]{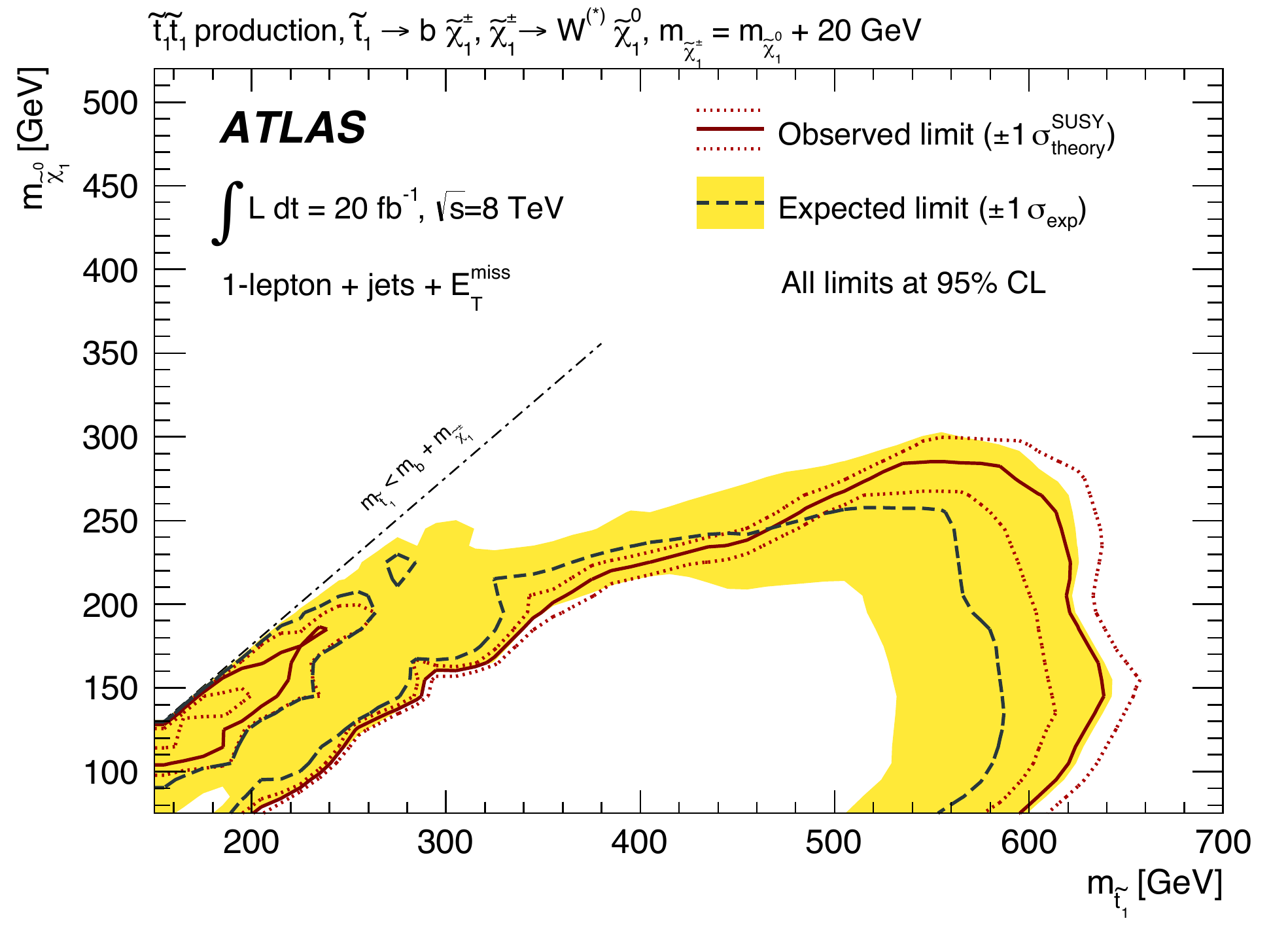}
\caption{Expected (black dashed) and observed (red solid) 95\% 
CL excluded region in the plane of $m_{\ninoone}$ vs. 
$m_{\stopone}$ , assuming $\mathcal{BR}(\bChargino) = 100$\%,
$\mathcal{BR}(\chinoonepm \to W^{\star} \ninoone) = 100$\% and 
$m_{\chinoonepm} = m_{\ninoone} + 20$\,\GeV.
\label{fig:exclusion_bC_plus20}}
\end{center}
\end{figure}

\begin{figure}
\begin{center}
\includegraphics[width=0.8\textwidth]{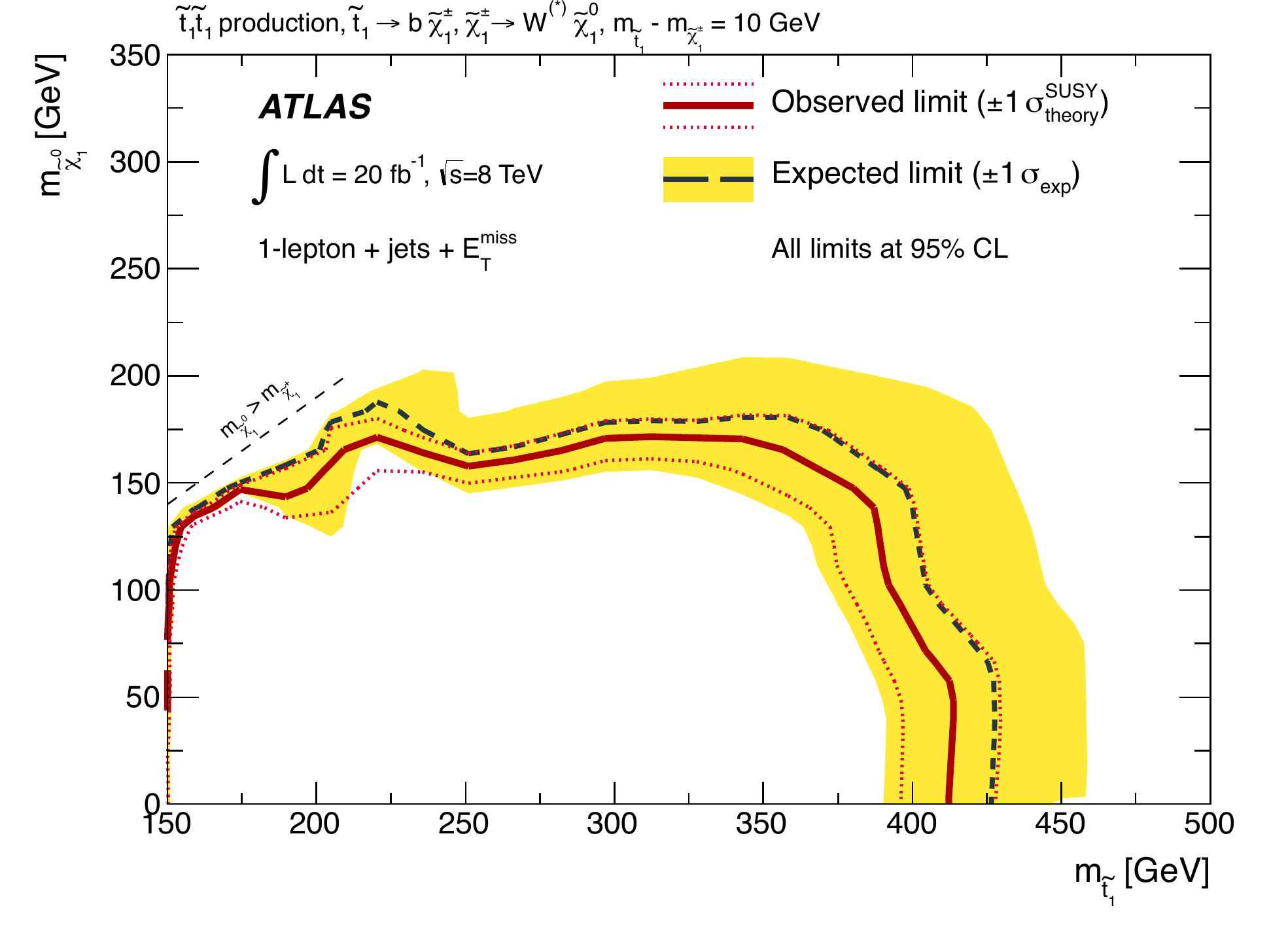}
\caption{Expected (black dashed) and observed (red solid) 95\% 
CL excluded region in the plane of $m_{\ninoone}$ vs. 
$m_{\stopone}$ , assuming $\mathcal{BR}(\bChargino) = 100$\%,
$\mathcal{BR}(\chinoonepm \to W^{(\star)} \ninoone) = 100$\% and
$m_{\chinoonepm} = m_{\tone} - 10\,\GeV$.
\label{fig:exclusion_bC_stch10}}
\end{center}
\end{figure}

\begin{figure}
\begin{center}
\includegraphics[width=0.8\textwidth]{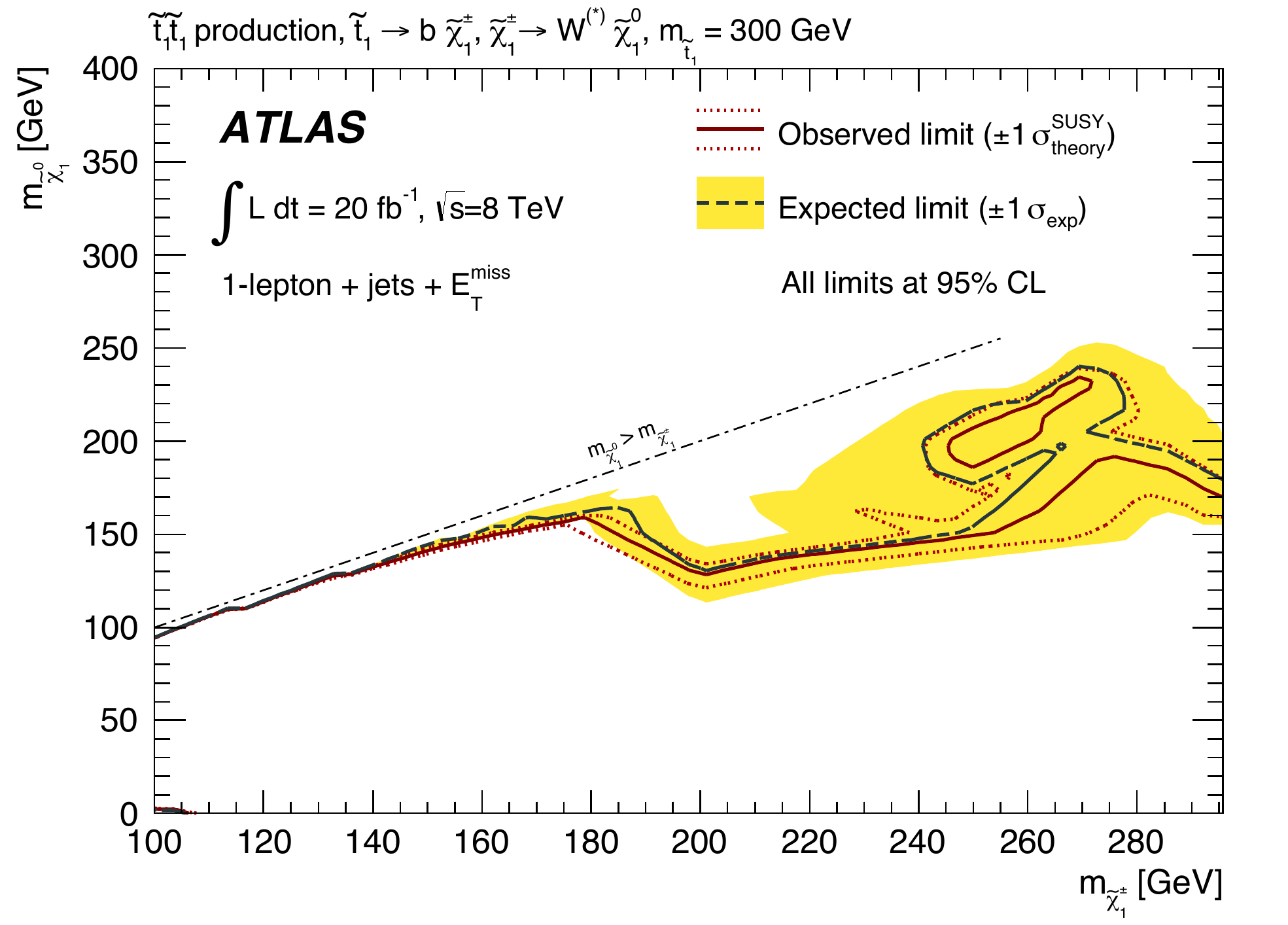}
\caption{Expected (black dashed) and observed (red solid) 95\% 
CL excluded region in the plane of $m_{\ninoone}$ vs. 
$m_{\chinoonepm}$ , assuming $\mathcal{BR}(\bChargino) = 100$\%, 
$\mathcal{BR}(\chinoonepm \to W^{(\star)} \ninoone) = 100$\% and
$m_{\tone} = 300$\,\GeV.
\label{fig:exclusion_stop300}}
\end{center}
\end{figure}

\begin{figure}
\begin{center}
\includegraphics[width=0.8\textwidth]{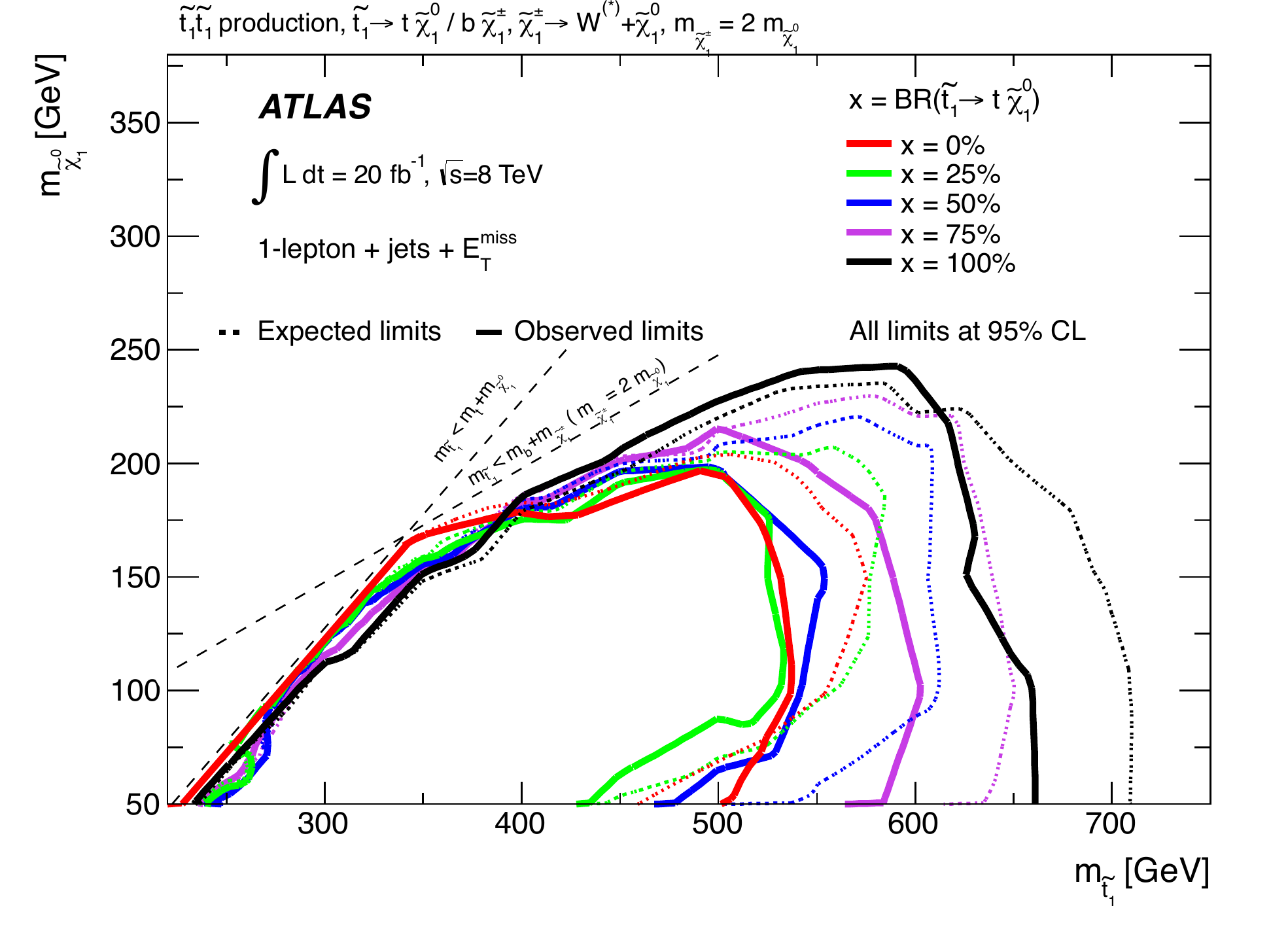} 
\caption{Expected (dashed) and observed (solid) 95\% CL excluded region 
in the plane of $m_{\ninoone}$ vs. $m_{\stopone}$, assuming 
$x = \mathcal{BR} (\topLSP) = 1- \mathcal{BR} (\bChargino)$, and $x$
varying from 0\% to 100\%.
\label{fig:exclusion_asym}}
\end{center}
\end{figure}

\begin{figure}
\begin{center}
\includegraphics[width=0.8\textwidth]{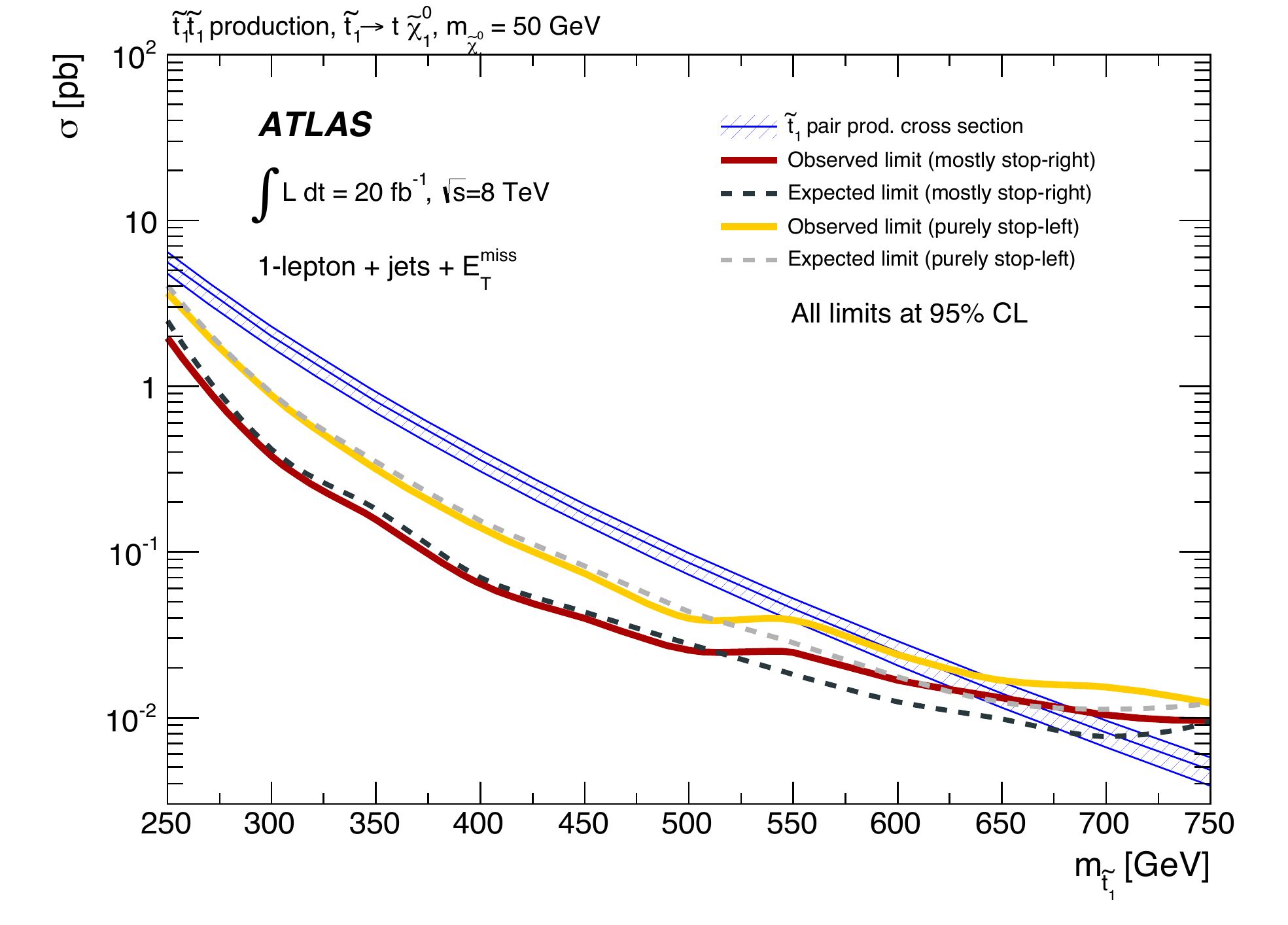}
\includegraphics[width=0.8\textwidth]{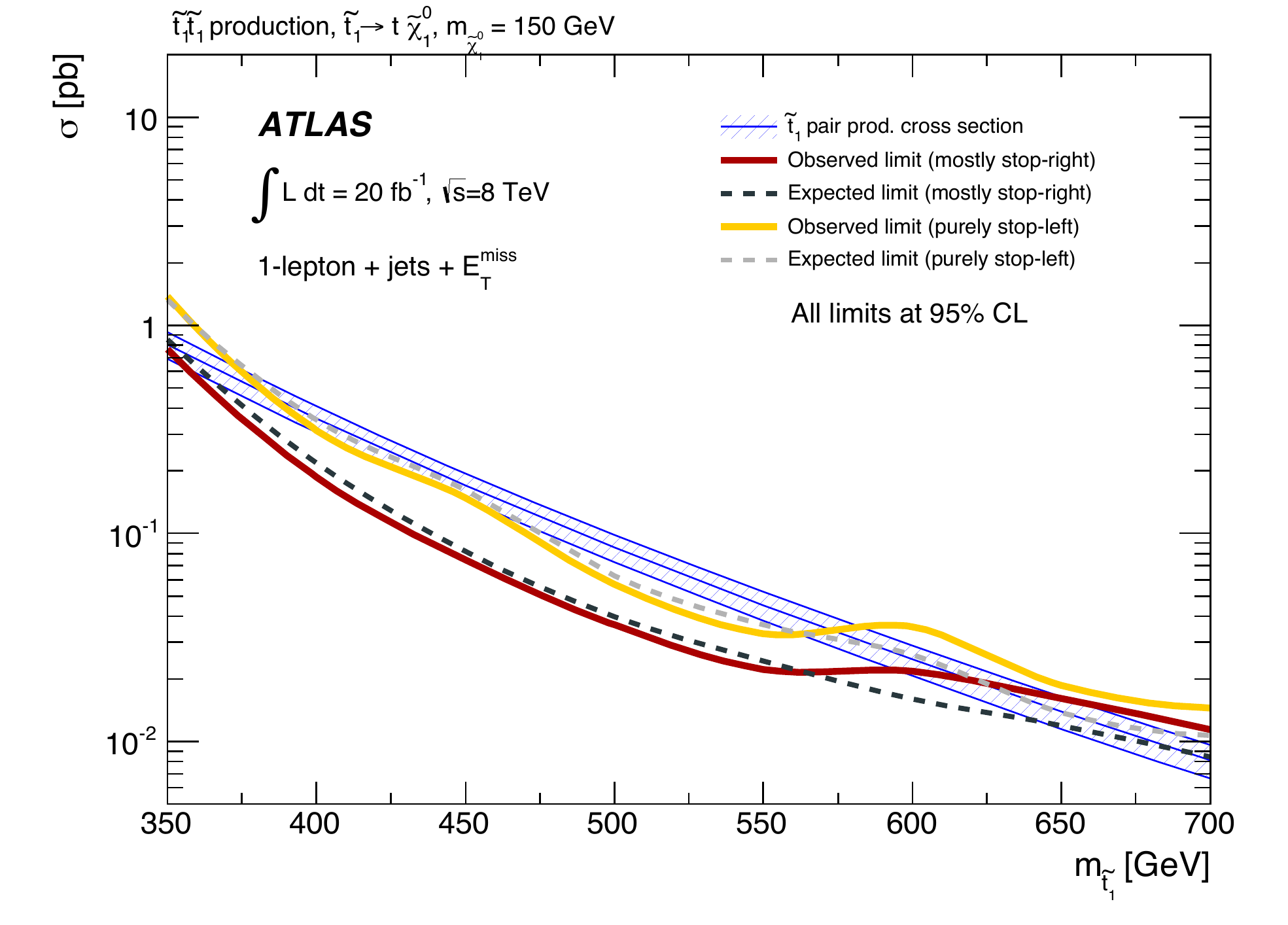}
\caption{
Expected and observed cross-section upper limits at $95\%$ CL
in the \topLSP\ decay mode with the LSP mass fixed to $50$\,\GeV\ (top) or $150$\,\GeV\ (bottom) for models with the \tone\ being a pure stop-left (\tleft) or mostly a stop-right (\tright).
The upper and lower blue lines correspond to the nominal signal cross-section scaled up and
down by the theoretical uncertainty.
\label{fig:leftright_limits}}
\end{center}
\end{figure}

\begin{figure}
\begin{center}
\includegraphics[width=0.8\textwidth]{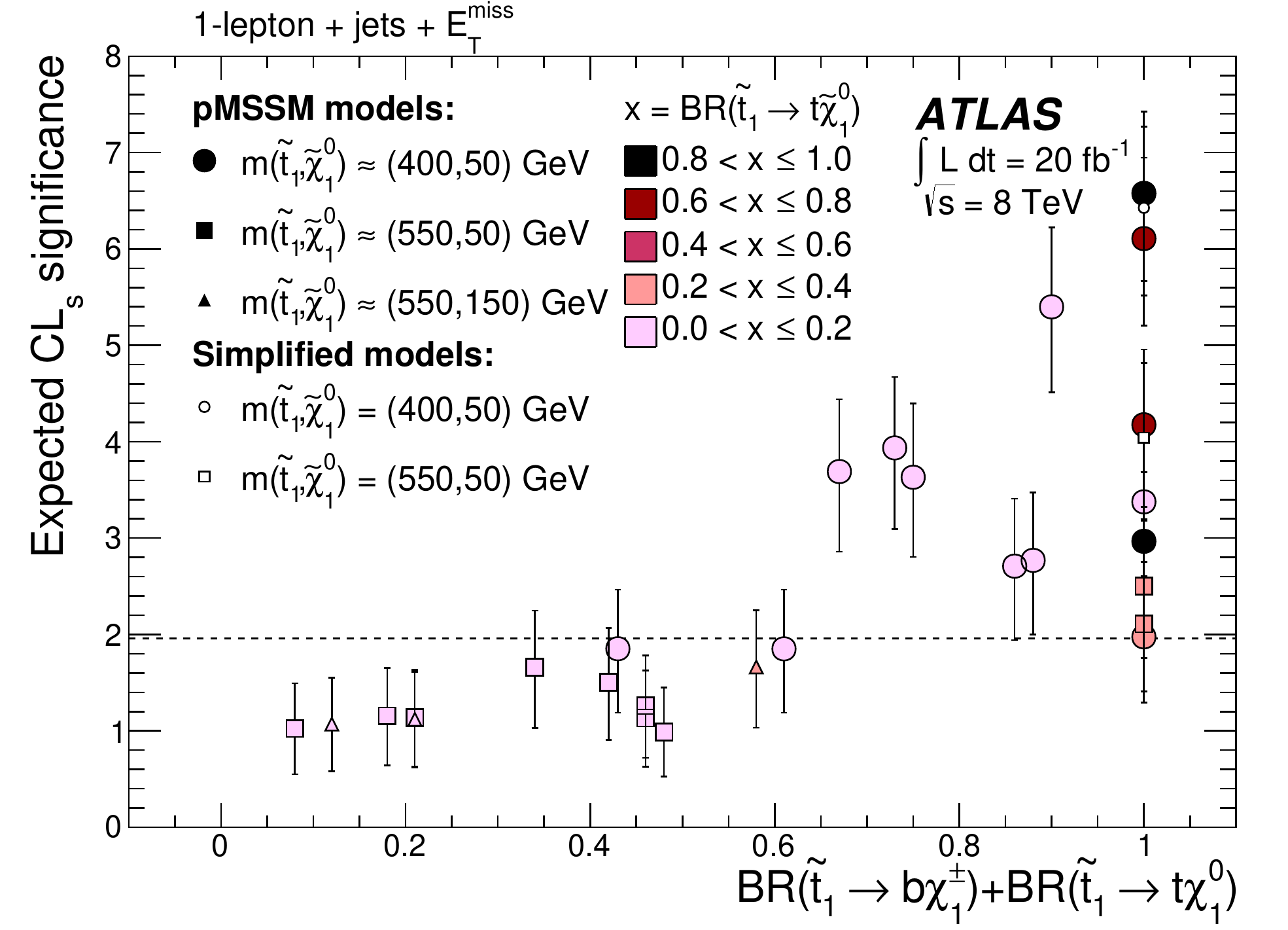}
\includegraphics[width=0.8\textwidth]{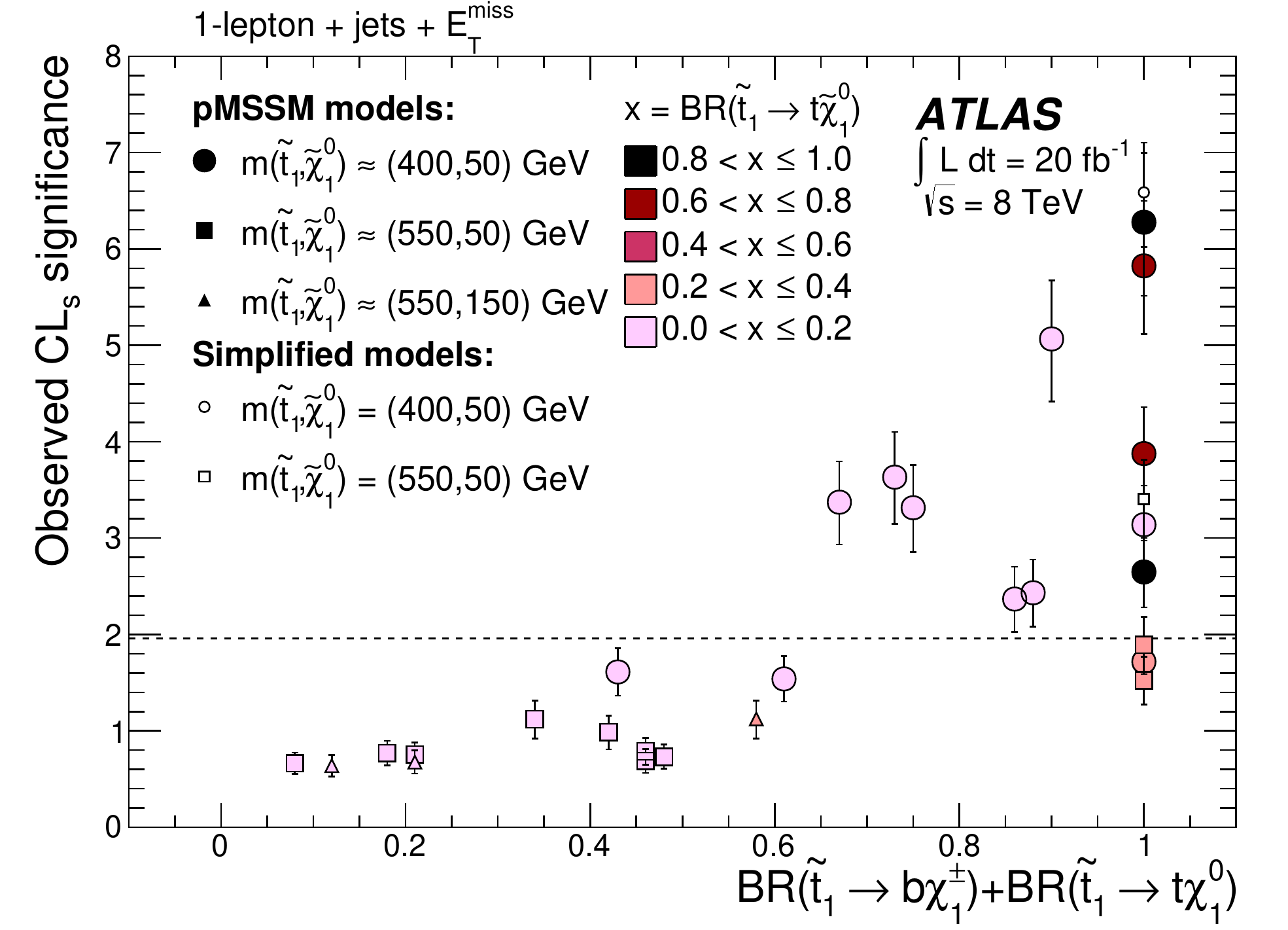}
\caption{The expected (top) and observed (bottom) \CLs\ significance values for the 27 pMSSM models described in 
\secref{sec:samples_signal} and for two simplified models.
Models above the dashed line, indicating the \CLs\ significance corresponding to $95\%$ CL, are excluded.
The results of the pMSSM models are displayed using filled markers where the marker symbol 
corresponds to the $m_{\tone}$, $m_{\ninoone}$ range and the colour represents
the branching ratio for the \topLSP\ decay, while the simplified models are shown using open markers.
The uncertainty on the expected \CLs\ significance includes all sources except the theoretical cross-section uncertainty,
while the uncertainty on the observed \CLs\ significance includes only the effect of scaling the nominal signal cross-section 
up and down by the theoretical cross-section uncertainty.
\label{fig:pMSSM_CLs}}
\end{center}
\end{figure}

 \section{Summary and conclusions}\label{sec:summary}

A search for stop pair production in final states with one isolated lepton, 
jets, and missing transverse momentum is presented.
Proton--proton collision data from the full 2012 data-taking period were analysed, corresponding 
to an integrated luminosity of $20$\,\ifb\ collected at $\sqrt{s}=8$\,\TeV\ by the ATLAS detector at the LHC.
Five decay modes are considered:
(1) each stop decays to a top quark and the 
LSP;
(2) each stop decays to a bottom quark and the lightest chargino (\chinoonepm),
where the \chinoonepm\ decays via an on- or off-shell $W$ boson to the LSP;
(3) each stop decays 
in a three-body process to a bottom quark, a $W$ boson, and the LSP;
(4) each stop decays in a four-body process to a bottom quark, the LSP and two light fermions;
(5) the two stops decay independently either as described in (1) or in (2).
In all scenarios, $R$-parity is conserved and the LSP is assumed to be the \ninoone.

The results are in agreement with predictions from the Standard Model, 
and are thus translated into 95\% CL upper limits on
the stop and \ninoone\ masses in various supersymmetric scenarios.
For models where the stop decays exclusively into a top quark and a \ninoone\ (scenario (1) above), stop masses
between $210$ and 
$640$\,\GeV\ are excluded for a massless LSP, and stop masses around $550$\,\GeV\ 
are excluded for LSP masses below $230$\,\GeV.
Limits are also derived in the three- and four-body scenarios.
For scenarios where the stop decays exclusively into a bottom quark and a \chinoonepm\ (scenario (2) above),
the excluded stop and \ninoone\ masses depend strongly on the mass of the \chinoonepm.
For models where the mass of the \chinoonepm\ is twice that of the LSP, 
stop masses up to $500$\,\GeV\ are excluded for an LSP mass in the range of $100$ to $150$\,\GeV.
For models in which the \chinoonepm\ mass is only 
$20$\,\GeV\ 
above the LSP mass, 
stop masses between $240$ and $600$\,\GeV\ 
are excluded for an 
LSP mass of $100$\,\GeV.
In scenarios where only the \topLSP\ and \bChargino\ decay modes are allowed,
the largest excluded stop mass for an LSP mass of $100$\,\GeV\ gradually increases from $530$\,\GeV\ to $660$\,\GeV\ as 
the branching ratio for \topLSP\ is increased from $0$\% to $100$\%.
Using a limited set of pMSSM models, the exclusion power is found to decrease with an increased 
branching ratio to decays other than \topLSP\ and \bChargino. 
These results supersede and significantly extend previous ATLAS limits.


\section{Acknowledgements}

We thank CERN for the very successful operation of the LHC, as well as the
support staff from our institutions without whom ATLAS could not be
operated efficiently.

We acknowledge the support of ANPCyT, Argentina; YerPhI, Armenia; ARC,
Australia; BMWFW and FWF, Austria; ANAS, Azerbaijan; SSTC, Belarus; CNPq and FAPESP,
Brazil; NSERC, NRC and CFI, Canada; CERN; CONICYT, Chile; CAS, MOST and NSFC,
China; COLCIENCIAS, Colombia; MSMT CR, MPO CR and VSC CR, Czech Republic;
DNRF, DNSRC and Lundbeck Foundation, Denmark; EPLANET, ERC and NSRF, European Union;
IN2P3-CNRS, CEA-DSM/IRFU, France; GNSF, Georgia; BMBF, DFG, HGF, MPG and AvH
Foundation, Germany; GSRT and NSRF, Greece; ISF, MINERVA, GIF, I-CORE and Benoziyo Center,
Israel; INFN, Italy; MEXT and JSPS, Japan; CNRST, Morocco; FOM and NWO,
Netherlands; BRF and RCN, Norway; MNiSW and NCN, Poland; GRICES and FCT, Portugal; MNE/IFA, Romania; MES of Russia and ROSATOM, Russian Federation; JINR; MSTD,
Serbia; MSSR, Slovakia; ARRS and MIZ\v{S}, Slovenia; DST/NRF, South Africa;
MINECO, Spain; SRC and Wallenberg Foundation, Sweden; SER, SNSF and Cantons of
Bern and Geneva, Switzerland; NSC, Taiwan; TAEK, Turkey; STFC, the Royal
Society and Leverhulme Trust, United Kingdom; DOE and NSF, United States of
America.

The crucial computing support from all WLCG partners is acknowledged
gratefully, in particular from CERN and the ATLAS Tier-1 facilities at
TRIUMF (Canada), NDGF (Denmark, Norway, Sweden), CC-IN2P3 (France),
KIT/GridKA (Germany), INFN-CNAF (Italy), NL-T1 (Netherlands), PIC (Spain),
ASGC (Taiwan), RAL (UK) and BNL (USA) and in the Tier-2 facilities
worldwide.

\clearpage
\clearpage
\bibliographystyle{JHEP}
\bibliography{paper}

\clearpage

\appendix
\section{Detailed description of the discriminating variables\label{sec:appendixeq}}
This section provides more detailed descriptions of the discriminating variables that are introduced
in  \secref{sec:object_defs}.

\begin{list}
{\textbf{\arabic{itemcounter2}.}}
{\usecounter{itemcounter}\leftmargin=1.4em}
\item[-] Stransverse Mass, \mtTwo \\
This variable targets decay topologies with two branches, referred to here as $a$ and $b$.
In each branch, there are some particles with fully measured momenta and some particles with momenta that are not measured directly. The sum of the four vectors of the measured momenta in branch $i\in\{a,b\}$ are denoted $p_i=(E_i,\vec{p}_{\text{T}i},p_{zi})$ and the sum of the four vectors of the unmeasured momenta are denoted $q_i=(F_i,\vec{q}_{\text{T}i},q_{zi})$.  With $m_{p_i}^2=E_i^2-{\vec{p}_i}^\text{  $2$}$ and $m_{q_i}^2=F_i^2-{\vec{q}_i}^\text{  $2$}$, the \mt\ of the particles in branch $i$ is given in general by

\begin{align*}
m_{\text{T}i}^2=\left(\sqrt{p_{\text{T}i}^2+m_{p_i}^2}+\sqrt{q_{\text{T}i}^2+m_{q_i}^2}\right)^2-\left(\vec{p}_{\text{T}i}+\vec{q}_{\text{T}i}\right)^2
\end{align*}

which in the case that $m_{q_i}=m_{p_i}=0$ is the same as the one given in \secref{sec:variables}.  A generalisation of \mt,~\mtTwo, is defined as a minimisation over the allocation of $\Ptmiss$ between $\vec{q}_{\text{T}a}$ and $\vec{q}_{\text{T}b}$ of the maximum of the corresponding $m_\text{Ta}$ or $m_\text{Tb}$: 

\begin{align*}
\mtTwo \equiv\underset{\vec{q}_{\text{T}a}+\vec{q}_{\text{T}b}= \Ptmiss}{\min} \{\max (m_{\text{T}a},m_{\text{T}b})\},
\end{align*}
where one must make an assumption of $m_{q_a}$ and $m_{q_b}$ in the computation of $m_{Ta}$ and $m_{Tb}$.  The result of the above minimisation is the minimum parent mass consistent with the observed kinematic distributions under the inputs $m_{q_a}$ and $m_{q_b}$.  The variants of $\mtTwo$ described below only differ in the measured particles, (assumed) unmeasured particles, and choices for the input masses, $m_{q_a}$ and $m_{q_b}$.

\item[-] Asymmetric \mtTwo, \amtTwo 
\begin{itemize}
\item[-] Measured particles: For branch $a$, this is one of the $b$-jets and for branch $b$ this is the second $b$-jet and the charged lepton.  The $b$-jets are identified based on the highest $b$-tagging weights.   Since there are two ways of assigning the $b$-tagged jets to branches $a$ and $b$, both $\mtTwo$ values are computed and the minimum kept for the final discriminant. 
\item[-] Unmeasured particles: For branch $a$, this is a $W$ boson that decays leptonically, with the charged lepton unidentified as such.  The unmeasured particle for branch $b$ is the neutrino associated with the measured charged lepton.
\item[-] Input masses: $m_{q_a}=m_{W} = 80$\,\GeV\ and $m_{q_b}=m_{\nu} = 0$\,\GeV.
\item[-] In cases in which the lost lepton is an electron and the corresponding energy deposit enters the \met\ calculation, for instance as a soft calorimeter cluster, \amtTwo\ can exceed the top mass boundary in \ttbar\ events, but the variable remains powerful at discriminating signal from background.
\end{itemize}

\item[-] $\tau$-based \mtTwo, \mtTwoTau 
\begin{itemize}
\item[-] Measured particles: For branch $a$, this is the $\tau$-jet, identified as the highest-$p_\text{T}$ jet excluding the selected two $b$-tagged jets.  The measured particle for branch $b$ is the charged lepton.
\item[-] Unmeasured particles: For branch $a$, this includes the two neutrinos associated with the $\tau$ production and hadronic decay.  The unmeasured particle for branch $b$ is the neutrino associated with the charged lepton.
\item[-] Input masses: $m_{q_a}= 0$\,\GeV\ and $m_{q_b}=m_{\nu} = 0$\,\GeV.
\end{itemize}

\item[-] Topness \\
The \topness\ event value is defined as $\ln(\min \hat{S})$, where $\hat{S}$ is the minimum of the $\chi^2$-type function $S$:
\begin{align*}
S(p_{W,x},p_{W,y},p_{W,z},p_{\nu,z})\quad=\quad&\frac{\left(m_W^2-\left(p_\ell+p_\nu\right)^2\right)^2}{a_W^4}+ \frac{\left(m_t^2-\left(p_{b_1}+p_\ell+p_\nu\right)^2\right)^2}{a_t^4}+\\
&\frac{\left(m_t^2-\left(p_{b_2}+p_W\right)^2\right)^2}{a_t^4}+
\frac{\left(4\,m_t^2-\left(\Sigma p\right)^2\right)^2}{a_{\rm CM}^4}.
\end{align*}

The first three arguments of $S$ are the components of the non-reconstructed $W$ boson
3-momentum ($p_{W,x}$, $p_{W,y}$, $p_{W,z}$). This $W$ is assumed to decay
leptonically, but the lepton is not reconstructed and is thus only noticeable in
the missing transverse momentum. The variable $p_{\nu,z}$ is the longitudinal momentum of the
neutrino from the other $W$ boson decay, for which the lepton was successfully
reconstructed. These four numbers are varied to find the minimum of $S$.

The momenta appearing on the right-hand side of the equation above are either 4-momenta of the
reconstructed objects (one lepton, $p_\ell$, and two $b$-jets, $p_{b_1}$ and
$p_{b_2}$) or 4-momenta assigned by the minimisation procedure ($p_W$ and
$p_\nu$). 
To find all four components, the neutrinos and the $W$ boson without reconstructed decay products 
are assumed to be on-shell.
Both combinations for $b_1$ and $b_2$ are evaluated during the
minimisation; if only one $b$-tagged jet is present, it is used together with the
leading or subleading jet (that means, a total of four possible jet assignments
is evaluated in this case).

The minimisation is constrained such that the observed missing transverse momentum is
attributed to the unobserved $W$ boson (decaying into a not-reconstructed lepton
and a neutrino) and a neutrino from the other top decay branch.

The constants $a_W$, $a_t$ and $a_{\rm CM}$ are set to the values suggested by the authors of $S$: $a_W=5\,\GeV$, $a_t=15\,\GeV$, $a_{\rm CM}=1\,\TeV$.

\item[-] Hadronic top mass, \mtophad: \\
This reconstructed top mass is constructed as $m_{j_1,j_2,b_i}$ by minimising

$$
\chi^2 = \frac{\bigl( m_{j_1,j_2,b_i} - m_{\mathrm top} \bigr)^2}{\sigma^2_{m_{j_1,j_2,b_i}}} + \frac{\bigl( m_{j_1,j_2} - m_W \bigr)^2}{\sigma^2_{m_{j_1,j_2}}},
$$

where $i =1$ or $2$; $b_1$ and $b_2$ are the two jets with the highest $b$-tagging weights; $j_1$, $j_2$ are the highest $p_\text{T}$ jets from the selected jets in the event excluding $b_1$ and $b_2$ and 

\begin{align*}
\sigma^2_{m_{j_1,j_2,b_i}}&=m_{j_1,j_2,b_i}^2(r_{j_1}^2+r_{j_2}^2+r_{b_i}^2)\\
\sigma^2_{m_{j_1,j_2}}&=m_{j_1,j_2}^2(r_{j_1}^2+r_{j_2}^2),
\end{align*}

where $r_i$ is the fractional jet energy uncertainty of the $p_\text{T}$ for jet $i$ determined by dedicated studies~\cite{Aad:2011he,Aad:2012ag}.

\item[-] $\tau$-veto:\\
For the construction of the $\tau$-veto, the reconstructed \tauh\ candidates are subject to further selection requirements. Candidates are required to have either one associated track (classified as one-prong $\tau$ decay), or two to three tracks (classified as three-prong $\tau$ decay, where one track can be missed).
The \tauh\ charge for candidates with one or three tracks is required to be $\pm 1$ and to be opposite to the charge of the selected electron or muon in the event.
For candidates with two tracks, the sign of the \tauh\ charge is required to be opposite to that of the selected lepton only if the \tauh\ charge is $\pm 2$.
Finally, three different BDT requirements are imposed on the candidates to define three $\tau$-veto working points:
loose, tight, and extra-tight. 
In simulated \ttbar\ events with one $W \rightarrow \ell \nu$ decay, signal- and background-like events are defined by requiring the other $W$ boson to either decay into quarks (signal) or into a \tauh (background). In these samples, the loose (tight) $\tau$-veto retains $99\%$ ($97\%$) of signal events, while for background events $81\%$ ($69\%$) with a one-prong and $75\%$ ($63\%$) with a three-prong \tauh\ decay survive the veto.

\item[-] Track-veto\\
Tracks are required to satisfy the following criteria:
$\pT >10$\,\GeV\ and $|\eta|< 2.5$,  transverse and longitudinal impact parameters
$|d_0|< 1$\,mm and $|z_0|<2$\,mm.
The track isolation requires that there are no additional tracks associated with the primary vertex with $\pT > 3$\,\GeV\ in a cone of $\Delta R=0.4$ around the track.
Events with at least one isolated track of opposite charge compared to that of the selected electron or muon in the event are rejected by the track-veto.

\item[-] \HTmissSig\ is an object-based missing transverse momentum, divided by the per-event resolution of the jets. It is defined by

$$
\HTmissSig = \frac{|\vec{H}_\text{T}^\text{miss}|-M}{\sigma_{|\vec{H}_\text{T}^\text{miss}|}},
$$

where $\vec{H}_\text{T}^\text{miss}$ is the negative sum of the jets and lepton vectors. The denominator is computed 
from the per-event jet energy uncertainties, while the lepton is assumed to be well-measured.
The parameter $M$ is chosen to be a characteristic `scale' of the background~\cite{Nachman:2013bia}, and is fixed at $100$\,\GeV\ in this analysis based on optimisation studies.

\end{list}

\section{Background fit results\label{sec:appendixbkg}}

This section contains the background fit results for all analyses. The model-dependent selection 
is used. The CR and SR bins (cut-and-count and one-dimensional shape-fits) or the full set of bins (two-dimensional shape-fits) are included in the likelihood. However, a potential signal contribution is neglected everywhere (the signal strength is fixed to zero). All background uncertainties are taken into account. 
This fit configuration is different from the background-only fit, which is used for the validation results in \secref{sec:backgrounds}, in that it includes more bins to constrain the likelihood.
The results of the cut-and-count analyses are given in \tabref{allBinsFit_cut_and_count}, and the results for the shape-fits are shown in \tabsref{allBinsFit_tN1p}--\ref{allBinsFit_bWN}.

\begin{table}[p]
\begin{center}
\setlength{\tabcolsep}{0.0pc}
{\small
  \begin{tabular}{|l|c|cc|cc|cc|cc|}
    \hline
    \multirow{2}{*}{Region} & \multirow{2}{*}{Obs.} & \multicolumn{8}{c|}{Fitted (estimated) background} \\
    \cline{3-10}
    & & \multicolumn{2}{c|}{Total} & \multicolumn{2}{c|}{$t\bar{t}$} & \multicolumn{2}{c|}{$W$+jets} & \multicolumn{2}{c|}{Other} \\
    \hline
    \SRtNtwox & & & & & & & & & \\
    TCR & $159$ & $159 \pm 12$ & $(155)$ & $124 \pm 14$ & $(108)$ & $19 \pm 4$ & $(32)$ & $15.8 \pm 2.6$ & $(16)$\\
    WCR & $161$ & $161 \pm 13$ & $(232)$ & $36 \pm 9$ & $(31)$ & $109 \pm 17$ & $(185)$ & $16 \pm 5$ & $(15)$\\
    TVR & $16$ & $26 \pm 6$ & $(26)$ & $20 \pm 5$ & $(17)$ & $3.4 \pm 1.0$ & $(6)$ & $3.1 \pm 0.7$ & $(3)$\\
    WVR & $25$ & $24 \pm 7$ & $(34)$ & $6.7 \pm 2.5$ & $(6)$ & $15 \pm 5$ & $(26)$ & $2.0 \pm 1.0$ & $(2)$\\
    SR & $12$ & $12.7 \pm 1.8$ & $(14)$ & $6.3 \pm 1.5$ & $(6)$ & $2.0 \pm 0.5$ & $(3)$ & $4.4 \pm 0.9$ & $(4)$\\
    \hline
    \SRtNthreep & & & & & & & & & \\ 
    TCR & $359$ & $359 \pm 19$ & $(361)$ & $287 \pm 22$ & $(274)$ & $39 \pm 7$ & $(55)$ & $32 \pm 6$ & $(32)$\\
    WCR & $483$ & $483 \pm 22$ & $(615)$ & $100 \pm 24$ & $(95)$ & $340 \pm 40$ & $(475)$ & $44 \pm 12$ & $(42)$\\
    TVR & $59$ & $79 \pm 18$ & $(79)$ & $64 \pm 15$ & $(61)$ & $6.8 \pm 2.1$ & $(10)$ & $8.3 \pm 1.2$ & $(8)$\\
    WVR & $74$ & $84 \pm 24$ & $(104)$ & $22 \pm 7$ & $(21)$ & $52 \pm 16$ & $(73)$ & $9.7 \pm 2.1$ & $(9)$\\
    SR & $5$ & $5.0 \pm 0.9$ & $(5)$ & $2.0 \pm 0.6$ & $(2)$ & $0.87 \pm 0.26$ & $(1)$ & $2.2 \pm 0.4$ & $(2)$\\
    \hline
    \SRtNboost & & & & & & & & & \\ 
    TCR & $117$ & $118 \pm 11$ & $(111)$ & $96 \pm 11$ & $(80)$ & $10.8 \pm 2.2$ & $(20)$ & $11.2 \pm 1.9$ & $(11)$\\
    WCR & $210$ & $210 \pm 14$ & $(332)$ & $35 \pm 10$ & $(29)$ & $149 \pm 20$ & $(278)$ & $26 \pm 7$ & $(24)$\\
    TVR & $20$ & $23 \pm 5$ & $(21)$ & $19 \pm 5$ & $(15)$ & $1.9 \pm 0.4$ & $(3)$ & $2.5 \pm 0.6$ & $(2)$\\
    WVR & $29$ & $35 \pm 7$ & $(51)$ & $7.7 \pm 2.3$ & $(6)$ & $21 \pm 5$ & $(38)$ & $6.3 \pm 1.2$ & $(6)$\\
    SR & $5$ & $3.5 \pm 0.7$ & $(3)$ & $1.2 \pm 0.4$ & $(0.9)$ & $0.31 \pm 0.14$ & $(0.5)$ & $2.1 \pm 0.4$ & $(2)$\\
    \hline
    \SRbCzero & & & & & & & & & \\ 
    TCR & $1650$ & $1650 \pm 40$ & $(1689)$ & $1240 \pm 60$ & $(1246)$ & $244 \pm 34$ & $(278)$ & $164 \pm 24$ & $(162)$\\
    WCR & $2162$ & $2160 \pm 50$ & $(2361)$ & $350 \pm 60$ & $(334)$ & $1670 \pm 90$ & $(1890)$ & $141 \pm 25$ & $(127)$\\
    TVR & $925$ & $890 \pm 100$ & $(876)$ & $770 \pm 90$ & $(752)$ & $46 \pm 7$ & $(51)$ & $73 \pm 7$ & $(71)$\\
    WVR & $693$ & $740 \pm 160$ & $(781)$ & $163 \pm 35$ & $(155)$ & $520 \pm 130$ & $(567)$ & $60 \pm 6$ & $(52)$\\
    SR & $493$ & $489 \pm 21$ & $(501)$ & $158 \pm 34$ & $(145)$ & $250 \pm 27$ & $(276)$ & $82 \pm 17$ & $(74)$\\
    \hline
    \SRbCfour & & & & & & & & & \\ 
    TCR & $218$ & $220 \pm 15$ & $(228)$ & $171 \pm 17$ & $(172)$ & $22 \pm 5$ & $(29)$ & $28 \pm 5$ & $(27)$\\
    WCR & $757$ & $758 \pm 28$ & $(1002)$ & $87 \pm 25$ & $(86)$ & $610 \pm 40$ & $(860)$ & $57 \pm 16$ & $(55)$\\
    TVR & $42$ & $53 \pm 13$ & $(53)$ & $43 \pm 11$ & $(41)$ & $4.5 \pm 1.3$ & $(6)$ & $6.3 \pm 1.2$ & $(6)$\\
    WVR & $166$ & $148 \pm 35$ & $(191)$ & $23 \pm 8$ & $(23)$ & $111 \pm 29$ & $(153)$ & $14.4 \pm 3.4$ & $(14)$\\
    SR & $16$ & $11.8 \pm 1.4$ & $(12)$ & $5.6 \pm 1.0$ & $(5)$ & $2.1 \pm 0.6$ & $(3)$ & $4.2 \pm 1.0$ & $(4)$\\
    \hline
    \SRbCfive & & & & & & & & & \\ 
    TCR & $129$ & $129 \pm 11$ & $(154)$ & $92 \pm 13$ & $(110)$ & $16 \pm 4$ & $(23)$ & $22 \pm 4$ & $(22)$\\
    WCR & $654$ & $654 \pm 26$ & $(911)$ & $56 \pm 19$ & $(67)$ & $550 \pm 40$ & $(792)$ & $53 \pm 16$ & $(52)$\\
    TVR & $28$ & $32 \pm 6$ & $(38)$ & $23 \pm 5$ & $(28)$ & $3.6 \pm 0.9$ & $(5)$ & $5.7 \pm 1.3$ & $(6)$\\
    WVR & $157$ & $135 \pm 29$ & $(185)$ & $16 \pm 6$ & $(19)$ & $106 \pm 25$ & $(152)$ & $13.7 \pm 2.9$ & $(13)$\\
    SR & $5$ & $4.5 \pm 0.8$ & $(5)$ & $1.8 \pm 0.4$ & $(2)$ & $0.73 \pm 0.32$ & $(1)$ & $2.0 \pm 0.5$ & $(2)$\\
    \hline
    \SRtNbC & & & & & & & & & \\ 
    TCR & $177$ & $178 \pm 13$ & $(168)$ & $139 \pm 15$ & $(125)$ & $20 \pm 4$ & $(25)$ & $18.5 \pm 3.5$ & $(18)$\\
    WCR & $387$ & $387 \pm 20$ & $(446)$ & $80 \pm 21$ & $(71)$ & $281 \pm 31$ & $(349)$ & $27 \pm 7$ & $(25)$\\
    TVR & $64$ & $77 \pm 19$ & $(70)$ & $63 \pm 16$ & $(54)$ & $7.4 \pm 1.8$ & $(9)$ & $7.4 \pm 1.8$ & $(7)$\\
    WVR & $118$ & $130 \pm 40$ & $(139)$ & $33 \pm 13$ & $(29)$ & $84 \pm 28$ & $(101)$ & $9.4 \pm 2.0$ & $(8)$\\
    SR & $10$ & $7.6 \pm 1.0$ & $(7)$ & $3.0 \pm 0.6$ & $(3)$ & $1.32 \pm 0.35$ & $(2)$ & $3.2 \pm 0.7$ & $(3)$\\
    \hline
  \end{tabular}
}
\end{center}
\caption{Results of the all-bins background fit for the cut-and-count analyses.
The numbers in parenthesis are the pre-fit background estimates using 
the most accurate theoretical cross-sections available (cf.  \secref{sec:samples}).
}
\label{allBinsFit_cut_and_count}
\end{table}

\begin{table}[p]
\begin{center}
\setlength{\tabcolsep}{0.05pc}
\renewcommand{\arraystretch}{1.5}
{\small
\begin{tabular}{|cc|rcl|c|cc|cc|cc|cc|}
\hline
\multicolumn{5}{|c|}{\multirow{2}{*}{\SRtNonep}} & \multirow{2}{*}{Obs.} & \multicolumn{8}{c|}{Fitted (estimated) background} \\
\cline{7-14}
\multicolumn{5}{|c|}{}  & & \multicolumn{2}{c|}{Total} & \multicolumn{2}{c|}{$t\bar{t}$} & \multicolumn{2}{c|}{$W$+jets} & \multicolumn{2}{c|}{Other} \\
\hline
\parbox[t]{3mm}{\multirow{5}{*}{\rotatebox[origin=c]{90}{$100 < \met $}}} & 
\parbox[t]{3mm}{\multirow{5}{*}{\rotatebox[origin=c]{90}{$<125 \GeV$}}} & 
   60  & $<$\mt$<$ & 90 \GeV + $b$-veto    & $1647$ & $1650 \pm 40$ & $(1896)$ & $550 \pm 130$ & $(619)$ & $990 \pm 150$ & $(1159)$ & $108 \pm 24$ & $(103)$\\ 
&& 60  & $<$\mt$<$ & 90 \GeV               & $3462$ & $3450 \pm 60$ & $(3402)$ & $3090 \pm 90$ & $(3005)$ & $180 \pm 40$ & $(212)$ & $180 \pm 50$ & $(184)$\\  
&& 90  & $<$\mt$<$ & 120 \GeV              & $1712$ & $1720 \pm 40$ & $(1693)$ & $1540 \pm 50$ & $(1487)$ & $96 \pm 23$ & $(115)$ & $84 \pm 27$ & $(89)$\\     
&& 120 & $<$\mt$<$ & 140 \GeV              & $313$ & $293 \pm 12$ & $(290)$ & $263 \pm 12$ & $(256)$ & $16 \pm 4$ & $(20)$ & $14 \pm 4$ & $(14)$\\             
&&     &    \mt$>$ & 140 \GeV              & $201$ & $222 \pm 11$ & $(229)$ & $202 \pm 12$ & $(206)$ & $7 \pm 4$ & $(10)$ & $12.8 \pm 3.1$ & $(13)$\\          
\hline
\parbox[t]{3mm}{\multirow{5}{*}{\rotatebox[origin=c]{90}{$125 < \met $}}} & 
\parbox[t]{3mm}{\multirow{5}{*}{\rotatebox[origin=c]{90}{$<150 \GeV$}}} & 
   60  & $<$\mt$<$ & 90 \GeV + $b$-veto    & $1081$ & $1081 \pm 33$ & $(1284)$ & $390 \pm 80$ & $(392)$ & $620 \pm 90$ & $(817)$ & $71 \pm 15$ & $(67)$\\     
&& 60  & $<$\mt$<$ & 90 \GeV               & $2018$ & $2020 \pm 40$ & $(1984)$ & $1800 \pm 60$ & $(1718)$ & $111 \pm 25$ & $(147)$ & $114 \pm 34$ & $(118)$\\ 
&& 90  & $<$\mt$<$ & 120 \GeV              & $768$ & $764 \pm 24$ & $(751)$ & $690 \pm 26$ & $(664)$ & $32 \pm 10$ & $(42)$ & $42 \pm 13$ & $(44)$\\          
&& 120 & $<$\mt$<$ & 140 \GeV              & $117$ & $130 \pm 8$ & $(134)$ & $117 \pm 8$ & $(118)$ & $6.0 \pm 2.5$ & $(9)$ & $6.7 \pm 1.9$ & $(7)$\\          
&&     &    \mt$>$ & 140 \GeV              & $163$ & $151 \pm 7$ & $(149)$ & $136 \pm 8$ & $(131)$ & $6.2 \pm 2.6$ & $(8)$ & $9.3 \pm 2.1$ & $(9)$\\          
\hline
\parbox[t]{3mm}{\multirow{5}{*}{\rotatebox[origin=c]{90}{$\met > 150$}}} & 
\parbox[t]{3mm}{\multirow{5}{*}{\rotatebox[origin=c]{90}{}}} & 
    60 & $<$\mt$<$ & 90 \GeV + $b$-veto    & $1742$ & $1740 \pm 40$ & $(2172)$ & $540 \pm 110$ & $(527)$ & $1080 \pm 130$ & $(1508)$ & $127 \pm 28$ & $(130)$\\ 
&&  60 & $<$\mt$<$ & 90 \GeV               & $2543$ & $2540 \pm 50$ & $(2554)$ & $2170 \pm 80$ & $(2099)$ & $200 \pm 40$ & $(271)$ & $170 \pm 50$ & $(182)$\\   
&&  90 & $<$\mt$<$ & 120 \GeV              & $647$ & $651 \pm 24$ & $(683)$ & $565 \pm 25$ & $(574)$ & $41 \pm 9$ & $(59)$ & $45 \pm 13$ & $(49)$\\             
&& 120 & $<$\mt$<$ & 140 \GeV              & $101$ & $95 \pm 6$ & $(97)$ & $82 \pm 6$ & $(82)$ & $4.5 \pm 1.4$ & $(6)$ & $8.3 \pm 2.0$ & $(9)$\\                
&&     & \mt$>$    & 140 \GeV              & $217$ & $223 \pm 12$ & $(234)$ & $192 \pm 13$ & $(195)$ & $8.8 \pm 2.8$ & $(14)$ & $23 \pm 5$ & $(24)$\\           
\hline
  \end{tabular}
}
\end{center}
\caption{Results of the all-bins background fit for the \SRtNonep\ analysis.
The numbers in parenthesis are the pre-fit background estimates using 
the most accurate theoretical cross-sections available (cf.  \secref{sec:samples}).
} 
\label{allBinsFit_tN1p}
\end{table}

\begin{table}[p]
\begin{center}
\setlength{\tabcolsep}{0.0pc}
{\small
  \begin{tabular}{|l|c|cc|cc|cc|cc|}
    \hline
    \multirow{2}{*}{\SRoneLoneBc} & \multirow{2}{*}{Obs.} & \multicolumn{8}{c|}{Fitted (estimated) background} \\
    \cline{3-10}
    & & \multicolumn{2}{c|}{Total} & \multicolumn{2}{c|}{$t\bar{t}$} & \multicolumn{2}{c|}{$W$+jets} & \multicolumn{2}{c|}{Other} \\
    \hline
TCR & $336$ & $338 \pm 18$ & $(409)$ & $192 \pm 35$ & $(250)$ & $74 \pm 21$ & $(87)$ & $72 \pm 18$ & $(72)$\\
 WCR & $1149$ & $1149 \pm 34$ & $(1333)$ & $62 \pm 17$ & $(78)$ & $940 \pm 50$ & $(1115)$ & $145^{+34}_{-29}$ & $(140)$\\
 VR & $74$ & $75 \pm 12$ & $(90)$ & $44 \pm 10$ & $(56)$ & $12.3 \pm 2.1$ & $(14)$ & $19^{+7}_{-7}$ & $(19)$\\
 \phantom{0}$6 < \pt ^{\ell} < 17$ \GeV & $6$ & $4.1 \pm 1.1$ & $(4)$ & $1.6 \pm 0.6$ & $(2)$ & $0.6 \pm 0.4$ & $(0.7)$ & $1.8^{+0.8}_{-0.7}$ & $(2)$\\
 $17 < \pt ^{\ell} < 28$ \GeV & $6$ & $3.7 \pm 0.9$ & $(4)$ & $1.7 \pm 0.7$ & $(2)$ & $0.76 \pm 0.24$ & $(0.9)$ & $1.2^{+0.5}_{-0.4}$ & $(1)$\\
 $28 < \pt ^{\ell} < 39$ \GeV & $4$ & $4.0 \pm 0.9$ & $(5)$ & $1.7 \pm 0.6$ & $(2)$ & $0.77 \pm 0.26$ & $(0.9)$ & $1.6^{+0.6}_{-0.6}$ & $(2)$\\
 $39 < \pt ^{\ell} < 50$ \GeV & $3$ & $2.8 \pm 0.7$ & $(3)$ & $1.2 \pm 0.7$ & $(1)$ & $0.82 \pm 0.34$ & $(0.9)$ & $0.80 \pm 0.29$ & $(0.8)$\\
    \hline
  \end{tabular}
}
\end{center}
\caption{Results of the all-bins background fit for the \SRoneLoneBc\ analysis.
The numbers in parenthesis are the pre-fit background estimates using 
the most accurate theoretical cross-sections available (cf.  \secref{sec:samples}).
}
\label{allBinsFit_SL1L1Bc}
\end{table}

\begin{table}[p]
\begin{center}
\setlength{\tabcolsep}{0.0pc}
{\small
  \begin{tabular}{|l|c|cc|cc|cc|cc|}
    \hline
    \multirow{2}{*}{\SRoneLoneBa} & \multirow{2}{*}{Obs.} & \multicolumn{8}{c|}{Fitted (estimated) background} \\
    \cline{3-10}
    & & \multicolumn{2}{c|}{Total} & \multicolumn{2}{c|}{$t\bar{t}$} & \multicolumn{2}{c|}{$W$+jets} & \multicolumn{2}{c|}{Other} \\
    \hline
 TCR & $136$ & $136 \pm 11$ & $(157)$ & $98 \pm 16$ & $(111)$ & $15 \pm 5$ & $(23)$ & $23^{+10}_{-9}$ & $(23)$\\
 WCR & $189$ & $189 \pm 14$ & $(259)$ & $26 \pm 7$ & $(29)$ & $121 \pm 19$ & $(188)$ & $42^{+11}_{-11}$ & $(42)$\\
 VR & $29$ & $28 \pm 5$ & $(31)$ & $19 \pm 4$ & $(22)$ & $1.7 \pm 0.5$ & $(3)$ & $6.7^{+2.4}_{-2.2}$ & $(7)$\\
 \phantom{0}$6 < \pt ^{\ell} < 17$ \GeV & $9$ & $9.2 \pm 2.0$ & $(10)$ & $5.3 \pm 1.3$ & $(6)$ & $0.61 \pm 0.21$ & $(0.9)$ & $3.3 \pm 1.5$ & $(3)$\\
 $17 < \pt ^{\ell} < 28 \GeV $ & $15$ & $10.9 \pm 1.7$ & $(12)$ & $7.1 \pm 1.5$ & $(8)$ & $1.00 \pm 0.32$ & $(1)$ & $2.8^{+1.1}_{-1.0}$ & $(3)$\\
 $28 < \pt ^{\ell} < 39 \GeV $ & $6$ & $10.9 \pm 1.7$ & $(12)$ & $7.1 \pm 1.4$ & $(8)$ & $0.87 \pm 0.31$ & $(1)$ & $2.9 \pm 1.0$ & $(3)$\\
 $39 < \pt ^{\ell} < 50$ \GeV & $9$ & $8.2 \pm 1.4$ & $(9)$ & $5.7 \pm 1.3$ & $(6)$ & $0.90 \pm 0.35$ & $(1)$ & $1.6^{+0.5}_{-0.5}$ & $(2)$\\
    \hline
  \end{tabular}
}
\end{center}
\caption{Results of the all-bins background fit for the \SRoneLoneBa\ analysis.
The numbers in parenthesis are the pre-fit background estimates using 
the most accurate theoretical cross-sections available (cf.  \secref{sec:samples}).
}
\label{allBinsFit_SL1L1Ba}
\end{table}

\begin{table}[p]
\begin{center}
\setlength{\tabcolsep}{0.0pc}
{\small
  \begin{tabular}{|l|c|cc|cc|cc|cc|}
    \hline
    \multirow{2}{*}{\SRoneLtwoBa} & \multirow{2}{*}{Obs.} & \multicolumn{8}{c|}{Fitted (estimated) background} \\
    \cline{3-10}
    & & \multicolumn{2}{c|}{Total} & \multicolumn{2}{c|}{$t\bar{t}$} & \multicolumn{2}{c|}{$W$+jets} & \multicolumn{2}{c|}{Other} \\
    \hline
 TCR & $390$ & $397 \pm 20$ & $(405)$ & $287 \pm 28$ & $(289)$ & $45 \pm 12$ & $(50)$ & $65^{+11}_{-10}$ & $(65)$\\
 WCR & $13724$ & $13720 \pm 130$ & $(14480)$ & $880 \pm 200$ & $(868)$ & $11740 \pm 330$ & $(12507)$ & $1100^{+190}_{-150}$ & $(1104)$\\
 VR & $76$ & $82 \pm 11$ & $(79)$ & $50 \pm 10$ & $(47)$ & $22.0 \pm 3.3$ & $(23)$ & $10 \pm 4$ & $(9)$\\
 \phantom{0}$83.3 < \amtTwo < 166.7$ \GeV & $111$ & $97 \pm 11$ & $(95)$ & $73 \pm 9$ & $(70)$ & $3.4 \pm 0.7$ & $(4)$ & $21^{+8}_{-5}$ & $(21)$\\
 $166.7 < \amtTwo < 250.0$ \GeV & $42$ & $35 \pm 4$ & $(33)$ & $20 \pm 4$ & $(18)$ & $7.8 \pm 1.2$ & $(8)$ & $6.8^{+2.3}_{-1.2}$ & $(7)$\\
 $250.0 < \amtTwo < 333.3$ \GeV & $2$ & $3.5 \pm 0.8$ & $(4)$ & $1.2 \pm 0.4$ & $(1)$ & $1.5 \pm 0.4$ & $(2)$ & $0.82^{+0.28}_{-0.22}$ & $(0.8)$\\
 $333.3 < \amtTwo < 416.7$ \GeV & $0$ & $0.58 \pm 0.28$ & $(0.6)$ & $0.21^{+0.22}_{-0.21}$ & $(0.2)$ & $0.31 \pm 0.12$ & $(0.3)$ & $0.064^{+0.028}_{-0.028}$ & $(0.1)$\\
 $416.7 < \amtTwo < 500.0$ \GeV & $0$ & $0.15 \pm 0.10$ & $(0.1)$ & $0.025^{+0.050}_{-0.025}$ & $(0.0)$ & $0.06 \pm 0.06$ & $(0.1)$ & $0.06 \pm 0.04$ & $(0.1)$\\
    \hline
  \end{tabular}
}
\end{center}
\caption{Results of the all-bins background fit for the \SRoneLtwoBa\ analysis.
The numbers in parenthesis are the pre-fit background estimates using 
the most accurate theoretical cross-sections available (cf.  \secref{sec:samples}).
}
\label{allBinsFit_SL1L2Ba}
\end{table}

\begin{table}[p]
\begin{center}
\setlength{\tabcolsep}{0.0pc}
{\small
  \begin{tabular}{|l|c|cc|cc|cc|cc|}
    \hline
    \multirow{2}{*}{\SRoneLtwoBc} & \multirow{2}{*}{Obs.} & \multicolumn{8}{c|}{Fitted (estimated) background} \\
    \cline{3-10}
    & & \multicolumn{2}{c|}{Total} & \multicolumn{2}{c|}{$t\bar{t}$} & \multicolumn{2}{c|}{$W$+jets} & \multicolumn{2}{c|}{Other} \\
    \hline
 TCR & $1111$ & $1108 \pm 34$ & $(1131)$ & $880 \pm 50$ & $(881)$ & $62 \pm 17$ & $(82)$ & $165^{+30}_{-28}$ & $(169)$\\
 WCR & $4089$ & $4090 \pm 60$ & $(4968)$ & $530 \pm 90$ & $(535)$ & $3140 \pm 130$ & $(4010)$ & $420^{+60}_{-50}$ & $(423)$\\
 VR & $12$ & $13.9 \pm 3.0$ & $(16)$ & $9.3 \pm 2.7$ & $(10)$ & $2.9 \pm 0.6$ & $(4)$ & $1.7^{+1.0}_{-0.5}$ & $(2)$\\
 \phantom{0}$83.3 < \amtTwo < 166.7$ \GeV & $11$ & $12.9 \pm 2.0$ & $(14)$ & $11.4 \pm 1.9$ & $(12)$ & $0.034 \pm 0.028$ & $(0.0)$ & $1.5 \pm 0.6$ & $(2)$\\
 $166.7 < \amtTwo < 250.0$ \GeV & $6$ & $5.8 \pm 1.2$ & $(6)$ & $4.0 \pm 1.1$ & $(4)$ & $0.78 \pm 0.26$ & $(1)$ & $1.07 \pm 0.31$ & $(1)$\\
 $250.0 < \amtTwo < 333.3$ \GeV & $4$ & $5.1 \pm 0.8$ & $(6)$ & $2.0 \pm 0.6$ & $(2)$ & $1.23 \pm 0.32$ & $(2)$ & $1.9 \pm 0.5$ & $(2)$\\
 $333.3 < \amtTwo < 416.7$ \GeV & $1$ & $1.7 \pm 0.6$ & $(2)$ & $0.47 \pm 0.28$ & $(0.6)$ & $0.29 \pm 0.15$ & $(0.4)$ & $0.9^{+0.4}_{-0.4}$ & $(0.9)$\\
 $416.7 < \amtTwo < 500.0$ \GeV & $0$ & $0.49 \pm 0.23$ & $(0.6)$ & $0.08^{+0.10}_{-0.08}$ & $(0.1)$ & $0.17 \pm 0.08$ & $(0.2)$ & $0.24^{+0.18}_{-0.17}$ & $(0.2)$\\
    \hline
  \end{tabular}
}
\end{center}
\caption{Results of the all-bins background fit for the \SRoneLtwoBc\ analysis.
The numbers in parenthesis are the pre-fit background estimates using 
the most accurate theoretical cross-sections available (cf.  \secref{sec:samples}).
}
\label{allBinsFit_SL1L2Bc}
\end{table}

\begin{table}[p]
\begin{center}
\setlength{\tabcolsep}{0.05pc}
\renewcommand{\arraystretch}{1.5}
{\small
\begin{tabular}{|cc|rcl|c|cc|cc|cc|cc|}
                     \hline
\multicolumn{5}{|c|}{\multirow{2}{*}{\SRbCvW}} & \multirow{2}{*}{Obs.} & \multicolumn{8}{c|}{Fitted (estimated) background} \\
\cline{7-14}
\multicolumn{5}{|c|}{}  & & \multicolumn{2}{c|}{Total} & \multicolumn{2}{c|}{$t\bar{t}$} & \multicolumn{2}{c|}{$W$+jets} & \multicolumn{2}{c|}{Other} \\
\hline
\parbox[t]{3mm}{\multirow{3}{*}{\rotatebox[origin=c]{90}{$\amtTwo >$}}} & 
\parbox[t]{3mm}{\multirow{3}{*}{\rotatebox[origin=c]{90}{80 \GeV}}} & 
&&&&&&&&&&&\\ 
&& 60&$<$\mt$<$&90 \GeV + $b$-veto  & $1069$ & $1067 \pm 33$ & $(1405)$ & $103 \pm 34$ & $(101)$ & $890 \pm 50$ & $(1222)$ & $76 \pm 24$ & $(80)$\\
&&&&&&&&&&&&&\\ 
\hline
\parbox[t]{3mm}{\multirow{3}{*}{\rotatebox[origin=c]{90}{$80 < \amtTwo $}}} & 
\parbox[t]{3mm}{\multirow{3}{*}{\rotatebox[origin=c]{90}{$<175 \GeV$}}} & 
60&$<$\mt$<$&90 \GeV          & $65$ & $61 \pm 6$ & $(60)$ & $57 \pm 6$ & $(55)$ & $1.1 \pm 0.4$ & $(2)$ & $3.1 \pm 1.1$ & $(4)$\\                
&& 90&$<$\mt$<$&120 \GeV      & $17$ & $16.8 \pm 2.1$ & $(17)$ & $15.8 \pm 2.1$ & $(15)$ & $0.26 \pm 0.29$ & $(0.4)$ & $0.76 \pm 0.30$ & $(0.9)$\\
&& &\mt$>$&120 \GeV           & $6$ & $7.5 \pm 1.1$ & $(8)$ & $6.6 \pm 1.2$ & $(7)$ & $0.11 \pm 0.11$ & $(0.2)$ & $0.71 \pm 0.27$ & $(0.8)$\\     
\hline
\parbox[t]{3mm}{\multirow{3}{*}{\rotatebox[origin=c]{90}{$175 < \amtTwo $}}} & 
\parbox[t]{3mm}{\multirow{3}{*}{\rotatebox[origin=c]{90}{$<250 \GeV$}}} & 
60&$<$\mt$<$&90 \GeV          & $33$ & $33 \pm 4$ & $(37)$ & $20.6 \pm 3.5$ & $(21)$ & $6.1 \pm 1.6$ & $(9)$ & $5.9 \pm 2.0$ & $(7)$\\    
&& 90&$<$\mt$<$&120 \GeV      & $10$ & $10.8 \pm 1.3$ & $(12)$ & $7.7 \pm 1.2$ & $(8)$ & $1.5 \pm 0.8$ & $(2)$ & $1.6 \pm 0.6$ & $(2)$\\  
&& &\mt$>$&120 \GeV           & $10$ & $7.0 \pm 1.0$ & $(7)$ & $4.6 \pm 0.9$ & $(4)$ & $0.45 \pm 0.19$ & $(0.7)$ & $1.9 \pm 0.6$ & $(2)$\\
\hline
\parbox[t]{3mm}{\multirow{3}{*}{\rotatebox[origin=c]{90}{$\amtTwo >$}}} & 
\parbox[t]{3mm}{\multirow{3}{*}{\rotatebox[origin=c]{90}{250 \GeV}}} & 
60&$<$\mt$<$&90 \GeV          & $65$ & $72 \pm 5$ & $(82)$ & $40 \pm 6$ & $(40)$ & $16 \pm 4$ & $(24)$ & $15 \pm 5$ & $(18)$\\            
&& 90&$<$\mt$<$&120 \GeV      & $16$ & $18.2 \pm 2.3$ & $(21)$ & $10.6 \pm 2.3$ & $(11)$ & $3.5 \pm 1.4$ & $(5)$ & $4.1 \pm 1.6$ & $(5)$\\
&& &\mt$>$&120 \GeV           & $9$ & $8.4 \pm 1.3$ & $(10)$ & $3.6 \pm 0.9$ & $(4)$ & $2.2 \pm 0.8$ & $(3)$ & $2.6 \pm 0.8$ & $(3)$\\    
\hline
  \end{tabular}
}
\end{center}
\caption{Results of the all-bins background fit for the \SRbCvW\ analysis.
The numbers in parenthesis are the pre-fit background estimates using 
the most accurate theoretical cross-sections available (cf.  \secref{sec:samples}).
}
\label{allBinsFit_bCvW}
\end{table}

\begin{table}[p]
\begin{center}
\setlength{\tabcolsep}{0.05pc}
\renewcommand{\arraystretch}{1.5}
{\small
  \begin{tabular}{|cc|rcl|c|cc|cc|cc|cc|}
    \hline
    \multicolumn{5}{|c|}{\multirow{2}{*}{\SRbCone}} & \multirow{2}{*}{Obs.} & \multicolumn{8}{c|}{Fitted (estimated) background} \\
    \cline{7-14}
    \multicolumn{5}{|c|}{}  & & \multicolumn{2}{c|}{Total} & \multicolumn{2}{c|}{$t\bar{t}$} & \multicolumn{2}{c|}{$W$+jets} & \multicolumn{2}{c|}{Other} \\
    \hline
    \parbox[t]{3mm}{\multirow{4}{*}{\rotatebox[origin=c]{90}{$80 < \amtTwo $}}} & 
    \parbox[t]{3mm}{\multirow{4}{*}{\rotatebox[origin=c]{90}{$<175 \GeV$}}} & 
    60&$<$\mt$<$&90 \GeV + $b$-veto   & $1345$ & $1340 \pm 40$ & $(1702)$ & $270 \pm 60$ & $(260)$ & $990 \pm 80$ & $(1354)$ & $86 \pm 24$ & $(86)$\\    
    && 60&$<$\mt$<$&90 \GeV           & $2229$ & $2220 \pm 90$ & $(2216)$ & $1920 \pm 90$ & $(1846)$ & $161 \pm 30$ & $(220)$ & $150 \pm 40$ & $(150)$\\ 
    && 90&$<$\mt$<$&120 \GeV          & $583$ & $590 \pm 40$ & $(602)$ & $516 \pm 31$ & $(518)$ & $30 \pm 6$ & $(42)$ & $41 \pm 12$ & $(42)$\\           
    && &\mt$>$&120 \GeV               & $361$ & $362 \pm 16$ & $(362)$ & $332 \pm 16$ & $(328)$ & $8.8 \pm 2.4$ & $(13)$ & $21 \pm 5$ & $(21)$\\         
    \hline
    \parbox[t]{3mm}{\multirow{4}{*}{\rotatebox[origin=c]{90}{$175 < \amtTwo $}}} & 
    \parbox[t]{3mm}{\multirow{4}{*}{\rotatebox[origin=c]{90}{$<250 \GeV$}}} & 
    60&$<$\mt$<$&90 \GeV + $b$-veto   & $705$ & $705 \pm 27$ & $(864)$ & $62 \pm 13$ & $(68)$ & $594 \pm 34$ & $(749)$ & $48 \pm 14$ & $(47)$\\   
    && 60&$<$\mt$<$&90 \GeV           & $547$ & $551 \pm 29$ & $(626)$ & $338 \pm 27$ & $(375)$ & $140 \pm 23$ & $(177)$ & $74 \pm 19$ & $(73)$\\ 
    && 90&$<$\mt$<$&120 \GeV          & $144$ & $141 \pm 13$ & $(152)$ & $93 \pm 10$ & $(99)$ & $30 \pm 7$ & $(37)$ & $17 \pm 5$ & $(16)$\\       
    && &\mt$>$&120 \GeV               & $78$ & $77 \pm 5$ & $(82)$ & $49 \pm 5$ & $(52)$ & $12.4 \pm 3.5$ & $(15)$ & $15.3 \pm 3.0$ & $(15)$\\    
    \hline
    \parbox[t]{3mm}{\multirow{4}{*}{\rotatebox[origin=c]{90}{$\amtTwo > 250 \GeV$}}} & 
    \parbox[t]{3mm}{\multirow{4}{*}{\rotatebox[origin=c]{90}{}}} & 
    60&$<$\mt$<$&90 \GeV+ $b$-veto    & $260$ & $260 \pm 16$ & $(344)$ & $17 \pm 5$ & $(21)$ & $222 \pm 19$ & $(302)$ & $21 \pm 7$ & $(20)$\\      
    && 60&$<$\mt$<$&90 \GeV           & $241$ & $239 \pm 13$ & $(285)$ & $117 \pm 19$ & $(136)$ & $75 \pm 14$ & $(102)$ & $47 \pm 12$ & $(47)$\\   
    && 90&$<$\mt$<$&120 \GeV          & $61$ & $65 \pm 5$ & $(78)$ & $32 \pm 6$ & $(38)$ & $20 \pm 4$ & $(27)$ & $13 \pm 4$ & $(13)$\\             
    && &\mt$>$&120 \GeV               & $29$ & $26.3 \pm 2.0$ & $(31)$ & $10.5 \pm 2.2$ & $(12)$ & $7.0 \pm 1.5$ & $(10)$ & $8.8 \pm 1.9$ & $(9)$\\
    \hline
  \end{tabular}
}
\end{center}
\caption{Results of the all-bins background fit for the \SRbCone\ analysis.
The numbers in parenthesis are the pre-fit background estimates using 
the most accurate theoretical cross-sections available (cf.  \secref{sec:samples}).
} 
\label{allBinsFit_bC1}
\end{table}

\begin{table}[p]
\begin{center}
\setlength{\tabcolsep}{0.05pc}
\renewcommand{\arraystretch}{1.5}
{\small
  \begin{tabular}{|cc|rcl|c|cc|cc|cc|cc|}
    \hline
    \multicolumn{5}{|c|}{\multirow{2}{*}{\SRbWN}} & \multirow{2}{*}{Obs.} & \multicolumn{8}{c|}{Fitted (estimated) background} \\
    \cline{7-14}
    \multicolumn{5}{|c|}{}  & & \multicolumn{2}{c|}{Total} & \multicolumn{2}{c|}{$t\bar{t}$} & \multicolumn{2}{c|}{$W$+jets} & \multicolumn{2}{c|}{Other} \\
    \hline
    \parbox[t]{3mm}{\multirow{4}{*}{\rotatebox[origin=c]{90}{$80 < \amtTwo $}}} & 
    \parbox[t]{3mm}{\multirow{4}{*}{\rotatebox[origin=c]{90}{$<90 \GeV$}}} & 
    60&$<$\mt$<$&90 \GeV + $b$-veto   & $8$ & $9.8 \pm 2.0$ & $(13)$ & $2.6 \pm 0.8$ & $(3)$ & $6.3 \pm 2.0$ & $(9)$ & $0.8 \pm 0.5$ & $(0.7)$\\      
    && 60&$<$\mt$<$&90 \GeV           & $3$ & $5.9 \pm 0.8$ & $(6)$ & $5.1 \pm 0.8$ & $(5)$ & $0.52 \pm 0.31$ & $(0.2)$ & $0.26 \pm 0.22$ & $(0.4)$\\ 
    && 90&$<$\mt$<$&120  \GeV         & $12$ & $14.6 \pm 1.7$ & $(14)$ & $12.0 \pm 1.7$ & $(11)$ & $1.6 \pm 0.7$ & $(2)$ & $1.0 \pm 0.6$ & $(1)$\\    
    && & \mt$>$&120      \GeV         & $8$ & $7.3 \pm 1.3$ & $(7)$ & $6.5 \pm 1.3$ & $(7)$ & $0.13 \pm 0.20$ & $(0.1)$ & $0.69 \pm 0.18$ & $(0.5)$\\ 
    \hline
    \parbox[t]{3mm}{\multirow{4}{*}{\rotatebox[origin=c]{90}{$90 < \amtTwo $}}} & 
    \parbox[t]{3mm}{\multirow{4}{*}{\rotatebox[origin=c]{90}{$<100 \GeV$}}} & 
    60&$<$\mt$<$&90 \GeV + $b$-veto   & $12$ & $16.8 \pm 2.9$ & $(21)$ & $4.1 \pm 1.1$ & $(4)$ & $11.8 \pm 2.8$ & $(16)$ & $1.0 \pm 0.6$ & $(0.8)$\\ 
    && 60&$<$\mt$<$&90 \GeV           & $14$ & $16.1 \pm 1.7$ & $(16)$ & $13.1 \pm 1.6$ & $(13)$ & $1.1 \pm 0.5$ & $(2)$ & $1.9 \pm 0.6$ & $(2)$\\   
    && 90&$<$\mt$<$&120 \GeV          & $29$ & $30.2 \pm 2.3$ & $(32)$ & $24.7 \pm 2.2$ & $(26)$ & $3.9 \pm 1.3$ & $(3)$ & $1.6 \pm 1.2$ & $(3)$\\   
    && &\mt$>$&120 \GeV               & $22$ & $26.0 \pm 2.1$ & $(26)$ & $22.9 \pm 2.3$ & $(23)$ & $1.1 \pm 0.6$ & $(1)$ & $2.0 \pm 0.7$ & $(2)$\\   
    \hline
    \parbox[t]{3mm}{\multirow{4}{*}{\rotatebox[origin=c]{90}{$100 < \amtTwo $}}} & 
    \parbox[t]{3mm}{\multirow{4}{*}{\rotatebox[origin=c]{90}{$<120 \GeV$}}} & 
    60&$<$\mt$<$&90 \GeV + $b$-veto   & $36$ & $35 \pm 4$ & $(34)$ & $7.0 \pm 1.9$ & $(7)$ & $26 \pm 4$ & $(26)$ & $2.1 \pm 0.7$ & $(2)$\\         
    && 60&$<$\mt$<$&90 \GeV           & $29$ & $32.4 \pm 2.3$ & $(32)$ & $27.9 \pm 2.2$ & $(28)$ & $3.0 \pm 1.2$ & $(3)$ & $1.5 \pm 0.9$ & $(1)$\\ 
    && 90&$<$\mt$<$&120 \GeV          & $57$ & $59 \pm 4$ & $(61)$ & $51 \pm 4$ & $(51)$ & $3.7 \pm 2.0$ & $(5)$ & $3.9 \pm 2.1$ & $(5)$\\         
    && &\mt$>$&120 \GeV               & $74$ & $70 \pm 5$ & $(66)$ & $62 \pm 6$ & $(58)$ & $4.4 \pm 1.4$ & $(4)$ & $3.9 \pm 1.2$ & $(4)$\\         
    \hline
    \parbox[t]{3mm}{\multirow{4}{*}{\rotatebox[origin=c]{90}{\amtTwo $> 120$ \GeV}}} & 
    \parbox[t]{3mm}{\multirow{4}{*}{\rotatebox[origin=c]{90}{}}} & 
    60&$<$\mt$<$&90 \GeV + $b$-veto   & $114$ & $112 \pm 10$ & $(115)$ & $14.4 \pm 3.3$ & $(13)$ & $90 \pm 11$ & $(95)$ & $8.0 \pm 1.9$ & $(5)$\\ 
    && 60&$<$\mt$<$&90 \GeV           & $108$ & $88 \pm 5$ & $(82)$ & $66 \pm 5$ & $(62)$ & $16 \pm 4$ & $(15)$ & $6.2 \pm 1.7$ & $(5)$\\         
    && 90&$<$\mt$<$&120 \GeV          & $160$ & $162 \pm 10$ & $(167)$ & $131 \pm 10$ & $(131)$ & $20 \pm 5$ & $(22)$ & $11 \pm 4$ & $(14)$\\     
    && &\mt$>$&120 \GeV               & $281$ & $281 \pm 12$ & $(277)$ & $216 \pm 15$ & $(211)$ & $33 \pm 8$ & $(32)$ & $32 \pm 7$ & $(33)$\\     
    \hline
  \end{tabular}
}
\end{center}
\caption{Results of the all-bins background fit for the \SRbWN\ analysis.
The numbers in parenthesis are the pre-fit background estimates using 
the most accurate theoretical cross-sections available (cf.  \secref{sec:samples}).
} 
\label{allBinsFit_bWN}
\end{table}

\onecolumn
\clearpage
\begin{flushleft}
{\Large The ATLAS Collaboration}

\bigskip

G.~Aad$^{\rm 84}$,
B.~Abbott$^{\rm 112}$,
J.~Abdallah$^{\rm 152}$,
S.~Abdel~Khalek$^{\rm 116}$,
O.~Abdinov$^{\rm 11}$,
R.~Aben$^{\rm 106}$,
B.~Abi$^{\rm 113}$,
M.~Abolins$^{\rm 89}$,
O.S.~AbouZeid$^{\rm 159}$,
H.~Abramowicz$^{\rm 154}$
% ...
\bigskip
\\
% ...
$^{11}$ Institute of Physics, Azerbaijan Academy of Sciences, Baku, Azerbaijan\\
% ...
$^{84}$ CPPM, Aix-Marseille Universit{\'e} and CNRS/IN2P3, Marseille, France\\
% ...
$^{89}$ Department of Physics and Astronomy, Michigan State University, East Lansing MI, United States of America\\
% ...
$^{106}$ Nikhef National Institute for Subatomic Physics and University of Amsterdam, Amsterdam, Netherlands\\
% ...
$^{112}$ Homer L. Dodge Department of Physics and Astronomy, University of Oklahoma, Norman OK, United States of America\\
$^{113}$ Department of Physics, Oklahoma State University, Stillwater OK, United States of America\\
% ...
$^{116}$ LAL, Universit{\'e} Paris-Sud and CNRS/IN2P3, Orsay, France\\
% ...
$^{152}$ Institute of Physics, Academia Sinica, Taipei, Taiwan\\
% ...
$^{154}$ Raymond and Beverly Sackler School of Physics and Astronomy, Tel Aviv University, Tel Aviv, Israel\\
% ...
$^{159}$ Department of Physics, University of Toronto, Toronto ON, Canada\\
% ...
\end{flushleft}

\end{document}